\documentclass[intlimits,a4paper,11pt]{article}
\usepackage{jheppub}
\usepackage{graphicx}
\usepackage{amssymb}
\usepackage{amsmath}
\usepackage[latin1]{inputenc} 
\usepackage[english]{babel}
\usepackage{axodraw4j}
\usepackage{color}

\title{Impact-parameter dependent nuclear parton distribution functions: \texttt{EPS09s} and \texttt{EKS98s} and their applications in nuclear hard-processes}

\author[a,b]{Ilkka Helenius,}
\author[a,b]{Kari J. Eskola,}
\author[a,c]{Heli Honkanen} 
\author[d,e]{Carlos A. Salgado}

\affiliation[a]{Department of Physics, P.O. Box 35, FI-40014 University of Jyv\"askyl\"a, Finland}
\affiliation[b]{Helsinki Institute of Physics, P.O. Box 64, FIN-00014 University of Helsinki, Finland}
\affiliation[c]{The Pennsylvania State University, 104 Davey Lab, University Park, PA 16802, USA}
\affiliation[d]{Departamento de F\'\i sica de Part\'\i culas and IGFAE, Universidade de Santiago de Compostela, Galicia-Spain}
\affiliation[e]{Physics Department, Theory Unit, CERN, CH-1211 Gen\`eve 23, Switzerland}

\emailAdd{ilkka.helenius@jyu.fi}
\emailAdd{kari.eskola@phys.jyu.fi}
\emailAdd{hmh17@psu.edu}
\emailAdd{carlos.salgado@usc.es}

\abstract{
We determine the spatial (impact parameter) dependence of nuclear parton distribution functions (nPDFs) using the $A$-dependence of the spatially independent (averaged) global fits EPS09 and EKS98. We work under the assumption that the spatial dependence can be formulated as a power series of the nuclear thickness functions $T_A$. 
To reproduce the $A$-dependence over the entire $x$ range we need terms up to $[T_A]^4$.
As an outcome, we release two sets, \texttt{EPS09s} (LO, NLO, error sets) and \texttt{EKS98s}, of spatially dependent nPDFs for public use. We also discuss the implementation of these into the existing calculations.
With our results, the centrality dependence of nuclear hard-process observables can be studied consistently with the globally fitted nPDFs for the first time. 
As an application, we first calculate the LO nuclear modification factor $R^{\rm 1jet}_{AA}$ for primary partonic-jet production in different centrality classes in Au+Au collisions at RHIC and Pb+Pb collisions at LHC. Also the corresponding central-to-peripheral ratios $R_{CP}^{\rm 1jet}$ are studied.
We also calculate the LO and NLO nuclear modification factors for single inclusive neutral pion production, $R_{\rm dAu}^{\pi^0}$, at mid- and forward rapidities in different centrality classes in d+Au collisions at RHIC. In particular, we show that our results are compatible with the PHENIX mid-rapidity data within the overall normalization uncertainties given by the experiment. 
Finally,  we show our predictions for the corresponding modifications $R_{\rm pPb}^{\pi^0}$ in the forthcoming p+Pb collisions at LHC. 
}

\arxivnumber{xxxx.yyyy}

\begin{document}
\maketitle
\setcounter{footnote}{0}
\renewcommand{\thefootnote}{\arabic{footnote}} 	

\section{Introduction}

In a high-energy hadronic or nuclear collision of particles $A$ and $B$ the inclusive cross sections for hard processes involving a large interaction scale $Q^2\gg\Lambda^2_{\rm QCD}$ can be computed using the QCD collinear factorization theorem \cite{Collins:1989gx,Brock:1993sz}, 
\begin{equation}
\mathrm{d} \sigma^{AB \rightarrow k + X} = \sum\limits_{i,j,X'} f_{i}^A(Q^2) \otimes f_{j}^B(Q^2) \otimes  \mathrm{d}\hat{\sigma}^{ij\rightarrow k + X'} + {\cal O}(1/Q^2), 
\label{eq:sigmaAB1}
\end{equation}
where $\mathrm{d}\hat{\sigma}$ represents the perturbatively computable partonic pieces (cross sections in lowest order), and $f_i^A$ ($f_j^B$) is the parton distribution function (PDFs) for a given parton flavor $i$ in the colliding particle
 $A$ (and correspondingly for the flavor $j$ in $B$). The PDFs are universal, process-independent functions of nonperturbative origin, whose evolution in the scale $Q^2$ can, however, be obtained from the DGLAP equations \cite{Dokshitzer:1977sg,Gribov:1972ri,Gribov:1972rt,Altarelli:1977zs} derived from perturbative QCD. 

A precise knowledge of the universal PDFs is thus vital for interpreting any hard-process results at the present colliders BNL-RHIC and CERN-LHC. This holds as well for proton-proton collisions as for proton-nucleus and nucleus-nucleus collisions. To determine the nonperturbative input in the PDFs, one has developed global analyses which exploit a multitude of experimental hard-process data and the DGLAP evolution.
Excellent fits for the free proton PDFs have been obtained, and sets like CT10 \cite{Lai:2010vv}, MSTW \cite{Martin:2009iq}, and NNPDF2.0 \cite{Ball:2010de} are nowadays available. 

\begin{figure}[htbp]
\centering
\includegraphics[width=15cm]{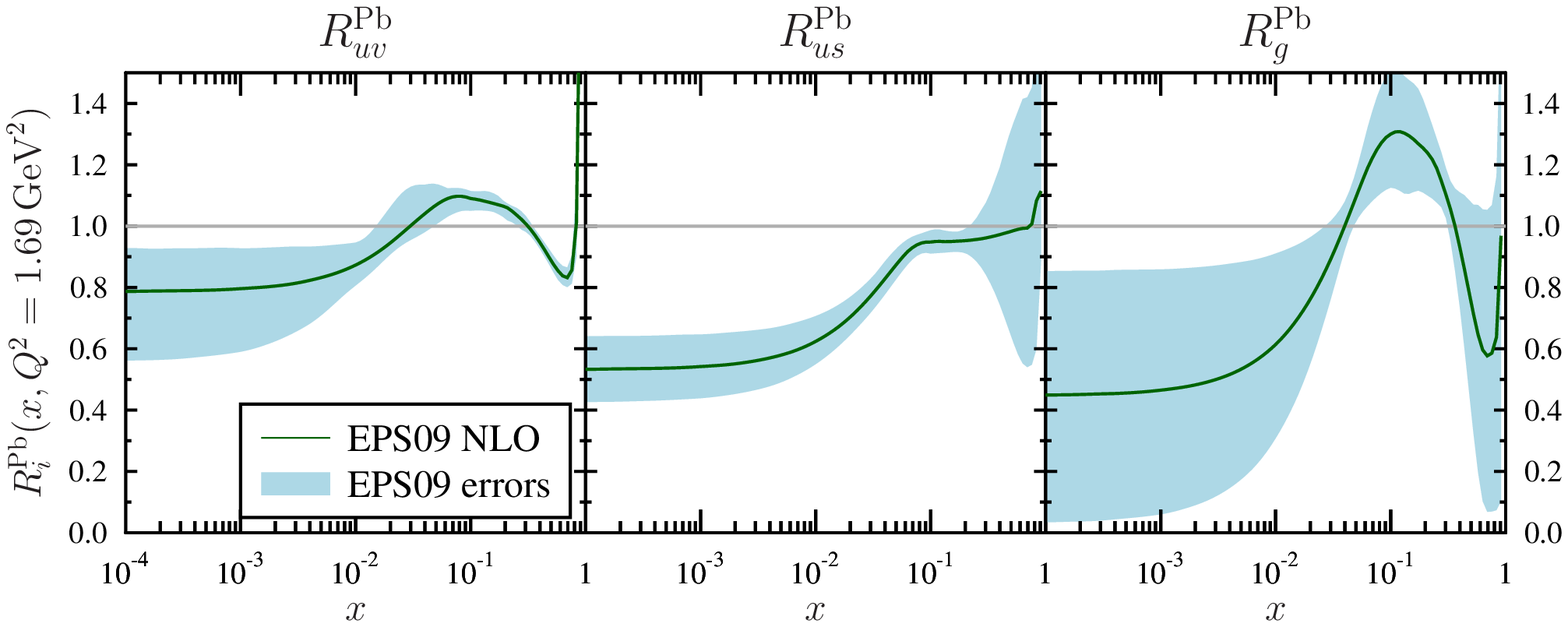}
\includegraphics[width=15cm]{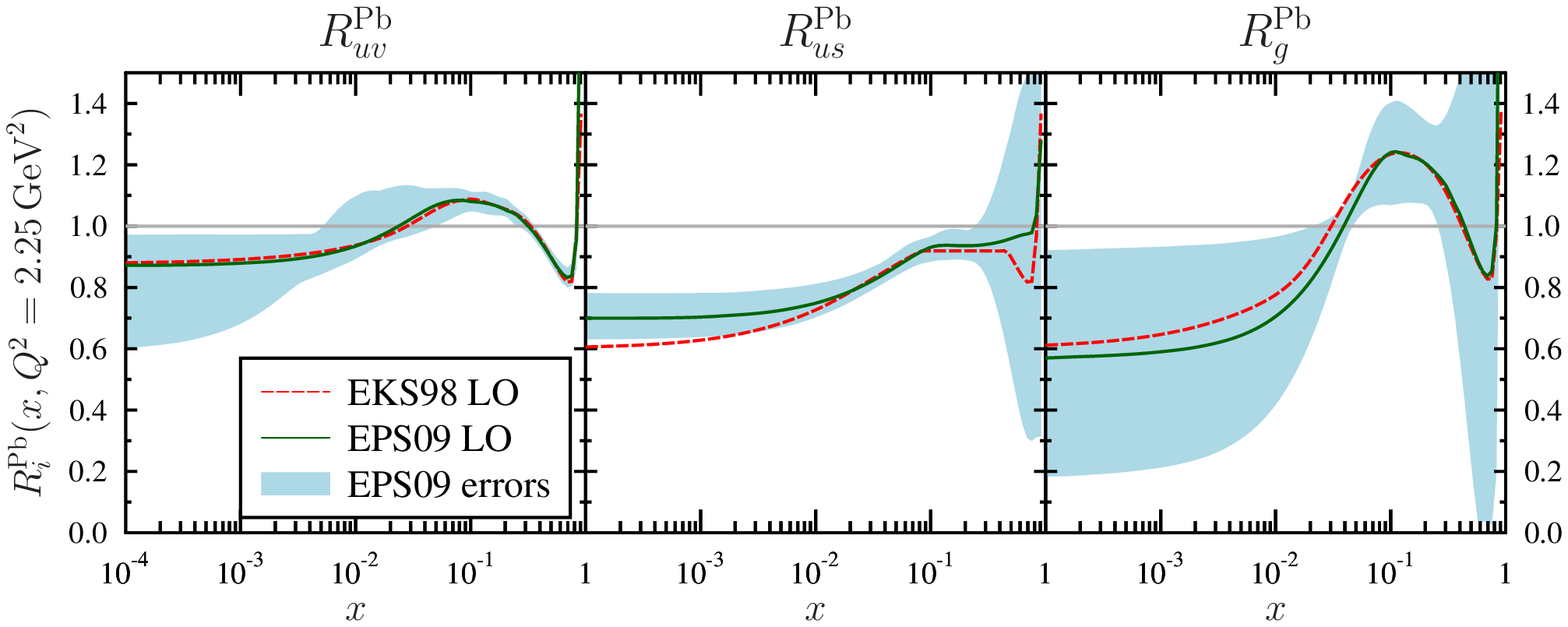}
\caption{The nuclear modifications and their uncertainties in a lead nucleus ($A=208$) for different parton flavors from EPS09NLO at the EPS09 initial scale $Q_0^2=1.69$~GeV$^2$ (upper panel), and from EPS09LO and EKS98 at the EKS98 initial scale $Q_0^2=2.25$~GeV$^2$ (lower panel).}
\label{fig:EPS09ebands}
\end{figure}

It is well known that the PDFs of nucleons bound to a nucleus, the nuclear PDFs (nPDFs), are modified relative to the free-nucleon PDFs. Analogously to the free-proton case, global DGLAP analyses have been developed also for the nPDFs. 
The further complication with these is that in addition to the usual $x$ and $Q^2$ dependences also the nuclear mass-number ($A$) dependence of the PDFs needs to be dealt with. The global nPDF fits have so far  resulted in leading-order (LO) nPDF sets EKS98 \cite{Eskola:1998df}, HKM \cite{Hirai:2001np} and HKN04 \cite{Hirai:2004wq}, and next-to-leading order (NLO) sets nDS \cite{deFlorian:2003qf}, HKN07 \cite{Hirai:2007sx}, EPS09 \cite{Eskola:2009uj}, nCTEQ \cite{Schienbein:2009kk,Kovarik:2010uv}, and DSZS \cite{deFlorian:2011fp}. Importantly, and similarly to the free proton case, with the error sets of EPS09 (and similar sets in DSZS), one can nowadays quantify how the uncertainties remaining in the nPDFs, illustrated in Fig.~\ref{fig:EPS09ebands}, are transmitted to the nuclear hard-process cross-sections. 

The global analyses mentioned above have all considered only the spatially averaged nPDFs, probed in minimum-bias nuclear collisions with no cuts on the collision centrality (impact parameter). 
In particular, as the modest amount of available nuclear hard-process data severely limits the number of possible fit parameters, it has so far been impossible to embed the spatial dependence, or the impact-parameter dependence, of the nPDFs directly into the global analysis. An obvious drawback with the globally analysed nPDFs then is that it has not been possible to consistently compute nuclear hard-process cross-sections in different centrality classes. 

The purpose of this study is to consider this problem by pinning down the spatial dependence of the nPDFs, i.e. the dependence of the nuclear modifications of the PDFs on the nucleon's position in a nucleus. 
We do this in a manner which is for the first time fully consistent with the nPDFs from a global analysis. Earlier attempts to this direction, lacking however such a consistency, can be found in Refs. \cite{Eskola:1991ec,Emel'yanov:1999bn,Klein:2003dj,Vogt:2004hf}. A further motivation for the current study is the Gribov-Glauber modeling of nuclear shadowing, reviewed lately in Ref. \cite{Frankfurt:2011cs}, whose output nPDFs are not a result of a global analysis like EPS09 but which have so far been the only ones where the spatial dependence arises in a self-consistent manner from modeling the physics origin of the nuclear effects.
On the experimental side, the current study is inspired by e.g. the measurements of single hadron production \cite{Arsene:2004ux,Adams:2004dv,Adler:2006wg,Adler:2006xd,Adams:2006nd,Adler:2007aa,Abelev:2009hx} and $J/\Psi$ production \cite{Adare:2010fn,Adare:2012qf} in different centrality classes in d+Au collisions at RHIC, as well as
by the hard-process measurements in the forthcoming p+Pb collisions at the LHC. Also the theoretical modeling of the $J/\Psi$
production discussed recently in Ref.~\cite{Nagle:2010ix} has motivated our study.

Our basic idea for uncovering the spatial dependence in the EKS98 and EPS09 nPDFs is straightforward: We first introduce the spatial dependence of the nuclear modification to the nPDF of each parton type $i$ in each nucleus $A$ at each $x$ and $Q^2$ in terms of a power series of the standard nuclear thickness functions $T_A$. Then, we determine the coefficients of each power of $T_A$ by exploiting the $A$-dependence of the EPS09 and EKS98 nPDFs (these sets, through the global fits, represent the experimental data here). As an output, we provide the numerical routines named \verb=EPS09s= (LO and NLO as well as error sets for both) and \verb=EKS98s= for computing the spatially dependent nPDFs which -- simultaneously for all nuclei considered -- normalize to the corresponding spatially independent EPS09 and EKS98 nPDFs. These new sets will be downloadable at the link \cite{downloadpage}.

As concrete examples of how to easily implement our spatially dependent nPDFs and the nuclear collision geometry in the computation of nuclear hard-process cross-sections in different centrality bins, we first discuss the centrality dependence of the LO nuclear modification ratios $R_{AA}^{\rm 1jet}(p_T)$ of primary partonic-jet production in Au+Au collisions at RHIC and Pb+Pb collisions at the LHC. We also study the nuclear modification factors of inclusive $\pi^0$ production, $R_{\rm dAu}^{\pi^0}$, in d+Au collisions at RHIC and $R_{\rm pPb}^{\pi^0}$ in p+Pb collisions at the LHC, both at mid- and forward rapidities, and considering both the NLO and LO cases. For $R_{\rm dAu}^{\pi^0}$ we also make, to our knowledge, a first comparison with the PHENIX centrality dependent data \cite{Adler:2006wg} where the overall normalization errors of the data are accounted for in detail. Due to the planned p+Pb program at the LHC, the ratio $R_{\rm pPb}^{h}(p_T)$ for single hadron production has been of growing interest recently \cite{QuirogaArias:2010wh,Albacete:2010bs,Salgado:2011wc,Arleo:2011gc,Barnafoldi:2011px,Xu:2012au}, and we will show also here how interesting and useful this ratio would be from the point of view of constraining the nPDFs further.

The paper is organized as follows: In Sec.~\ref{sec:framework} we define the model framework and explain the fitting procedure. In Sec.~\ref{sec:results} we show the results for the spatially dependent nuclear modifications of PDFs. Also a comparison with selected other works is presented here. Applications of our results are discussed in Sec.~\ref{sec:applications}. For clarity, a summary of the standard elements used in the applications here, the formulation of the nuclear collision geometry, different overlap functions and the optical Glauber model, is given in the Appendix \ref{sec:geometry}.  

\section{The Analysis Framework}
\label{sec:framework}

\subsection{Definitions of the Nuclear Modifications}
\label{sec:def_nucl_mod}

First we need to define how we introduce the spatial dependence to the nPDFs in terms of the hard-process cross-sections. 
Let us start with the usual spatially averaged nPDFs. The number distribution of an observable $k$ produced in a collision of nuclei $A$ and $B$ at an impact parameter \textbf{b} is given by 
\begin{equation}
\mathrm{d}N^{AB\rightarrow k + X}(\mathbf{b}) = T_{AB}(\mathbf{b})\mathrm{d} \sigma^{AB \rightarrow k + X},
\label{eq:dNAB}
\end{equation}
where $T_{AB}(\mathbf{b})$ is the standard nuclear overlap function normalized to $AB$ (cf. Eq.~(\ref{eq:taa}) in App.~\ref{sec:glauber}, see the nuclear collision geometry in Fig.~\ref{fig:transversePlane}), and $\mathrm{d} \sigma^{AB \rightarrow k + X}$ is the \textbf{b}-independent inclusive hard cross-section of Eq.~(\ref{eq:sigmaAB1}) containing the nPDFs and perturbative pieces. 
The spatially averaged nPDFs in a nucleus $A$ with $Z$ protons and $A-Z$ neutrons are now given by
\begin{equation}
f_i^A(x,Q^2) = \frac{Z}{A}f_i^{p/A}(x,Q^2) + \frac{A-Z}{A}f_i^{n/A}(x,Q^2),
\label{eq:fullnPDFs}
\end{equation}
where the nPDFs of a bound neutron, $f_i^{n/A}$, may be (approximately) obtained from those of the bound proton, $f_i^{p/A}$, by using the isospin symmetry (see \cite{Eskola:2009uj}). As in EKS98 and EPS09, we define the nPDF for each parton flavor in terms of the spatially averaged nuclear modification $R_i^A(x,Q^2)$ and the corresponding free proton PDF $f_i^p(x,Q^2)$, 
\begin{equation}
f_i^{p/A}(x,Q^2) \equiv R_i^{A}(x,Q^2) f_i^p(x,Q^2).
\end{equation}
To lighten the notations, we express the nPDFs in Eq.~(\ref{eq:fullnPDFs}) as 
\begin{equation}
f_i^A(x,Q^2) = \frac{1}{A}\sum\limits_{N} R_i^{N/A}(x,Q^2) f_i^N(x,Q^2),
\end{equation}
where the sum runs over all the nucleons $N=1,\ldots,A$.

Decomposing the $T_{AB}$ into the standard nuclear thickness functions (cf. Eq.~(\ref{eq:taa})), and using Eq.~(\ref{eq:sigmaAB1}) we may write 
\begin{equation}
\begin{split}
\mathrm{d} & N^{AB\rightarrow k + X}(\mathbf{b}) = \sum\limits_{i,j,X'} \frac{1}{AB}\sum\limits_{N_A,N_B} \int \mathrm{d}^2\mathbf{s_1} \, T_A(\mathbf{s_1}) \, R_i^{N_A/A}(x_1,Q^2) \, f_i^{N_A}(x_1,Q^2) \, \otimes \\ 
& \int \mathrm{d}^2\mathbf{s_2} \, T_B(\mathbf{s_2}) \, R_j^{N_B/B}(x_2,Q^2) \, f_{j}^{N_B}(x_2,Q^2) \otimes \mathrm{d}\hat{\sigma}^{ij\rightarrow k + X'} \delta(\mathbf{s_2} - \mathbf{s_1} - \mathbf{b}).
\label{eq:sigmaAB2_ave}
\end{split}
\end{equation}
From this, we see that a suitable definition of the spatially dependent nuclear modification $r_i^A(x,Q^2,\mathbf{s})$ for the PDF of parton flavor $i$ (per nucleon) is 
\begin{equation}
R_i^A(x,Q^2) \equiv \frac{1}{A}\int \mathrm{d}^2 \mathbf{s} \,T_A(\mathbf{s}) \,r_i^A(x,Q^2,\mathbf{s}),
\label{eq:normalization}
\end{equation}
where the thickness function $T_A$ is normalized to $A$ and where the case of no nuclear effects corresponds to $R_i^A=r_i^A=1$. Using these definitions, we can now generalize Eq.~(\ref{eq:sigmaAB2_ave}) to include the spatially dependent nuclear modifications,  
\begin{equation}
\begin{split}
\mathrm{d} & N^{AB\rightarrow k + X}(\mathbf{b}) =  \sum\limits_{i,j,X'} \frac{1}{AB} \sum\limits_{N_A,N_B} \int \mathrm{d}^2\mathbf{s_1} \, T_A(\mathbf{s_1}) \, r_i^A(x_1,Q^2,\mathbf{s_1}) \, f_i^{N_A}(x_1,Q^2) \, \otimes \\ & \int \mathrm{d}^2\mathbf{s_2} \, T_B(\mathbf{s_2}) \, r_j^B(x_2,Q^2,\mathbf{s_2}) \, f_{j}^{N_B}(x_2,Q^2) \otimes  \mathrm{d}\hat{\sigma}^{ij\rightarrow k + X'}
\delta(\mathbf{s_2} - \mathbf{s_1} - \mathbf{b}).
\label{eq:sigmaAB2}
\end{split}
\end{equation}
As a consistency check, we note that the definition in Eq.~(\ref{eq:normalization}) guarantees that the minimum-bias cross sections, which are obtained by integrating Eq.~(\ref{eq:sigmaAB2}) over the whole \textbf{b} space, become simply $AB$ times the hard cross-section computed with the spatially averaged nPDFs,
\begin{equation}
 \mathrm{d}{\sigma}^{AB\rightarrow k + X}_{\rm MB} = 
\int \mathrm{d}^2\mathbf{b}\, \mathrm{d}N^{AB\rightarrow k + X}(\mathbf{b}) = 
AB\sum\limits_{i,j,X'} f_{i}^A(x,Q^2) \otimes f_{j}^B(x,Q^2) \otimes  \mathrm{d}\hat{\sigma}^{ij\rightarrow k + X'}.
\label{eq:mianbias}
\end{equation}

The key assumption in the present analysis is that the spatial dependence of $r_i^A(x,Q^2,\mathbf{s})$ is a function of the nuclear thickness $T_A(\mathbf{s})$. The motivation for this comes mainly from the shadowing region at small $x$, where the partons of sufficiently small values of $x$ may  interact with partons from any other nucleon near enough in the transverse direction. Also in the Gribov-Glauber modeling \cite{Frankfurt:2011cs} of the initial state nPDFs the nuclear effects become essentially functions of $T_A$. The functional form we choose to use and test here is a simple power series of the thickness functions,
\begin{equation}
r_i^A(x,Q^2,\mathbf{s}) = 1 + \sum\limits_{j=1}^{n} c^i_j(x,Q^2)\left[T_A(\mathbf{s})\right]^j.
\label{eq:ta_series}
\end{equation}

Here we would like to emphasize the following points: First, all the $A$ dependence is now in the thickness functions which are fully known, and all the coefficients $c^i_j(x,Q^2)$ which will be our fit parameters, depend on $x$ and $Q^2$ but not on $A$. Second, the power series of the form $1 + \ldots$ also fixes by construction that $r_i^A(x,Q^2,\mathbf{s}) \rightarrow 1$ when $|\mathbf{s}| \rightarrow \infty$, which means that the nucleons at the very edge of the nucleus are essentially regarded as free nucleons. Third, what is known from the EKS98 and EPS09 -types of analyses, are only the spatially averaged nuclear modifications and their $A$ systematics, i.e. $T_A$-weighted integrals of Eq.~(\ref{eq:ta_series}) over $\mathbf{s}$ for each nucleus. Fourth, since the EKS98 and EPS09 global analyses have not been constructed to reproduce any specific theoretically motivated $A$ dependence of the nPDFs, we can test the validity of the assumption of Eq.~(\ref{eq:ta_series}), as well as the number of terms needed, only a posteriori.

\begin{figure}[htbp]
\centering
\includegraphics[width=8cm]{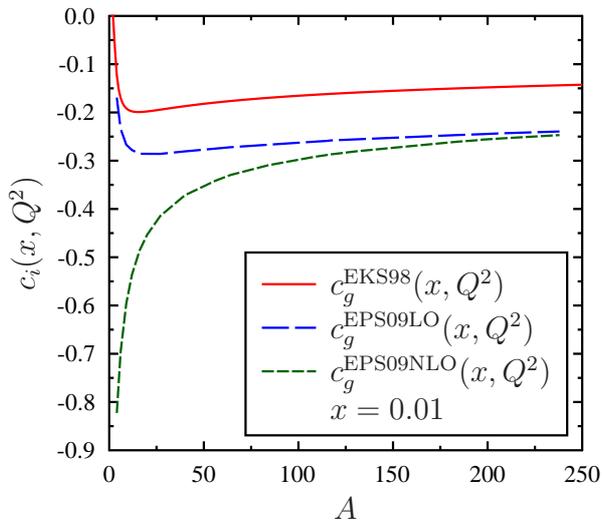}
\caption{The problematic $A$ dependence of the parameter $c^g(x,Q^2) = [R_g^A(x,Q^2)-1]A/T_{AA}(0)$ 
for EPS09NLO1 and EPS09LO1 (EKS98) gluons at $x = 0.01$ and $Q^2= 1.69\, (2.25) \textrm{ GeV}^2$
in the 1-parameter approach where one includes only the first nontrivial term in the power series in Eq.~(\ref{eq:ta_series}).}
\label{fig:fi1rx}
\end{figure}

Using the definitions above, we can see why the simplest 1-parameter approach with $n=1$ in Eq.~(\ref{eq:ta_series})
(which is used e.g. in \cite{Eskola:1991ec,Emel'yanov:1999bn,Klein:2003dj,Vogt:2004hf}
as well as in e.g. the HIJING event generator \cite{Gyulassy:1994ew}) is not fully consistent with the observed $A$ systematics of the nuclear data. In this case, $r_i^A(x,Q^2,\mathbf{s}) = 1 + c^i(x,Q^2)T_A(\mathbf{s})$, and from the definition in Eq.~(\ref{eq:normalization}), one obtains $c^i(x,Q^2) = [R_i^A(x,Q^2)-1]A/T_{AA}(0)$, where $R_i^A$ is given by the globally analysed nPDFs (i.e. nuclear data).
The problem then is that the coefficient $c^i(x,Q^2)$ may depend in fact quite strongly on $A$, which indicates that the simplest assumption of terminating the power series at the first nontrivial term does not correctly capture the spatial dependence of the measured nuclear structure functions. This redundant $A$ dependence is illustrated in Fig.~\ref{fig:fi1rx} for gluons in a lead nucleus at $x=0.01$ at the initial scales of the sets EKS98, EPS09LO1 and EPS09NLO1. We can see that especially for the NLO set the problem is more serious.
One of the driving motivations for the present study is to solve the problem of recovering the $A$ systematics in the spatially dependent nPDFs. 

\subsection{Fitting Procedure}

To extract the $A$-independent coefficients $c^i_j(x,Q^2)$, we need to introduce a fitting procedure, where we
utilize the definition (\ref{eq:normalization}) and the $A$ dependence of the EKS98 and EPS09 nuclear modifications at different values of $x$ and $Q^2$ for each parton flavor $i$. To reproduce the $A$ systematics in the spatially independent nuclear modifications with the power-series ansatz of Eq.~(\ref{eq:ta_series}), we minimize the $\chi^2$ defined as
\begin{equation}
\chi^2_i(x,Q^2) \equiv \sum_A \left[\frac{R_i^A(x,Q^2) - \frac{1}{A}\int \mathrm{d}^2 \mathbf{s}\, T_A(\mathbf{s}) \, r_i^A(x,Q^2,\mathbf{s})}{W_i^A(x,Q^2)} \right]^2,
\end{equation}
where the spatially averaged modifications $R_i^A(x,Q^2)$ from EKS98 and EPS09 now represent the "experimental" data. The weight factors $W_i^A(x,Q^2)$ are artificial errors which control the quality of the fit and which are set by hand. Our numerical observation is that for good fits we need 4th-order polynomials in $T_A$, i.e. $n=4$ in Eq.~(\ref{eq:ta_series}). Furthermore, best fits were obtained with the weight $W_i^A(x,Q^2) = R_i^A(x,Q^2) - 1$ for EKS98 (this corresponds to fitting the deviations from unity within a constant relative error) and $W_i^A(x,Q^2)=1$ for EPS09 (corresponds to fitting the modifications within a constant error). 

By construction, both EKS98 and EPS09 give no nuclear modifications for deuterium. This cannot be reproduced with the functional form we selected for $r_i^A(x,Q^2)$, and we do not expect the fit form of Eq.~(\ref{eq:ta_series}) work for the smallest values of $A$, either. Consequently, we exclude the nuclei $A < 16$ from the fit. Thus for EKS98 the sum runs over $A = 16, 20, \ldots, 300$ (i.e. emphasising the large nuclei) and for EPS09 we use all the $A \ge 16$ values for which these sets are currently available.

\section{Results}
\label{sec:results}

\subsection{Quality of the fit}

\begin{figure}[htbp]
\centering
\begin{tabular}{cc}
\includegraphics[width=7cm]{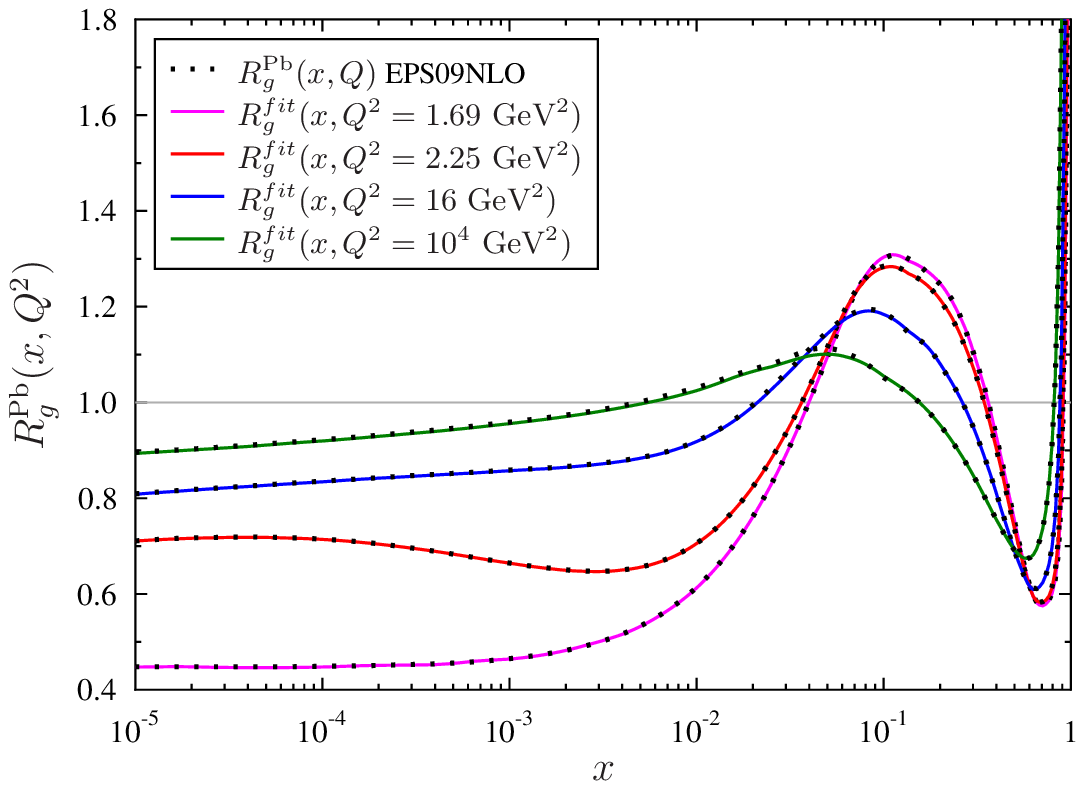}
&
\centering
\includegraphics[width=7cm]{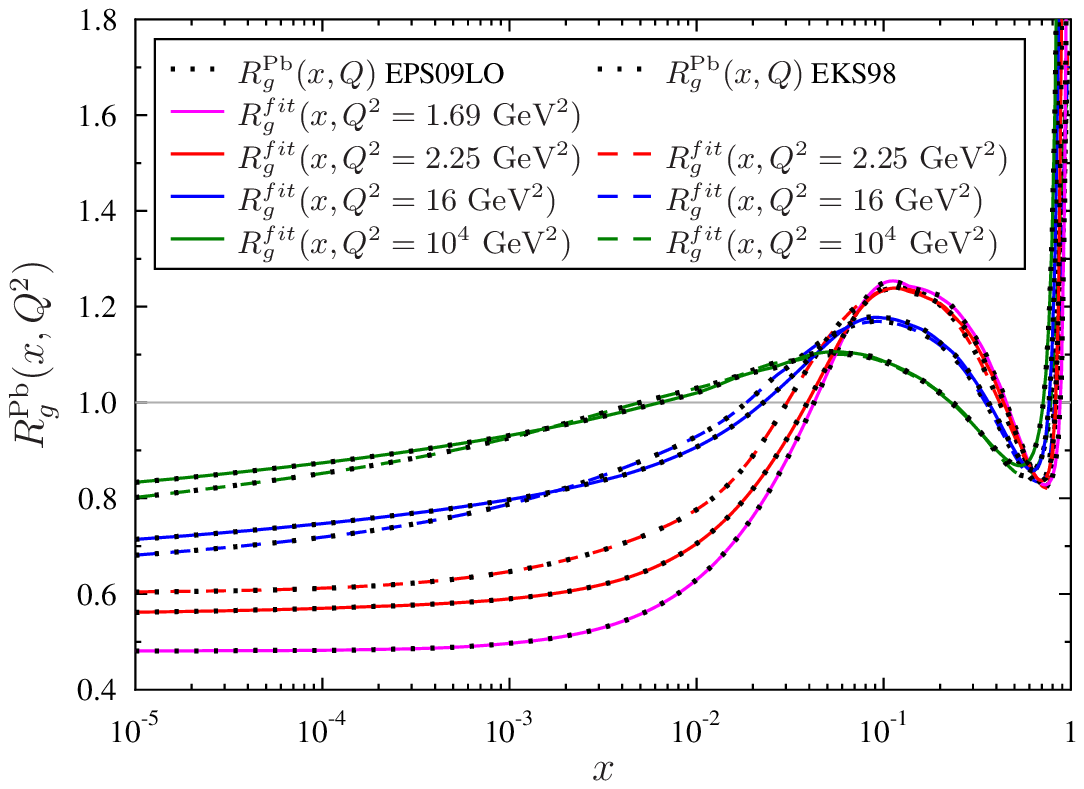}
\end{tabular}
\caption{\textbf{Left:} The spatially averaged nuclear modification $R_g^A(x,Q^2)$ for a lead nucleus ($A=208$) from the NLO set EPS09NLO1 (dotted lines) and from the EPS09sNLO1 spatial fit presented here (solid lines) at four different scales. \textbf{Right:} The same with the LO sets EKS98 and EPS09LO1 (dotted) and with the spatial fits EKS98s (dashed) for three different scales and EPS09sLO1 (solid) for four different scales.
}
\label{fig:R_g_fits}
\end{figure}

First, we demonstrate that our fit framework manages to reproduce the spatially averaged nuclear modifications and especially their $A$ dependence indeed very well. Figure~\ref{fig:R_g_fits} shows the obtained spatially dependent gluon modifications integrated over the transverse plane according to Eq.~(\ref{eq:normalization}), and the corresponding input modifications at different fixed values of $Q^2$ from the  NLO set EPS09NLO1 (left panel), and from the LO sets EPS09LO1 and EKS98 (right panel), for a lead nucleus. 
In what follows, we refer to these cases as "EPS09sNLO1", "EPS09sLO1" and "EKS98s" where "s" is for "spatial" and "1" for the central sets. As seen in the figure, the match with the input and output distributions is very good; for all parton flavors and the nuclei included in our fits it is within  2 \% at $x<0.75$ for EPS09NLO, 1 \% at $x<0.85$ for EPS09LO, and 0.2 \% at $x<0.95$ for EKS98. Importantly, the key-feature here, the $A$ dependence of EPS09 and EKS98, is similarly well reproduced, as is demonstrated by Fig.~\ref{fig:R_g_A_fits} below.

\begin{figure}[htbp]
\centering
\begin{tabular}{cc}
\includegraphics[width=7cm]{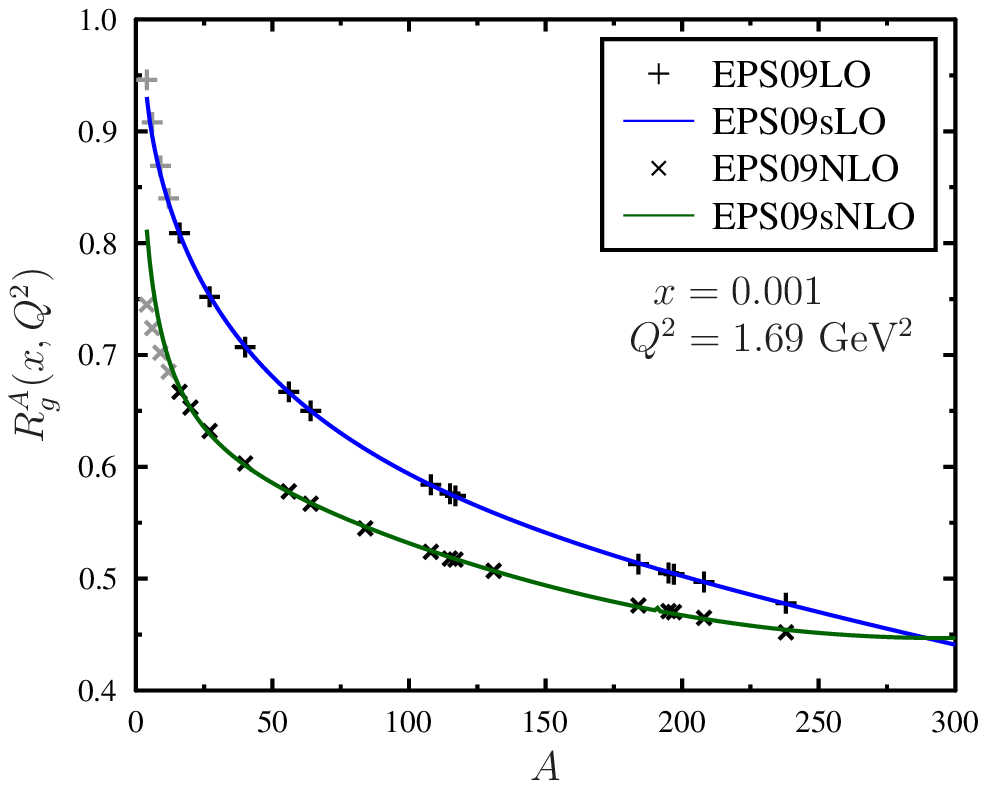}
&
\centering
\includegraphics[width=7cm]{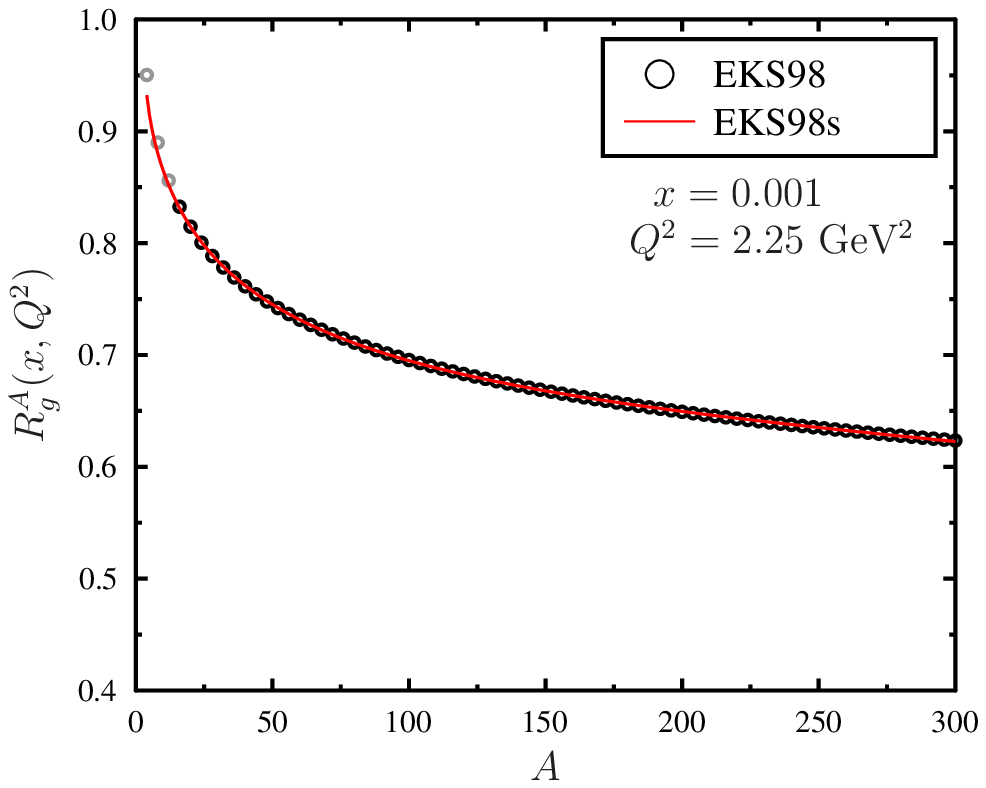}
\end{tabular}
\caption{\textbf{Left:} The $A$ dependence of the spatially averaged nuclear modification $R_g^A(x,Q^2)$ at fixed values  $x=0.001$ and $Q^2=1.69$~GeV$^2$ from the sets EPS09NLO1 (crosses) and EPS09LO1 (pluses) and from the corresponding spatial fits EPS09sNLO1 (solid green line) and EPS09sLO1 (solid blue line). 
\textbf{Right:} The same but with the LO set EKS98 (circles) and the corresponding spatial fit EKS98s (solid red) at $Q^2=2.25$~GeV$^2$.  
The small nuclei shown with gray markers in both panels were not used in our spatial fits. 
}
\label{fig:R_g_A_fits}
\end{figure}

Recall also that in the EPS09 global analysis in addition to the best fit there are also 30 error sets, which enables one to compute how the uncertainties of the nPDFs propagate into physical observables. The above fitting and determination of the spatial dependence are done also for each of these error sets, both in LO and in NLO, and the fit quality is similar as in Figs.~\ref{fig:R_g_fits} and \ref{fig:R_g_A_fits}. Thus the error propagation calculations (as instructed in \cite{Eskola:2009uj}) for centrality-dependent nuclear hard cross-sections can now be done as before, using the EPS09s sets. 

\subsection{Spatial Dependence}

\begin{figure}[ht]
\begin{center}
\hspace{-1.0cm}\includegraphics[width=8.5cm]{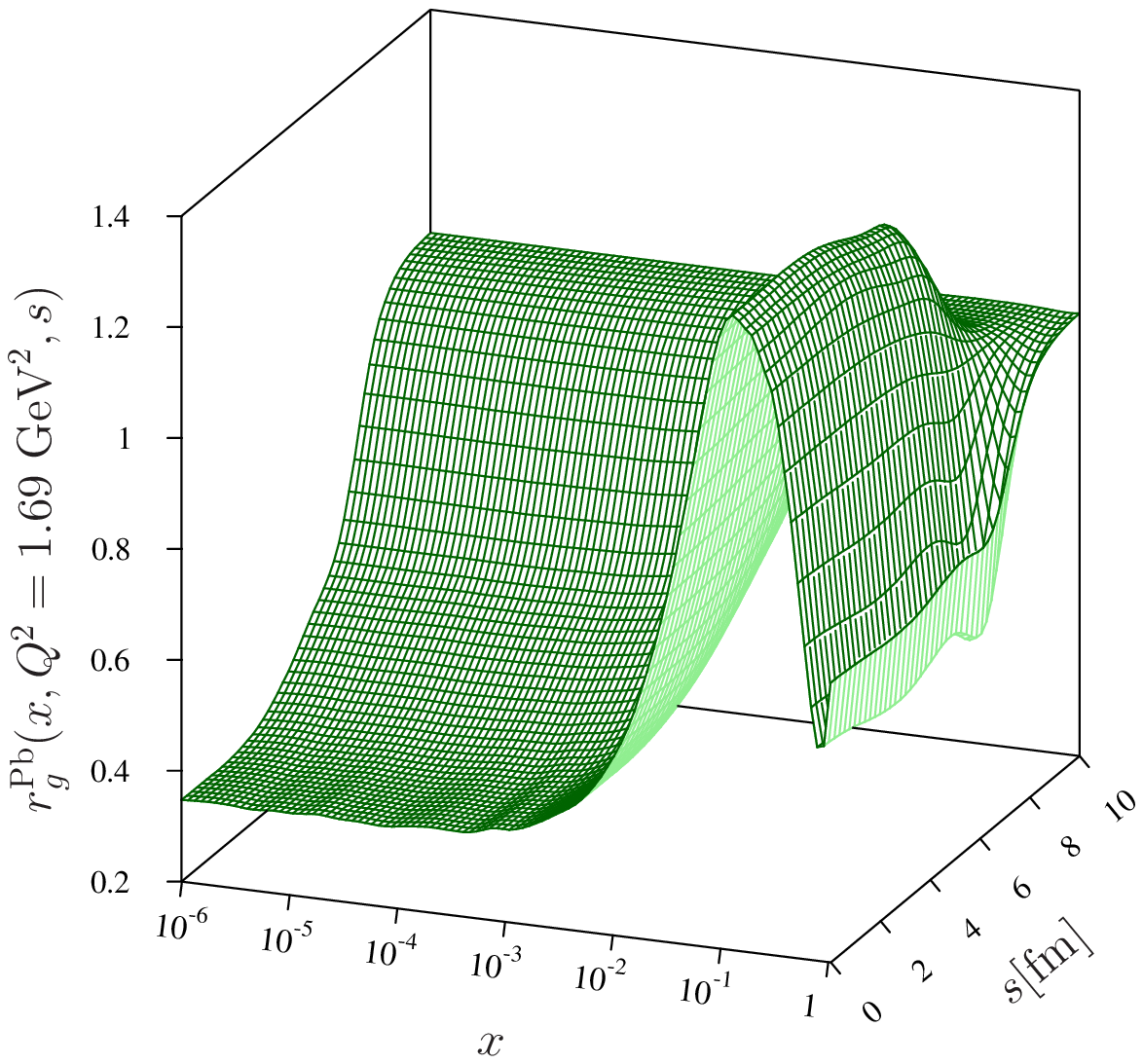}
\hspace{-1.0cm}\includegraphics[width=8.5cm]{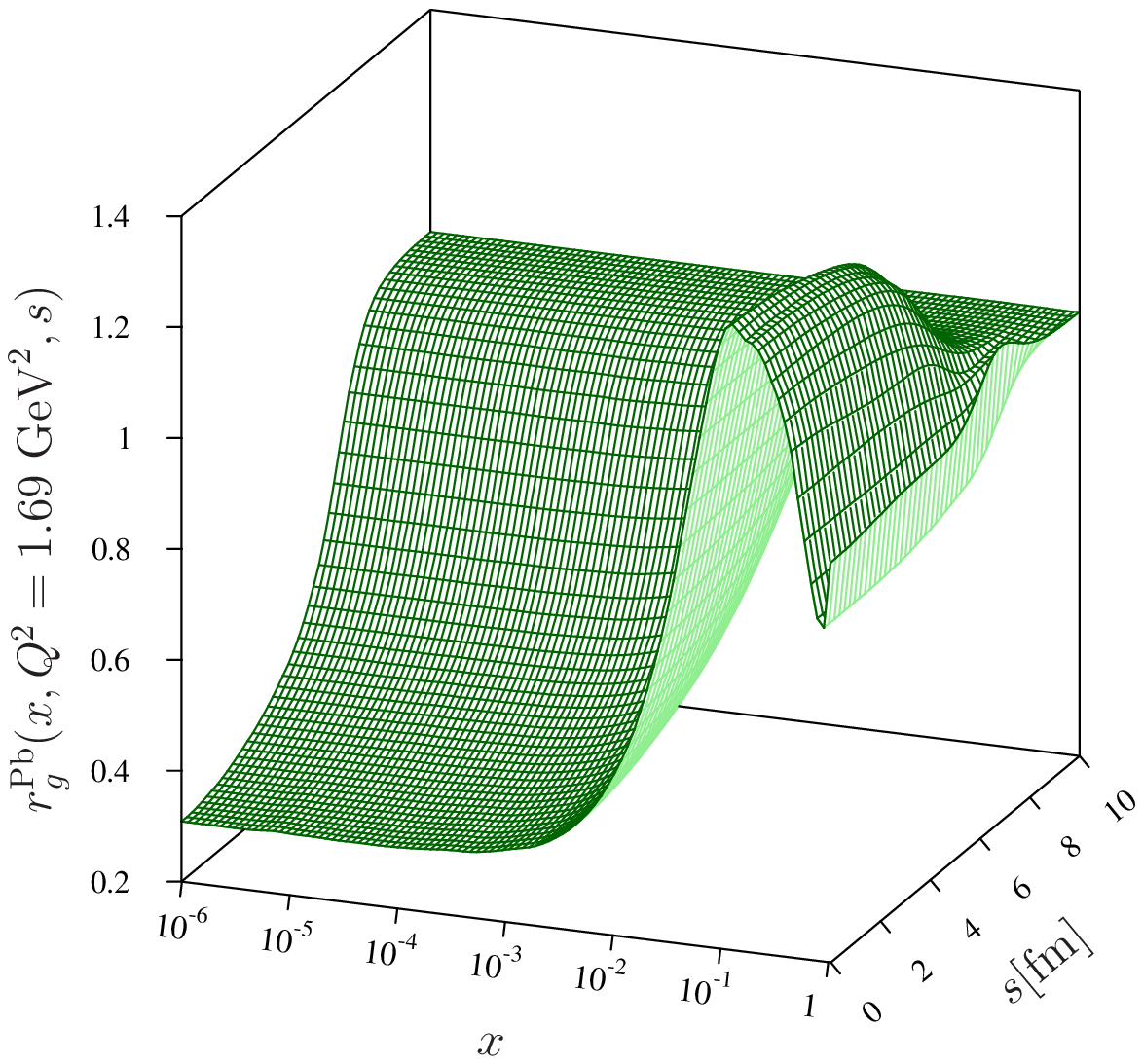}
\includegraphics[width=8.5cm]{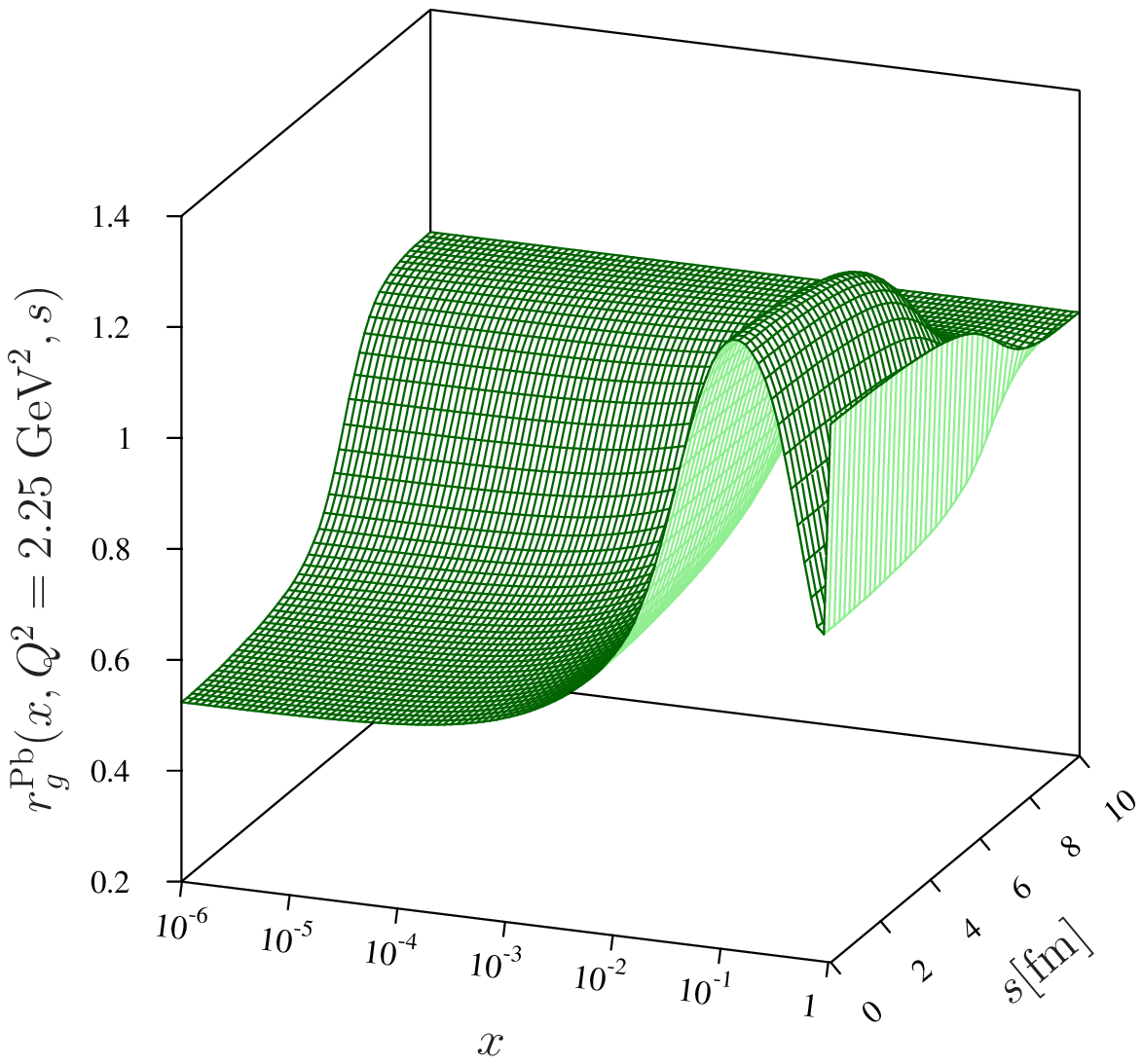}
\caption{The spatially dependent modification of gluon distribution in a lead nucleus, $r_g^{\rm Pb}(x,Q^2,s)$, from 
EPS09sNLO1 (upper left), EPS09sLO1 (upper right) and EKS98s(lower plot) as a function of $x$ and $s$ at the initial scale $Q^2 = 1.69 (2.25) \textrm{ GeV}^2$ of EPS09 (EKS98). For examples of the corresponding plots of other parton flavors, see App.~\ref{sec:ruvus}.}
\label{fig:rxsg3d}
\end{center}
\end{figure}

After the consistency checks above, let us next discuss the spatial dependence obtained for the nuclear modifications of the PDFs. In Fig.~\ref{fig:rxsg3d} we present the nuclear modification $r_g^{\rm Pb}(x,Q^2,s)$ at the initial scale $Q^2 = 1.69 \textrm{ GeV}^2$ as a function of $x$ and $s$, as obtained from the fitting to the sets EPS09 NLO and LO, as well as the LO set EKS98. The three main observations are
\begin{itemize}
\item The overall $x$-shape of the nuclear modification away from the edge of the nucleus, at $|\mathbf{s}|<R_A$, is similar as in the input distribution. This confirms that our fit does not generate any unwanted extra curvature.
\item In the center of the nucleus, $|\mathbf{s}|\approx 0$, the nuclear modification is only slightly larger than the input average modification. This also confirms the earlier similar findings in \cite{Eskola:1991ec,Emel'yanov:1999bn,Vogt:2004hf}.
\item The nuclear modification dies out as expected, by construction, when $|\mathbf{s}| > R_A$. This feature arises from the vanishing $T_A(\mathbf{s})$ at the edge of the nucleus.
\end{itemize}
The observations for the spatial dependence of the sea and valence quarks nuclear modifications are the same.
Examples of these can be found in App. \ref{sec:ruvus}. 

\subsection{Comparison with other approaches}

Next, we compare our EPS09s and EKS98s fits with the 1-parameter approach described in the end of Sec.~\ref{sec:def_nucl_mod}. \cite{Emel'yanov:1999bn,Vogt:2004hf}.\footnote{In \cite{Emel'yanov:1999bn} the spatial dependence
enters through the first nontrivial power of the nuclear density $\rho_A(\mathbf{r})$ or the thickness function $T_A(\mathbf{s})$. The latter scenario corresponds to what we refer to as "1-parameter approach" here.}
This model has been used to study the centrality dependence of the $J/\Psi$ suppression e.g. in Refs.~\cite{Ferreiro:2009ur,Adare:2010fn,Ferreiro:2012zb,Nagle:2010ix}\footnote{In \cite{Nagle:2010ix} one studies also other types of spatial dependences.} and inclusive hadron production in d+Au collisions at RHIC in Ref.~\cite{Vogt:2004hf}. 
We also compare with the leading-twist formulation \cite{Alvero:1998bz,Frankfurt:2011cs} of nuclear shadowing which is based on the generalization of Gribov-Glauber theory, QCD factorization and diffractive PDFs measured at HERA.
For the spatially averaged nuclear modifications, this model typically predicts a stronger smallest-$x$ shadowing than what is implemented in the parametrizations of EKS98 and EPS09 (see e.g. Ref.~\cite{Frankfurt:2011cs}). For the comparison, we consider the FGS10\_L set \cite{Guzey:2009jr,Frankfurt:2011cs}, and choose the value of $x$ not too small, so that the spatially averaged FGS10\_L  nuclear gluon modification is close to that in EPS09 or EKS98. 

\begin{figure}[htbp]
\centering
\includegraphics[width=7cm]{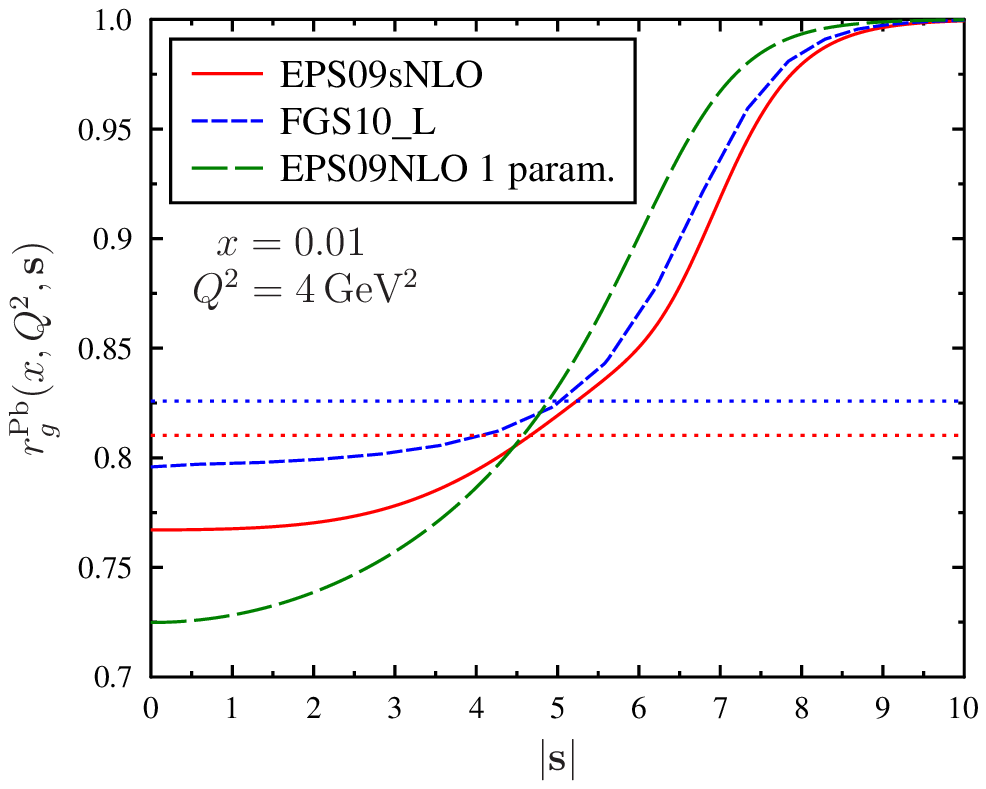}
\includegraphics[width=7cm]{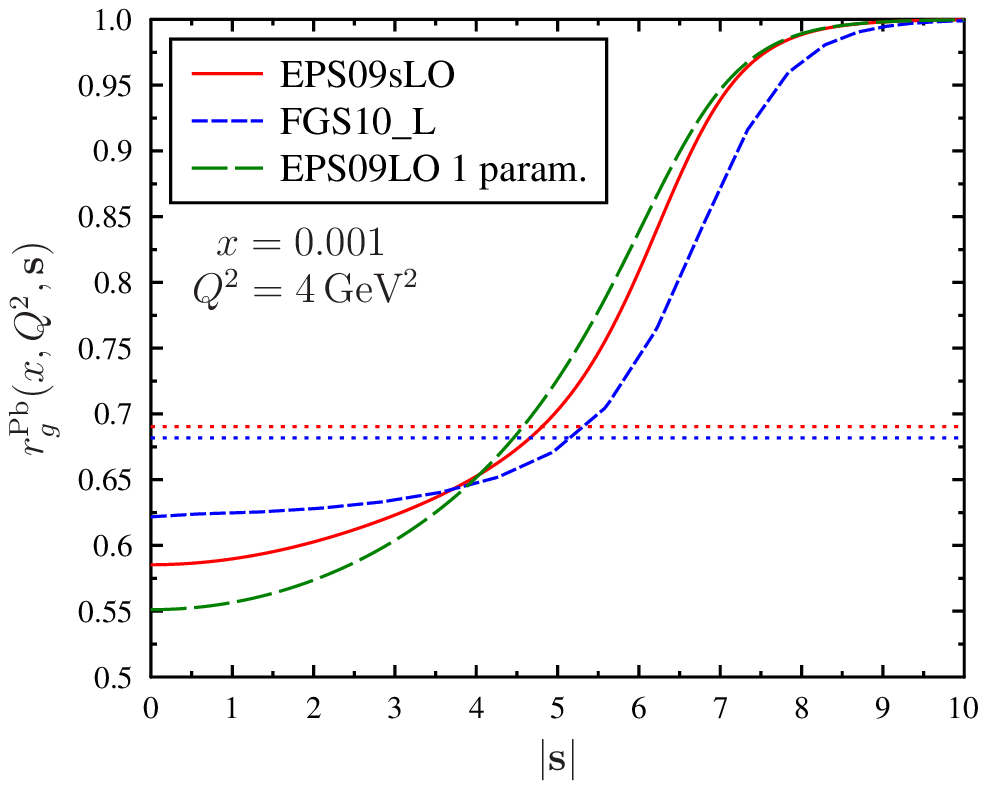}
\includegraphics[width=7cm]{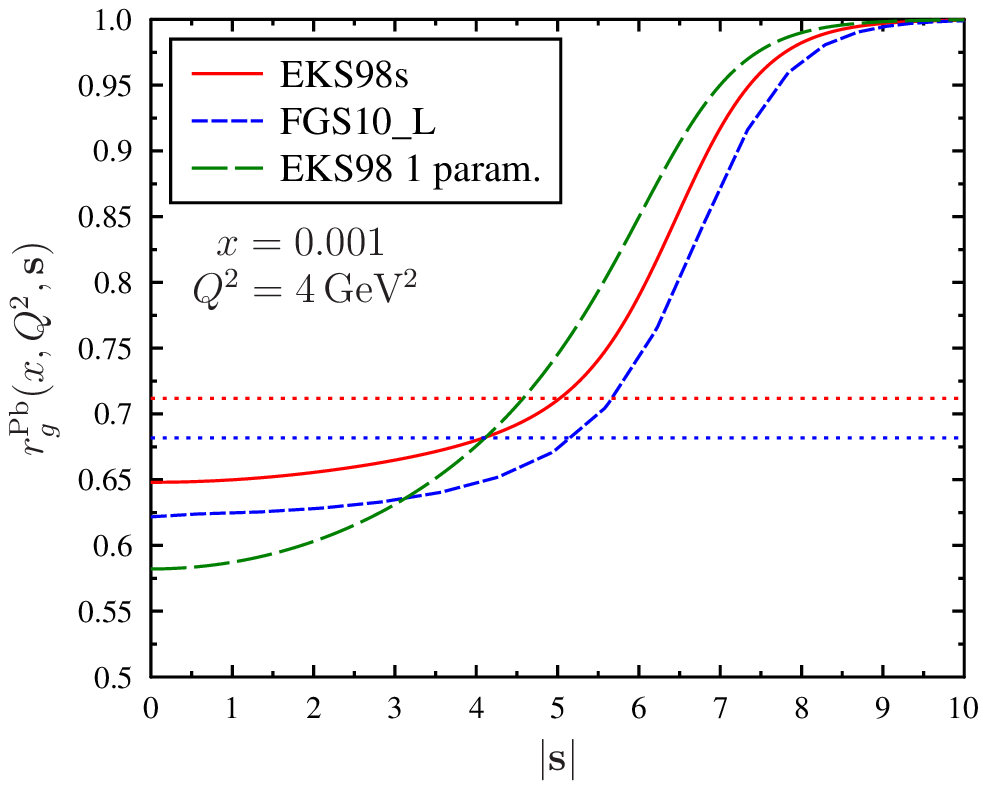}
\caption{Comparison of the spatial dependence of the gluon modification in a lead nucleus, $r_g^{\rm Pb}(x,Q^2,s)$, 
between FGS10\_L (short-dashed blue curves), 1-parameter approach (long-dashed green) and our spatial fits (solid red) EPS09sNLO1 (upper left), EPS09sLO1 (upper right) and EKS98s (lower plot). The scale $Q^2=4$~GeV$^2$ for all plots but the values of $x$ have been chosen so that the spatially averaged $R_g^{\rm Pb}(x,Q^2)$ (dotted horizontal red lines) approximately coincides with FGS10\_L (dotted blue).}
\label{fig:rsg208_eks_eps}
\end{figure}

In Fig.~\ref{fig:rsg208_eks_eps} we plot the nuclear modification for gluons at fixed values of $x$ and scale $Q^2 = 4 \text{ GeV}^2$ for $A=208$ as a function of $|\mathbf{s}|$ from our EPS09sNLO1, EPS09sLO1 and EKS98s fits, from the 1-parameter approach using the averaged sets EPS09NLO1, EPS09LO1 and EKS98, and from FGS10\_L.
Although numerically the differences are not very large, we notice that while both the EPS09sNLO and EKS98s results are close to FGS10\_L, the 1-parameter approach leads to a too steep transverse profile for the modifications in all cases. 

\section{Applications}
\label{sec:applications}

Next, we consider some concrete examples of computing the nuclear hard-process cross-sections in different centrality classes using the spatially dependent nPDFs. First, we discuss the centrality dependence of primary partonic-jet production in $A$+$A$ collisions at RHIC and LHC. Then, we consider neutral pion production in d+Au collisions at RHIC and in p+Pb collisions at the LHC.

\subsection{Implementation of EKS98s and EPS09s}
For defining the centrality classes we use the optical Glauber model specified in App.~\ref{sec:glauber}. In this case, each centrality class corresponds simply to a certain impact-parameter interval $|\mathbf{b}| \in [b_1,b_2]$. 
The generic average number distribution of a hard-process observable $k$ in this centrality class of an $A$+$B$ collision is 
\begin{equation}
\left\langle \mathrm{d} N_{AB}^{k}\right\rangle_{b_1,b_2} = 
\dfrac{\int_{b_1}^{b_2} \mathrm{d}^2 \mathbf{b}\, \mathrm{d}N_{AB}^{k}(\mathbf{b})}{\int_{b_1}^{b_2} \mathrm{d}^2 \mathbf{b} \,p_{AB}^{inel}(\mathbf{b})}, 
\label{eq:dNcentrality}
\end{equation}
where $p_{AB}^{inel}(\mathbf{b})=1-\exp[-T_{AB}(\mathbf{b})\sigma_{NN}^{inel}]$ from Eq.~(\ref{eq:p_inelAB}), and   
$\mathrm{d}N_{AB}^{k}(\mathbf{b})$ is obtained from Eq.~(\ref{eq:sigmaAB2}). 
Using the expansion of $r_i^A(x,Q^2,\mathbf{s})$ in powers of $T_A$ from Eq.~(\ref{eq:ta_series}), the integrals over the impact parameter for the spatially dependent parts can be conveniently separated from the spatially independent fit coefficients, free nucleon PDFs and pQCD parts as follows:   
\begin{align}
\int_{b_1}^{b_2} \mathrm{d}^2 \mathbf{b}\, \mathrm{d}N_{AB}^{k}(\mathbf{b})
 =&\sum\limits_{n,m=0}^4 T_{AB}^{nm}(b_1,b_2) \sum\limits_{i,j,X'} \frac{1}{AB}\sum\limits_{N_A,N_B} c^i_n(x_1,Q^2)f_i^{N_A}(x_1,Q^2) \, \otimes \notag\\
 & c^j_m(x_2,Q^2)f_{j}^{N_B}(x_2,Q^2) \otimes  \mathrm{d}\hat{\sigma}^{ij\rightarrow k + X'} 
\label{eq:dNintb1b2}
\end{align}
where $f_{i,j}^{N_A,N_B}$ are the free nucleon PDFs, and we have defined $c_0^{i,j}(x,Q^2) \equiv 1$ and  
\begin{equation}
T_{AB}^{nm}(b_1,b_2) \equiv \int_{b_1}^{b_2} \mathrm{d}^2\mathbf{b} \int \mathrm{d}^2\mathbf{s} \, \left[T_A(\mathbf{s}-\mathbf{b}/2)\right]^{n+1} \, \left[T_B(\mathbf{s}+\mathbf{b}/2)\right]^{m+1}.
\label{eq:TAnBm}
\end{equation}

From Eq.~(\ref{eq:dNintb1b2}) we see the most straightforward implementation of the spatially dependent nPDFs. The purely geometric integrals, the $T_{AB}^{nm}(b_1,b_2)$ in Eq.~(\ref{eq:TAnBm}) for each pair of the powers $n$ and $m$, can be computed independently of the kinematic variables $x_1$, $x_2$, and $Q^2$ and also independently of the parton flavors $i,j$. Thus, in total we have 25 different geometric integrals to do (or 15 if $A=B$) but we need to do them only once. In comparison with the spatially averaged case, the fit parameters $c_{n}^{i}(x,Q^2)$ thus play the role of the nuclear modifications $R^i_A(x,Q^2)$ for each of the 25 pairs $n,m$. To arrive at the final \textbf{b}-integrated result for the number distribution of $k$, we thus need to repeat the computation of the kinematic parts 25 times, each with different sets of the coefficient pairs $\{c_{n}^{i}\}, \{c_{m}^{i}\}$ and a different geometric weight $T_{AB}^{nm}(b_1,b_2)$. The \verb=EKS98s= and \verb=EPS09s= routines which we provide in \cite{downloadpage}, give in addition to the fit coefficients $\{c_{n}^{i}(x,Q^2)\}$ also the thickness functions $T_A(s)$ (used in the fits here) for the computation of $T_{AB}^{nm}(b_1,b_2)$, as well as the combination $T_A(s)r_i^A(x,Q^2,s)$ for other possible implementations. 
Note also that for the \textbf{b} integral in Eq.~(\ref{eq:TAnBm}) the angular part is trivial, giving just $2\pi$. 

\subsection{The Nuclear Modification Factors $R_{AA}^{1\rm jet}$ and $R_{CP}^{1\rm jet}$}

Let us now consider the centrality dependence of primary inclusive high-$p_T$ parton production in $A$+$A$ collisions at RHIC and LHC. Following the generic discussion above, we define the nuclear modification ratio 
$R_{AA}^{1\rm jet}(p_T)$ relative to the p+p case for each centrality class as 
\begin{equation}
R_{AA}^{1\rm jet}(p_T,y; b_1,b_2) \equiv \dfrac{\left\langle\dfrac{\mathrm{d}^2 N_{AA}^{1\rm jet}}{\mathrm{d}p_T \mathrm{d}y}\right\rangle_{b_1,b_2}}{\langle N_{bin}^{AA} \rangle_{b_1,b_2} \dfrac{1} {\sigma^{NN}_{inel}}\dfrac{\mathrm{d}^2\sigma_{\rm pp}^{1\rm jet}}{\mathrm{d}p_T \mathrm{d}y}} 
= \dfrac{\int_{b_1}^{b_2} \mathrm{d}^2 \mathbf{b} \dfrac{\mathrm{d}^2 N_{AA}^{1\rm jet}(\mathbf{b})}{\mathrm{d}p_T \mathrm{d}y} }{ \int_{b_1}^{b_2} \mathrm{d}^2 \mathbf{b} \,T_{AA}(\mathbf{b})\dfrac{\mathrm{d}^2\sigma_{\rm pp}^{1\rm jet}}{\mathrm{d}p_T \mathrm{d}y}},
\label{eq:R_AA}
\end{equation}
where $\langle N_{bin}^{AA} \rangle_{b_1,b_2}$ is the average number of binary collisions in this centrality class given by Eq.~(\ref{eq:Nbin_ave})  and $\sigma^{NN}_{inel}$ is the inelastic nucleon-nucleon cross section. Apart from the (small) isospin effect, this ratio yields unity if there are no nuclear effects in the nPDFs. Thus, for peripheral enough centrality bins, this ratio should approach unity. For the details of the partonic cross sections, bookkeeping and kinematics, we refer to \cite{Eskola:2002kv}.

The nuclear mofication factor in the minimum-bias collisions is obtained from above by setting $b_1=0$ and $b_2\rightarrow \infty$, in which case we have
\begin{equation}
\left\langle R_{AA}^{1\rm jet}(p_T,y)\right\rangle = 
\dfrac{1}{A^2}
\dfrac{\mathrm{d}^2{\sigma}^{1{\rm jet}}_{AA,\rm MB}}{\mathrm{d}p_T \mathrm{d}y} 
\Big/
\dfrac{\mathrm{d}^2\sigma_{\rm pp}^{1\rm jet}}{\mathrm{d}p_T \mathrm{d}y},
\label{eq:R_AA_MB}
\end{equation}
where $\mathrm{d}{\sigma}^{1\rm jet}_{AA,\rm MB}$, which contains only the spatially averaged nPDFs, is obtained from Eq.~(\ref{eq:mianbias}) by setting $B=A$, and the p+p baseline $\mathrm{d}\sigma_{\rm pp}^{1\rm jet}$ from the same equation by setting $A=B={\rm p}$.

In addition to the centrality dependence of $R_{AA}^{1\rm jet}$, we are interested in the central-to-peripheral ratios, defined as 
\begin{equation}
R^{1\rm jet}_{CP} \equiv \dfrac{\left\langle \dfrac{\mathrm{d}^2 N_{AA}^{1\rm jet}}{\mathrm{d}p_T \mathrm{d}y}\right\rangle_{C} \dfrac{1}{\langle N_{bin}^{AA} \rangle}_{C}}
{\left\langle \dfrac{\mathrm{d}^2 N_{AA}^{1\rm jet}}{\mathrm{d}p_T \mathrm{d}y}\right\rangle_{P} 
\dfrac{1}{\langle N_{bin}^{AA} \rangle}_{P}} 
= 
\dfrac{\int_{b^C_1}^{b^C_2} \mathrm{d}^2 \mathbf{b}\, \dfrac{\mathrm{d}^2 N_{AA}^{1\rm jet}(\mathbf{b})}{\mathrm{d}p_T \mathrm{d}y} \Big/ \int_{b^C_1}^{b^C_2} \mathrm{d}^2 \mathbf{b}\, T_{AA}(\mathbf{b})}{\int_{b^P_1}^{b^P_2} \mathrm{d}^2 \mathbf{b}\, \dfrac{\mathrm{d}^2 N_{AA}^{1\rm jet}(\mathbf{b})}{\mathrm{d}p_T \mathrm{d}y} \Big/ \int_{b^P_1}^{b^P_2} \mathrm{d}^2 \mathbf{b}\, T_{AA}(\mathbf{b})},
\end{equation}
where $C$ and $P$ refer to the central and peripheral bins, correspondingly. The advantage of this ratio (in the experiments) is that the information of the proton-proton baseline is not required. In particular, we would like to see exactly how much $R^{1\rm jet}_{CP}$ differs from the modification $R_{AA}^{1\rm jet}$ which is computed with the spatially averaged nPDFs. We perform these example-calculations for both RHIC and LHC but for simplcity only to LO pQCD, since without jet quenching these ratios do not directly correspond to observables. They illustrate, however, the points we wish to make with the spatially dependent nPDFs, and also serve as (LO) pQCD baselines for the observed suppression of high-$p_T$ particles.  

The two different centrality classes we consider here for Au+Au collisions at RHIC and Pb+Pb collisions at the LHC, are the central 0-5\% and peripheral 60-80\% bins. The Glauber model input and the resulting impact parameter intervals and average numbers of binary collisions in these centrality classes are summarized in Table~\ref{tab:ogmparams}. 
\begin{table}[tbh]
\caption{The centrality classes as impact parameter intervals, and average number of binary collisions from the optical Glauber model in $A$+$A$ collisions for RHIC and LHC.}
\begin{center}
\begin{tabular}{|lcc|ccc|ccc|}
\hline
&$\sqrt{s_{NN}}$&$\sigma^{NN}_{inel}$& \multicolumn{3}{|c|}{Central = 0-5 \%} & \multicolumn{3}{c|}{ Peripheral = 60-80 \% }\\
&[GeV]& [mb]& $b_1 \textrm{ [fm]}$ & $b_2 \textrm{ [fm]} $ & $\langle N_{bin} \rangle $ & $b_1 \textrm{ [fm]}$ & $b_2 \textrm{ [fm]}$ & $\langle N_{bin} \rangle $\\
\hline
Au+Au &200  & 42 & 0.0 & 3.355 & 1083 & 11.62 & 13.42 & 15.10\\
Pb+Pb &2760 & 64 & 0.0 & 3.478 & 1771 & 12.05 & 13.91 & 19.08\\
\hline
\end{tabular}
\end{center}
\label{tab:ogmparams}
\end{table}

\begin{figure}[thbp]
\centering
\begin{tabular}{cc}
\includegraphics[width=7cm]{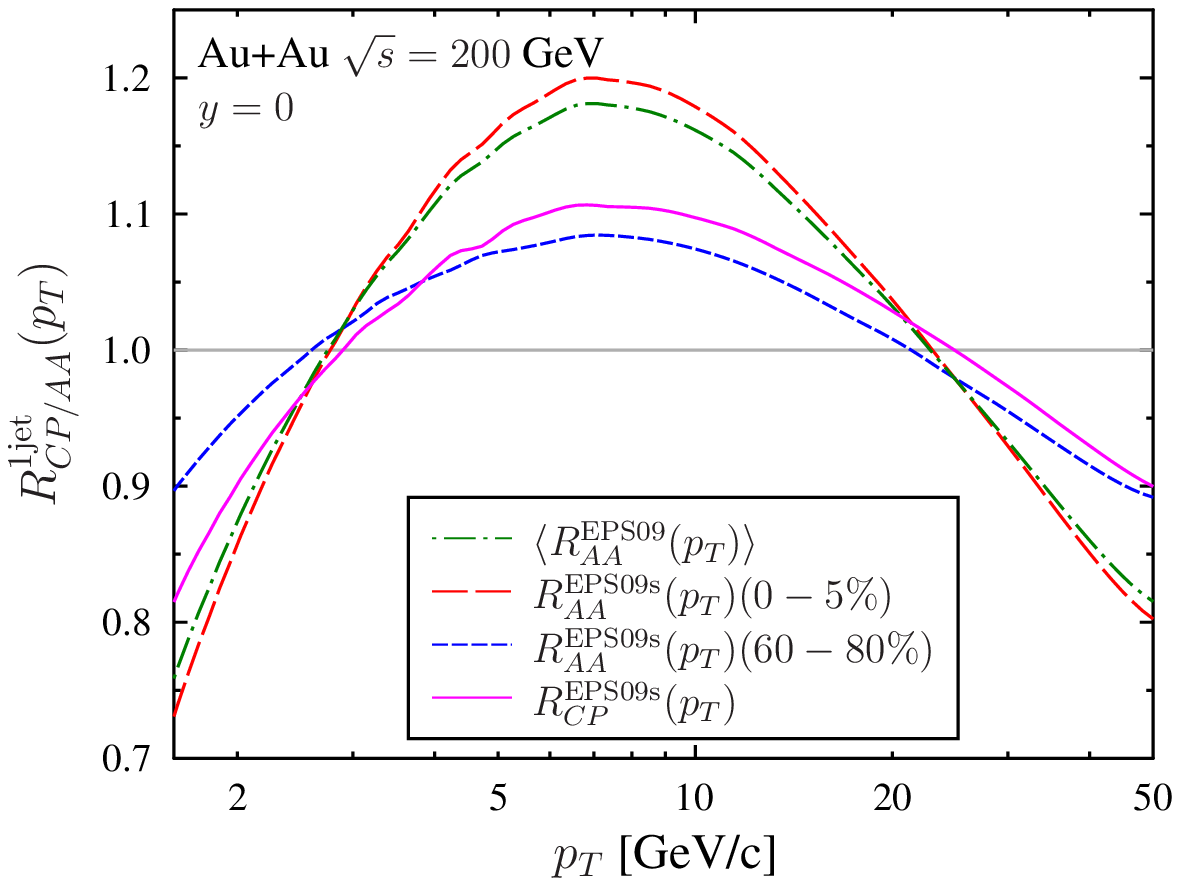}
&
\centering
\includegraphics[width=7cm]{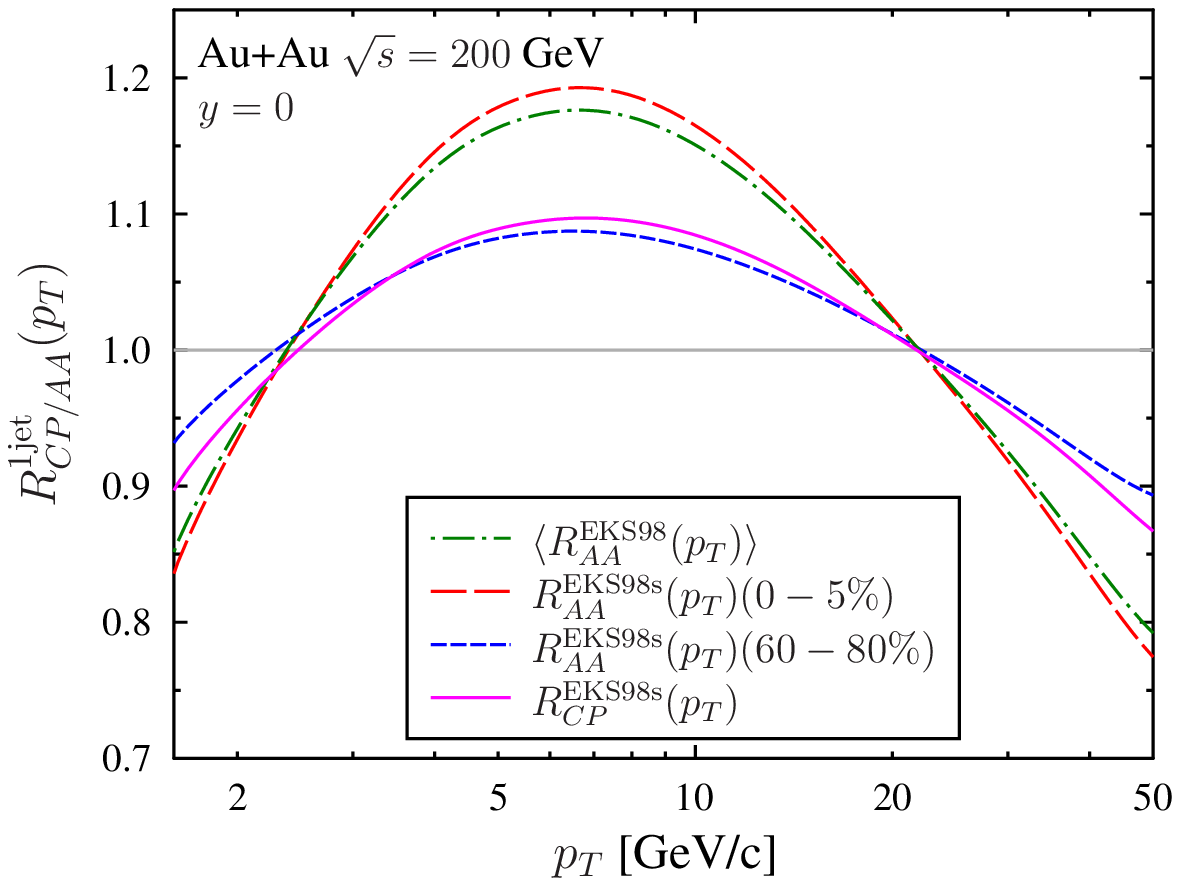}
\end{tabular}
\caption{The LO nuclear modification $R^{1\rm jet}_{AA}$ as a function of partonic transverse momentum for central (red long-dashed), peripheral (blue dashed) and minimum-bias (green dot dashed) collisions, and $R^{1\rm jet}_{CP}$ (solid magenta) for Au+Au collisions at $\sqrt{s_{NN}}=200\textrm{ GeV}$ and $y=0$ using EPS09sLO1  (left panel) and EKS98s  (right panel).}
\label{fig:rcpaaRHIC}
\end{figure}

\begin{figure}[thbp]
\centering
\begin{tabular}{cc}
\includegraphics[width=7cm]{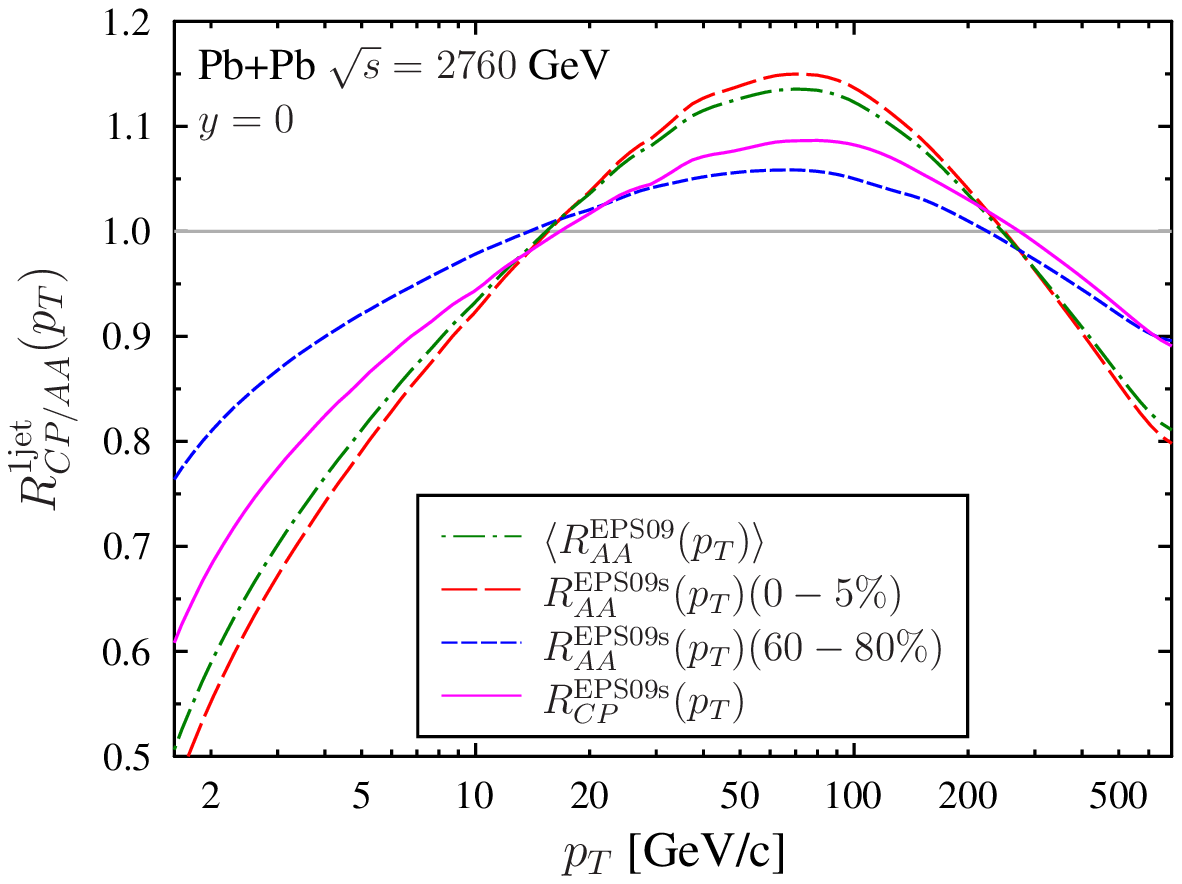}
&
\centering
\includegraphics[width=7cm]{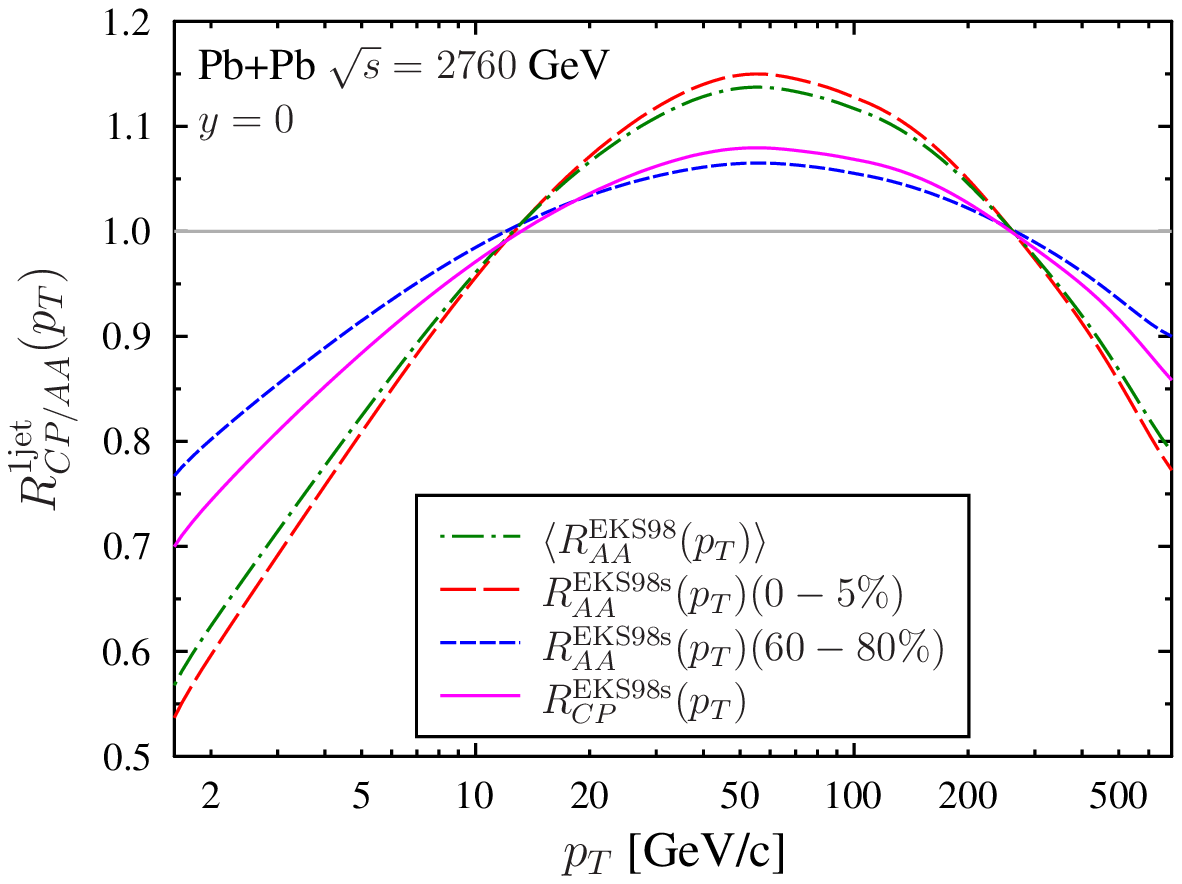}
\end{tabular}
\caption{The same as Fig.~\ref{fig:rcpaaRHIC} but for Pb+Pb collisions at  $\sqrt{s_{NN}}=2.76\textrm{ TeV}$}.
\label{fig:rcpaaLHC}
\end{figure}

In Fig.~\ref{fig:rcpaaRHIC} we plot the ratio $R^{1\rm jet}_{AA}(p_T,y=0)$ for central, peripheral and minimum-bias collisions, as well as $R^{1\rm jet}_{CP}(p_T)$ in Au+Au collisions at RHIC. Figure~\ref{fig:rcpaaLHC} shows the same quantities for the LHC Pb+Pb case. The central and peripheral $R^{1\rm jet}_{AA}$ and $R^{1\rm jet}_{CP}$ have been obtained with the spatially dependent nPDFs EPS09sLO1 (left) and EKS98s (right), and the average $\langle R^{1\rm jet}_{AA} \rangle$ in minimum bias collisions with the spatially independent EPS09 and EKS98 nuclear modifications. For the free proton PDFs we have used CTEQ6.1L \cite{Pumplin:2002vw}. The renormalization scale $\mu$ and factorization scale $Q$ has been set to be the transverse momentum, $p_T$, of the parton.

The main observations from the figures are:
\textit{(i)} The central $R^{1\rm jet}_{AA}$ is quite close to the average $R^{1\rm jet}_{AA}$, which is expected since the nuclear modifications at small $s$ are close to the average modifications.
\textit{(ii)} The peripheral $R^{1\rm jet}_{AA}$ is clearly not unity but there appear almost 10\% antishadowing effects at mid-$p_T$ at RHIC and even more than 20\% shadowing effects at small $p_T$ at the LHC, and up to 10\% EMC effects at large $p_T$ both at RHIC and LHC.
\textit{(iii)} Consequently, the ratio $R^{1\rm jet}_{CP}$ differs significantly from the average $R^{1\rm jet}_{AA}$. 
The results suggest that in a precision theory-analysis of the centrality dependence of jet quenching, one needs to account also for the spatial dependence of nPDFs. Finally, regarding the differences between the different nPDF sets applied here, we observe in Figs.~\ref{fig:rcpaaRHIC} and \ref{fig:rcpaaLHC} how the stronger shadowing in the EPS09 case (cf. Fig.\ref{fig:R_g_fits}) translates into steeper $p_T$ slopes of $R^{1\rm jet}_{AA}$ at small $p_T$ than in the EKS98 case.

\subsection{Centrality dependence of $R_{\rm dAu}^{\pi^0}(p_T)$ at RHIC -- comparison with data}

While the above ratios $R^{1\rm jet}_{AA}$ and $R^{1\rm jet}_{CP}$ mainly serve as theoretical pQCD baselines for jet quenching studies, it is important to test our spatially-dependent nPDF framework against some measured centrality-dependent observables. To avoid the complications of hot QCD matter modeling, we turn to the highest-energy d+Au collisions at RHIC and p+Pb at the LHC. For our purposes a promising published data set is the nuclear modification factor $R_{\rm dAu}^{\pi^0}(p_T)$ for single inclusive neutral-pion production at mid-rapidity $|\eta| < 0.35$, measured by PHENIX \cite{Adler:2006wg} at different centrality classes in d+Au collisions at $\sqrt{s_{NN}}=200$~GeV. Since the minimum-bias data precisely from this data set was among the constraints in the global EPS09 fits, it is now very interesting to study, for the first time consistently with EPS09, how well we can reproduce the measured centrality dependence of this ratio. 

Analogously with Eq.~(\ref{eq:R_AA}), we define the centrality-dependent nuclear modification factors as
\begin{equation}
R_{{\rm d}A}^{\pi^0}(p_T,y; b_1,b_2) \equiv \dfrac{\left\langle\dfrac{\mathrm{d}^2 N_{{\rm d}A}^{\pi^0}}{\mathrm{d}p_T \mathrm{d}y}\right\rangle_{b_1,b_2}}{\langle N_{bin}^{{\rm d}A} \rangle_{b_1,b_2} \dfrac{1} {\sigma^{NN}_{inel}}\dfrac{\mathrm{d}^2\sigma_{\rm pp}^{\pi^0}}{\mathrm{d}p_T \mathrm{d}y}} 
= \dfrac{\int_{b_1}^{b_2} \mathrm{d}^2 \mathbf{b} \dfrac{\mathrm{d}^2 N_{{\rm d}A}^{\pi^0}(\mathbf{b})}{\mathrm{d}p_T \mathrm{d}y} }{ \int_{b_1}^{b_2} \mathrm{d}^2 \mathbf{b} \,T_{{\rm d}A}(\mathbf{b})\dfrac{\mathrm{d}^2\sigma_{\rm pp}^{\pi^0}}{\mathrm{d}p_T \mathrm{d}y}},
\label{eq: R_dA}
\end{equation}
where the number distribution now involves a further folding over the fragmentation functions,  
\begin{equation}
\mathrm{d} N^{\pi^0}_{{\rm d}A}(\mathbf{b}) = \sum_k \mathrm{d} N^{k}_{{\rm d}A}(\mathbf{b}) \otimes D_{\pi^0/k}(z,Q_F^2)
\label{eq:dNdA}
\end{equation}
and where $\mathrm{d} N^{k}_{{\rm d}A}(\mathbf{b})$ is obtained from Eq.~(\ref{eq:sigmaAB2}) by setting $A={\rm d}$ and $B=A$, and (as we do not assign any nuclear effects for the deuterium PDFs) also $r_i^{\rm d}\equiv 1$. 
For obtaining a realistic thickness function for deuterium, we use the Hulthen wavefunction formulation \cite{Hulthen} given in App.~\ref{sec:deuterium}. The impact parameter ranges and average numbers of binary collisions at the corresponding centrality classes are obtained again from the optical Glauber model. 

\subsubsection{Minimum-bias $R_{\rm dAu}^{\pi^0}(p_T)$}

Setting the spatial integrals in Eq.~(\ref{eq: R_dA}) over the whole impact-parameter space gives again the minimum-bias ratios, 
\begin{equation}
\left\langle R_{{\rm d}A}^{\pi^0}(p_T,y)\right\rangle = 
\dfrac{1}{2A}
\dfrac{\mathrm{d}^2{\sigma}^{\pi^0}_{{\rm d}A,\rm MB}}{\mathrm{d}p_T \mathrm{d}y} 
\Big/
\dfrac{\mathrm{d}^2\sigma_{\rm pp}^{\pi^0}}{\mathrm{d}p_T \mathrm{d}y},
\label{R_dA_MB}
\end{equation}
where $\mathrm{d}{\sigma}^{\pi^0}_{{\rm d}A,\rm MB}= \sum_k \mathrm{d}{\sigma}^{k}_{{\rm d}A,\rm MB} \otimes D_{\pi^0/k}(z,Q_F^2)$ again contains only the spatially averaged nPDFs in $\mathrm{d}{\sigma}^{k}_{{\rm d}A,\rm MB}$ which is obtained from Eq.~(\ref{eq:mianbias}) by setting $A={\rm d}$, $B=A$. As noted earlier, in the EKS98 and EPS09 frameworks there are no nuclear modifications to the deuterium PDFs. The p+p baseline $\mathrm{d}\sigma_{\rm pp}^{\pi^0}$ is computed correspondingly, but without any nuclear effects.

\begin{figure}[thbp]
\begin{center}
\includegraphics[width=7cm]{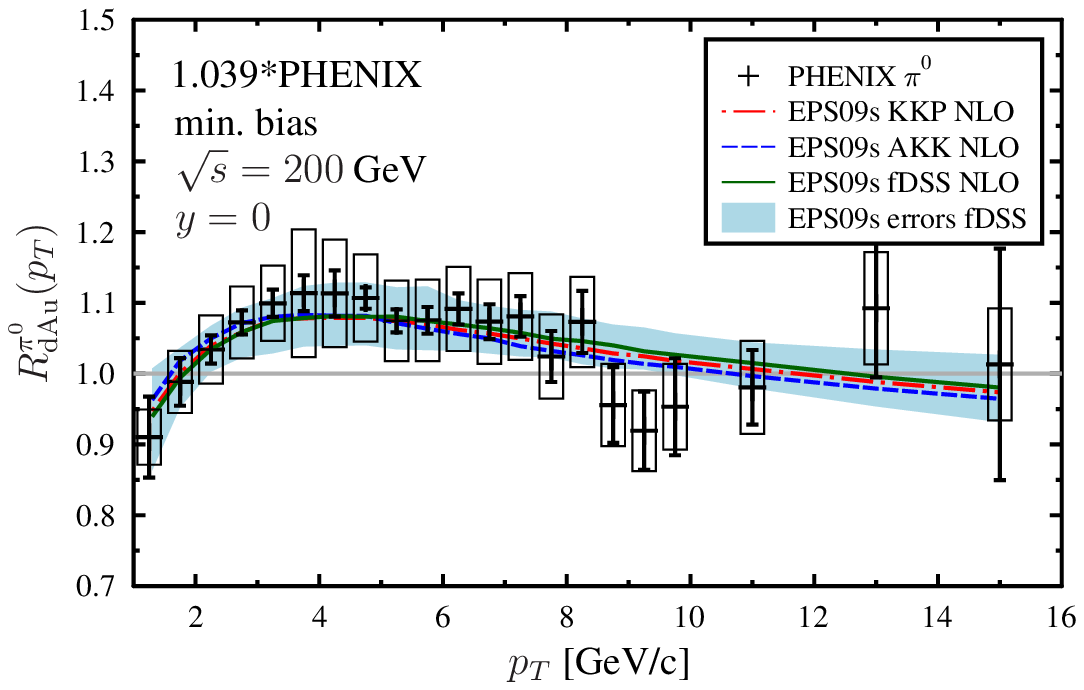}
\includegraphics[width=7cm]{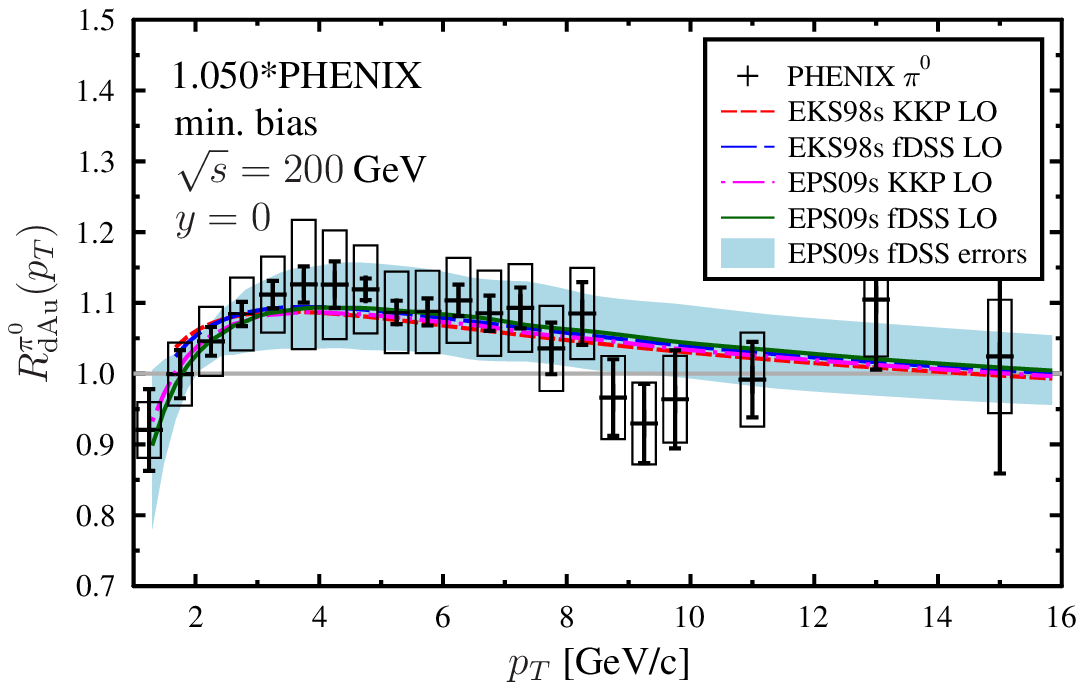}
\caption{The nuclear modification factor $R^{\pi^0}_{\rm dAu}(p_T)$ for $\sqrt{s_{NN}} = 200 \text{ GeV}$ at $y=0$ for minimum bias collisions. Calculations are for the NLO pQCD using EPS09NLO1 with three different fragmentation functions
(left), and LO using EPS09LO1 and EKS98 with two different fragmentation functions (right). The blue bands are computed using the EPS09 error sets and fDSS fragmentation functions. The experimental PHENIX data \cite{Adler:2006wg} are shown by markers, and their error bars (boxes) stand for the point-to-point statistical (systematic) errors. Notice that the data points and their errors have been multiplied by a factor 1.039 (1.050) for the NLO (LO) case, which is well within the 9.7~\%  overall normalization error quoted by the experiment (see text for details). 
}
\label{fig:R_dAu_mb}
\end{center}
\end{figure}

Figure~\ref{fig:R_dAu_mb} shows the PHENIX data \cite{Adler:2006wg} and our NLO (left) and LO (right) results for the nuclear modification factor $\langle R_{\rm dAu}^{\pi^0}(p_T,y=0)\rangle$ in minimum-bias collisions. For the NLO calculation with EPS09sNLO1 (equivalently one may use EPS09NLO1, since the spatial dependence is here irrelevant) we used the NLO fragmentation functions from KKP \cite{Kniehl:2000fe}\footnote{The KKP set was also used in the EPS09 global analysis.}, AKK \cite{Albino:2008fy} and fDSS \cite{deFlorian:2007aj}. For the free proton PDFs, we use CTEQ6M \cite{Pumplin:2002vw}.
Correspondingly, the LO case was computed with EPS09sLO1 and EKS98s, using the KKP and fDSS LO fragmentation functions and CTEQ6.1L PDFs \cite{Pumplin:2002vw}. The renormalization scale $\mu$, factorization scale $Q$ and fragmentation scale $\mu_F$ are all fixed to $p_T$, the transverse momentum of the produced hadron. For details of the LO calculation, we again refer to Ref.~\cite{Eskola:2002kv}, while the NLO computation was performed by using the \verb=INCNLO= code \cite{incnlopage,Aversa:1988vb}.

From Fig.~\ref{fig:R_dAu_mb}, we notice the following: 
\textit{(i)} 
The EPS09 uncertainty bands for the NLO results are slightly smaller than in the LO case, reflecting the fact that the EPS09NLO gluons are somewhat better constrained in the antishadowing region than those of EPS09LO. 
\textit{(ii)}
In the small $p_T$ region there is a difference in the $p_T$ slopes between the EKS98 and EPS09LO1 results. This is caused by the weaker shadowing in EKS98. However, also the EKS98 results remain within the EPS09 error bars. 
\textit{(iii)}
The uncertainty caused by the differences in the fragmentation functions remains conveniently small in all cases. 

Regarding the data comparison in Fig.~\ref{fig:R_dAu_mb}, we emphasize the following important point:
In addition to the the statistical uncertainties (error bars) and point-to-point systematic errors (boxes), PHENIX quotes a 9.7~\% overall uncertainty which originates from the p+p reference and which is not included in the statistical error bars shown. Consequently, allowing for a shift of the data points and their errors by less than 9.7~\% and requiring the best possible overall fit to the data (using the fDSS FFs and by minimizing the $\chi^2$ with the point-to-point statistical and systematic errors added in quadrature), we have multiplied the data by a factor 1.039 (NLO) and 1.050 (LO). Such a few-percent shift is well within the uncertainty given by the experiment.
As already noticed in the EPS09 analysis \cite{Eskola:2009uj}, the resulting agreement with the data is quite good, both in LO and in NLO.

\begin{figure}[htbp]
\begin{center}
\includegraphics[width=7cm]{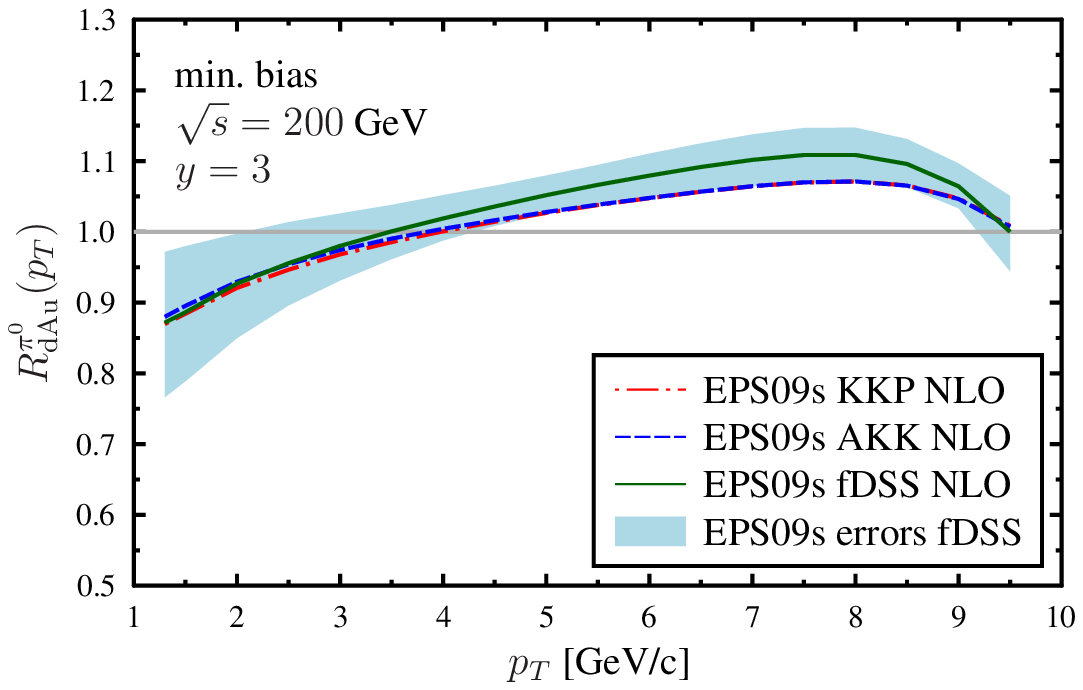}
\includegraphics[width=7cm]{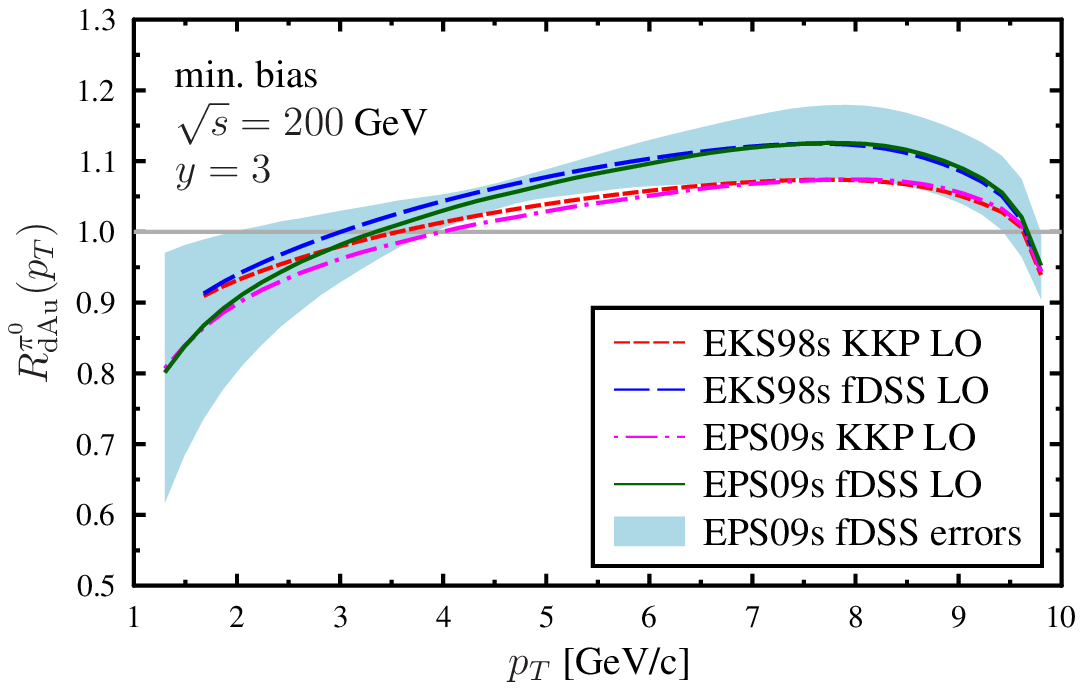}
\caption{The same as Fig.~\ref{fig:R_dAu_mb} but for neutral pions at a forward rapidity $y=3$. 
}
\label{fig:R_dAu_mby3}
\end{center}
\end{figure}

Figure \ref{fig:R_dAu_mby3} shows the ratio $\langle R_{\rm dAu}^{\pi^0}(p_T,y=3)\rangle$ at the forward region in minimum-bias collisions. Again, we show both the NLO (left) and LO (right) results with the same set-up as in Fig.~\ref{fig:R_dAu_mb} above. Now the differences between the fragmentation functions start to be visible, as one is probing their larger-$z$ tails where the uncertainties are larger: for the same pion $p_T$, the differences in the large-$z$ fragmentation functions map to different values of $x$ in the nPDFs.
We also notice that the LO calculation gives a stronger small-$p_T$ suppression than the NLO case, which is partly due to the different pace of the scale evolution with the NLO and LO nPDFs (cf. Fig.~\ref{fig:R_g_fits}) and partly because the NLO computation probes slightly higher values of $x$ than the LO case. Like in Fig.~\ref{fig:R_dAu_mb}, the EPS09 error band is smaller for NLO than for LO.\footnote{We note that there exist experimental data for $R_{\rm dAu}^{\pi^0}$ at larger rapidities \cite{Adams:2006uz,Adare:2011sc} which suggest a more substantial small-$p_T$ suppression than what the EPS09 and EKS98 predictions could accommodate. This deviation calls for a more detailed investigation which, however, 
is clearly beyond the scope of the present paper.}

\subsubsection{Centrality dependent $R_{\rm dAu}^{\pi^0}(p_T)$}

\begin{figure}[thbp]
\begin{center}
\includegraphics[width=15.0cm]{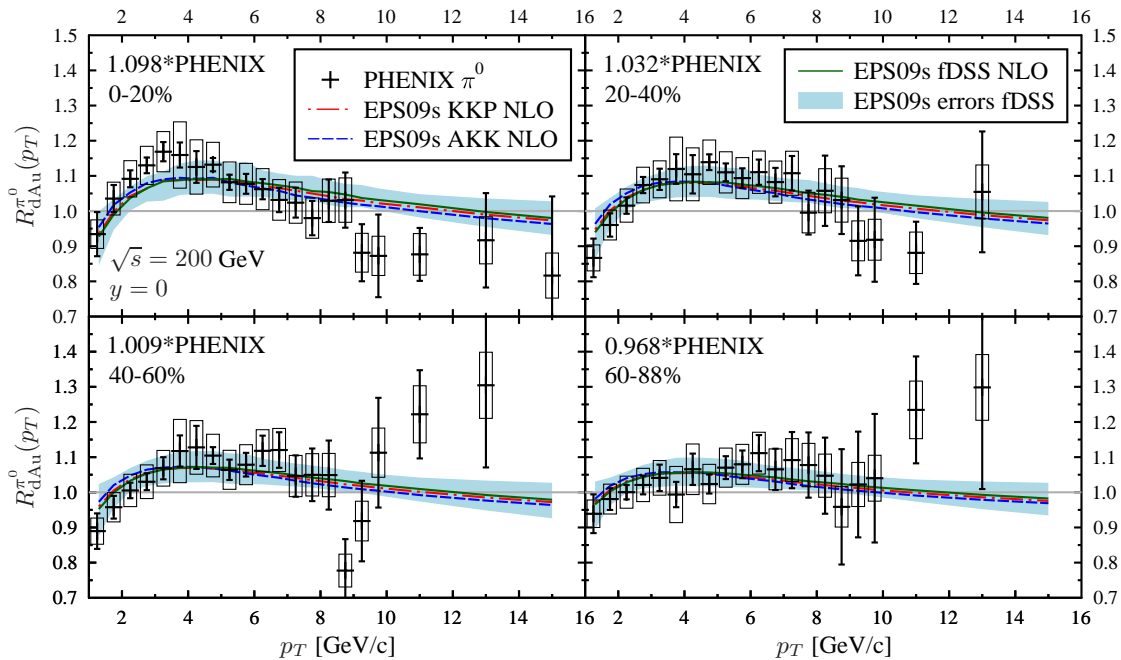} 
\caption{The nuclear modification factor $R^{\pi^0}_{\rm dAu}(p_T)$ for $\sqrt{s_{NN}} = 200 \text{ GeV}$ at $y=0$ for different centrality classes. Calculations are in NLO pQCD using EPS09sNLO1 and three different fragmentation functions. The blue error bands are computed with the error sets EPS09sNLOx (x=2,...,31) and fDSS, and the data are from PHENIX~\cite{Adler:2006wg}. The set-up and labeling are the same as in the left panel of Fig.~\ref{fig:R_dAu_mb}. Notice that the experimental data have been multiplied by a different factor in each panel, which all are well within the total overall normalization uncertainties given by the experiment (6.6, 6.7, 8.5, 9.6~\% for the four centrality bins from the Glauberization and 9.7~\% from the p+p baseline.)
}
\label{fig:R_dAuNLO}
\end{center}
\end{figure}

\begin{figure}[hbtp]
\begin{center}
\includegraphics[width=15.0cm]{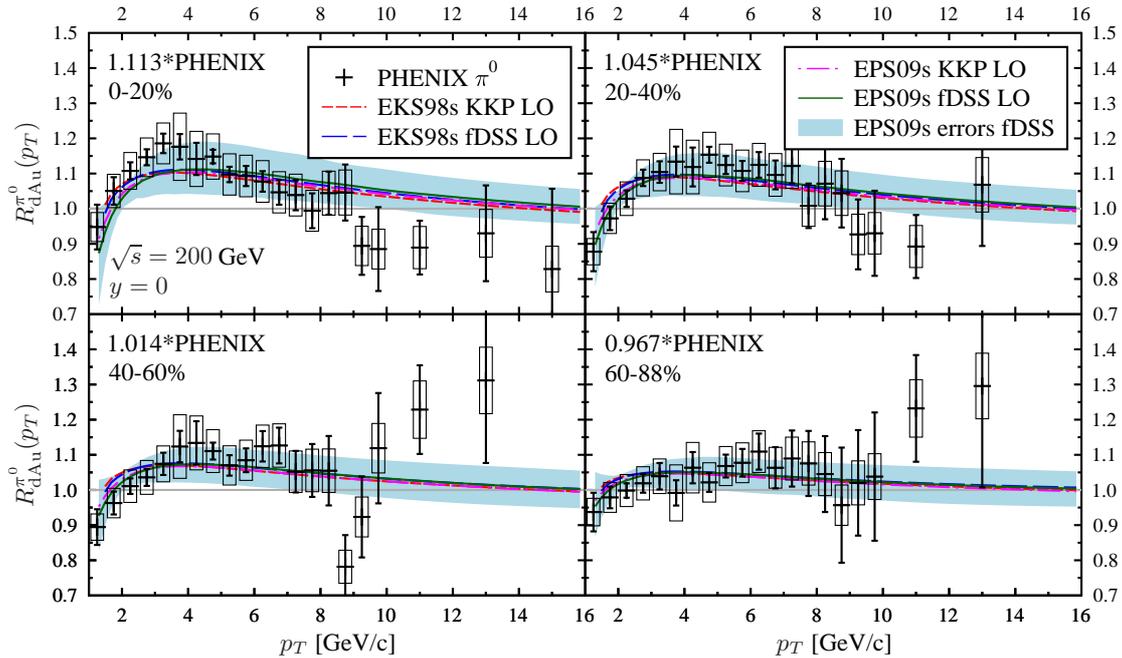}
\caption{The same as in Fig.~\ref{fig:R_dAuNLO} but with LO calculations using EPS09sLO1 and EKS98s and for two different fragmentation functions, and the blue error band is computed with the EPS09sLOx (x=2,...,31) and fDSS. Notice again the overall multiplicative factors for the experimental data.
}
\label{fig:R_dAuLO}
\end{center}
\end{figure}

Let us then look at the centrality dependence of the ratio $R_{\rm dAu}^{\pi^0}$. Our NLO and LO results for $R_{\rm dAu}^{\pi^0}$ in the centrality bins $0-20\%$, $20-40\%$, $40-60\%$ and $60-88\%$ are plotted in Figs.~\ref{fig:R_dAuNLO} and \ref{fig:R_dAuLO}, correspondingly, together with the PHENIX data. Table ~\ref{tab:dAuparams} lists the 
impact parameter ranges and average number of binary collisions for each centrality class, obtained from the optical Glauber model with $\sigma_{inel}^{NN} = 42 \text{ mb}$. 

\begin{table}[tbh]
\caption{The centrality classes as impact parameter intervals, and average number of binary collisions from optical Glauber model for d+Au collisions at $\sqrt{s_{NN}} = 200 \text{ GeV}$ using  $\sigma_{inel}^{NN} = 42 \text{ mb}$.}
\begin{center}
\begin{tabular}{rccc}
\hline
& $b_1 \textrm{ [fm]}$ & $b_2 \textrm{ [fm]} $ & $\langle N_{bin} \rangle $\\
\hline
$0-20\%$ & 0.0 & 3.798 & 15.57 \\
$20-40\%$ & 3.798 & 5.371 & 10.95\\
$40-60\%$ & 5.371 & 6.583 & 6.013\\
$60-88\%$ & 6.583 & 8.336 & 2.353\\
\hline
\end{tabular}
\end{center}
\label{tab:dAuparams}
\end{table}

Again, it is important to consider the different overall normalization errors in the experimental data. For the centrality-dependent ratios plotted in Figs.~\ref{fig:R_dAuNLO} and \ref{fig:R_dAuLO} there still is the 9.7~\% overall systematic error due to the p+p baseline discussed above. In addition to this, an overall normalization error of 6.6--9.6~\% arising from the determination of the average number of binary collisions, is quoted separately for each centrality bin. Following again the same procedure as for Fig.~\ref{fig:R_dAu_mb}, we multiply the data and their point-to-point errors by a factor which minimizes the difference to our calculation. Even the largest upwards shift, 11.3~\% for the centralmost bin in the LO case, is well within the acceptable total overall normalization error quoted by the experiment. Note also the systematic decrease of the multiplication factor from central to peripheral collisions, which we believe is due to the difference in the experimental and Glauber-model definitions of the centrality classes.


From Figs.~\ref{fig:R_dAuNLO} and \ref{fig:R_dAuLO} we observe that
within the experimental and theoretical uncertainties our calculations are consistent with the measurements. Especially the centrality systematics obtained from our spatially-dependent nPDFs agrees quite well with the data: the nuclear modifications are strongest in the most central collisions and systematically weaken when going to more peripheral collisions. This is especially nicely reflected in the region $1.3\le p_T\le 4$~GeV, where the $p_T$ slopes (which are not affected by the overall multiplications) become steeper towards more central collisions.   
We also see that, like in the minimum-bias case, the EPS09 error bands are slighty smaller for the NLO than for the LO case, and that the uncertainties arising from the fragmentation functions remain small.

\begin{figure}[thbp]
\begin{center}
\includegraphics[width=15cm]{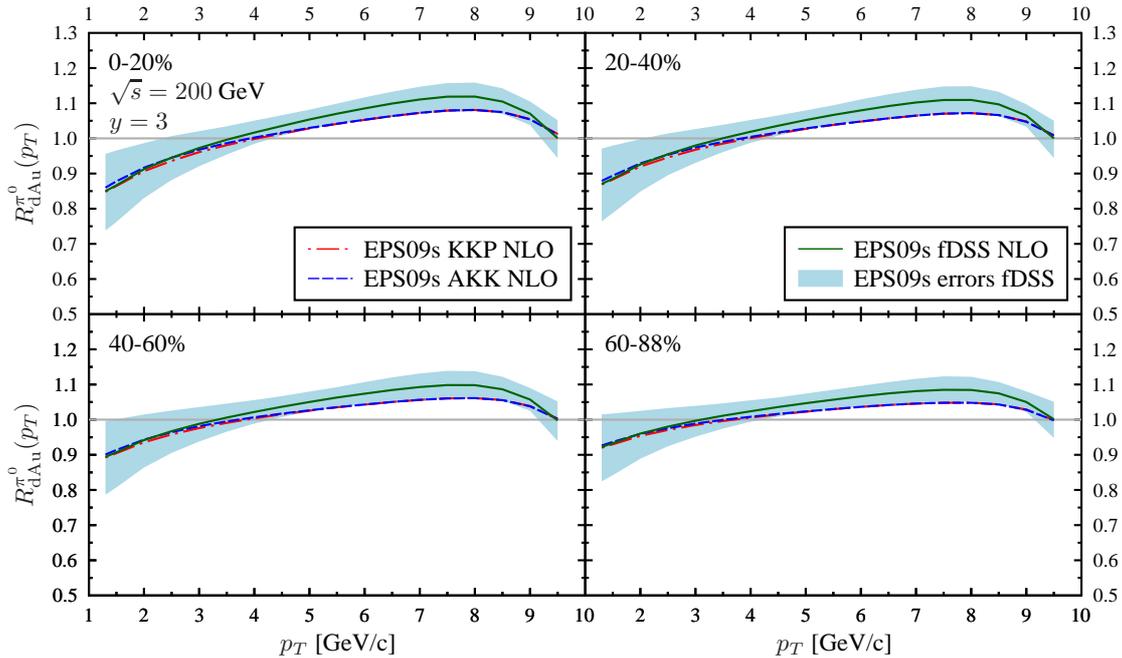}
\caption{The nuclear modification factor $R^{\pi^0}_{\rm dAu}(p_T)$ for $\sqrt{s_{NN}} = 200 \text{ GeV}$ at $y=3$ in different centrality classes. The computation is done in NLO pQCD using EPS09sNLO1 and three different fragmentation functions. The error bands are computed with the EPS09sNLOx (x=2,...,31) and fDSS. 
}
\label{fig:R_dAu_y3NLO}
\end{center}
\end{figure}

\begin{figure}[thbp]
\begin{center}
\includegraphics[width=15cm]{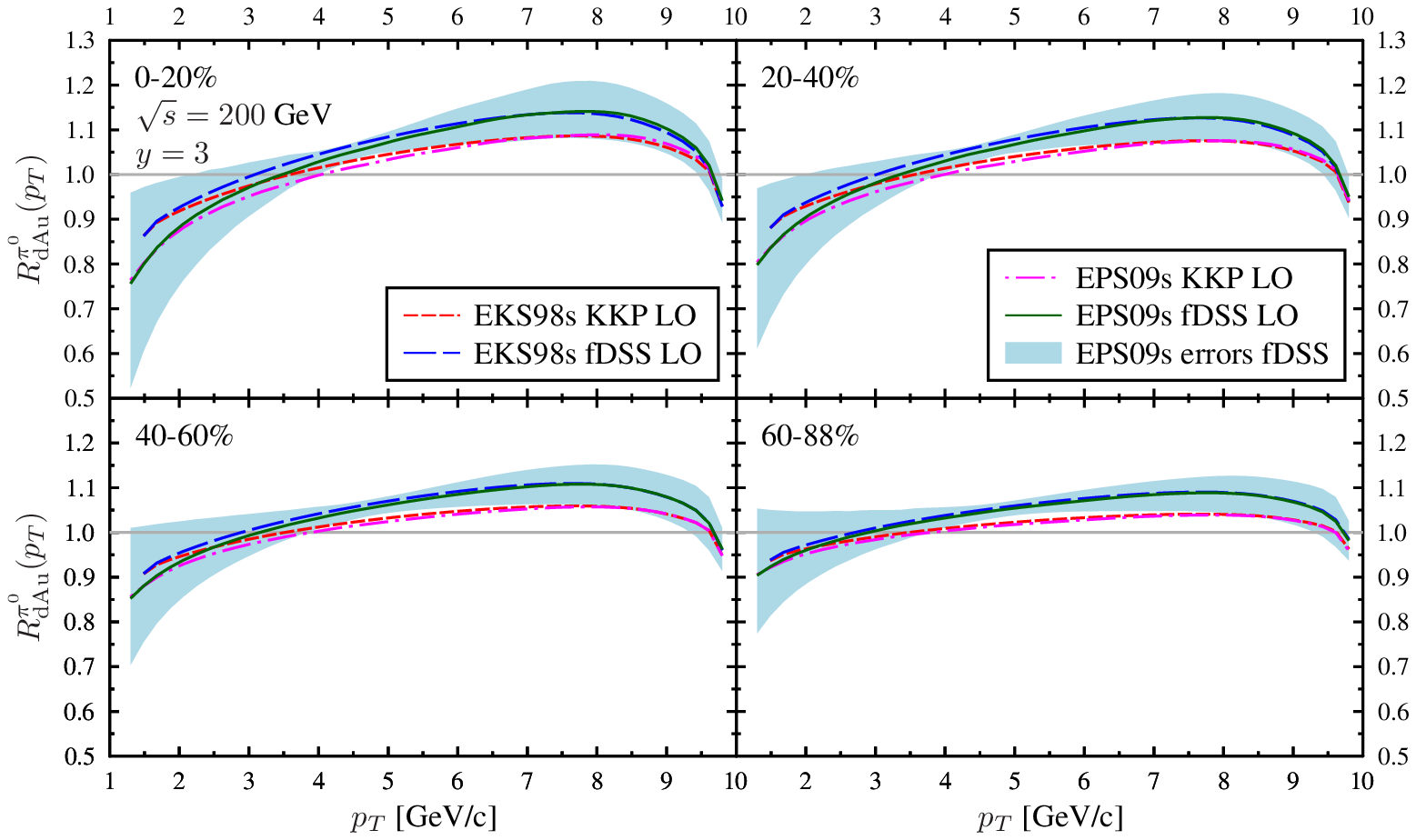}
\caption{The same as Fig.~\ref{fig:R_dAu_y3NLO} but for LO pQCD using EPS09sLO1 and EKS98s with two different fragmentation functions. The error bands are computed with the EPS09sLOx (x=2,...,31) and fDSS.
}
\label{fig:R_dAu_y3LO}
\end{center}
\end{figure}

In Figs.~\ref{fig:R_dAu_y3NLO} and \ref{fig:R_dAu_y3LO} we plot also our NLO and LO results for $R_{\rm dAu}^{\pi^0}$ at $\sqrt{s_{NN}} = 200 \text{ GeV}$ at a forward rapidity, $y=3$, in the different centrality bins. In the forward region the nuclear modifications are larger since we now are probing smaller $x$ values in the nPDFs than in the mid-rapidity region. Like in the minimum-bias case, we notice that the difference between the fragmentation function sets we use, becomes noticeable in the forward region. Again the small-$p_T$ suppression is stronger and nPDF-orginating uncertainties are larger for the LO case. 

\subsection{Predictions for p+Pb collisions at LHC}

In the heavy-ion program of the LHC at CERN, there are now plans to collide protons with lead nuclei. Such collisions would be very useful for testing the QCD factorization and the universality of nPDFs, as well as for constraining the nuclear PDF modifications further especially at small values of $x$. Also the centrality dependence of nPDFs could be examined in these collisions via inclusive hadron production, similarly to the RHIC d+Au collisions discussed above but without the theoretical uncertainties arising from modeling the deuterium geometry. Thus, it is interesting to see what are the predictions from our spatially dependent nPDFs for these collisions. 

\begin{figure}[t!]
\begin{center}
\includegraphics[width=14.5cm]{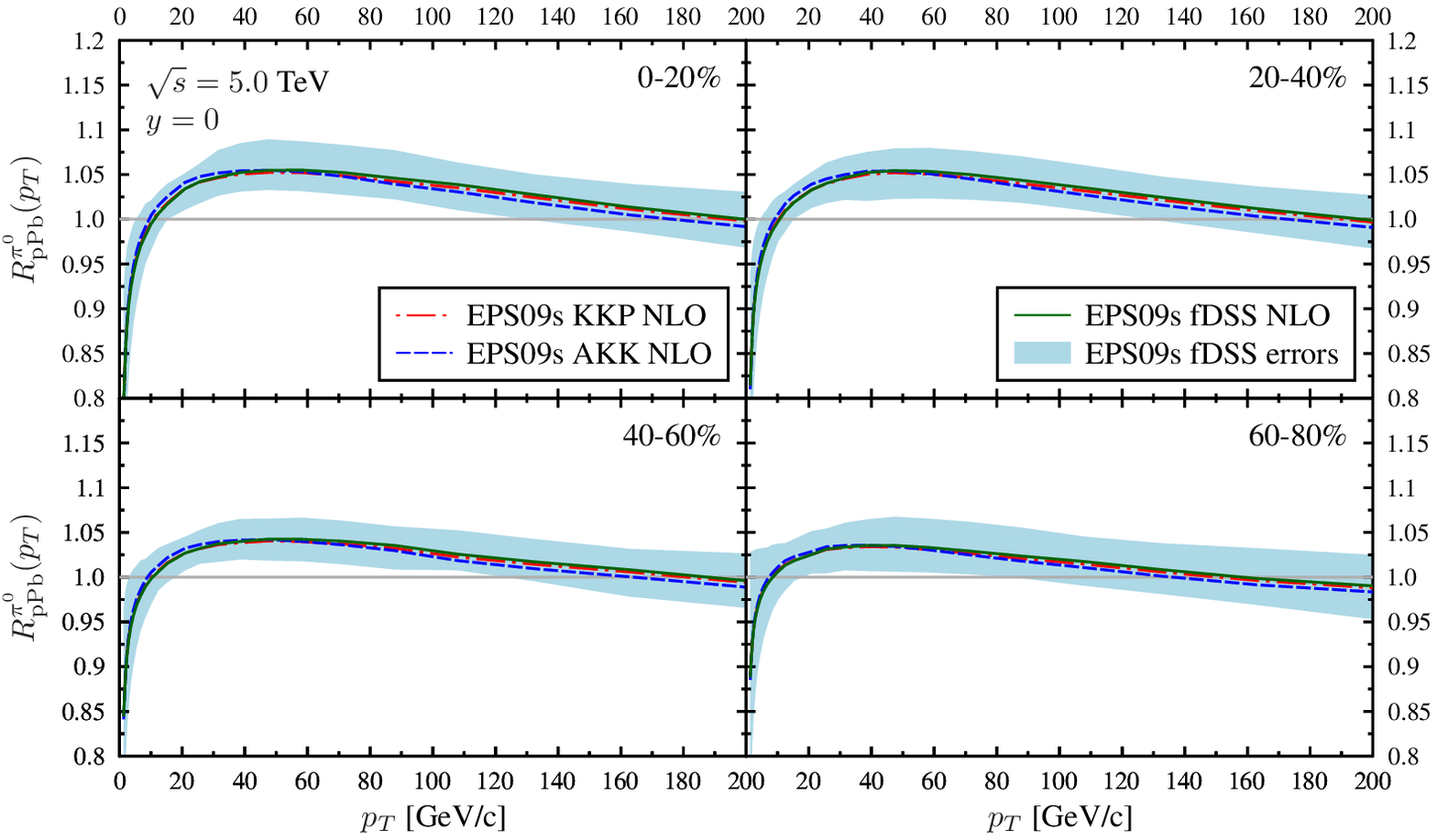}
\caption{The nuclear modification factor $R^{\pi^0}_{\rm pPb}(p_T)$ for $\sqrt{s} = 5.0 \text{ TeV}$ at $y=0$ for four different centrality classes, computed in NLO pQCD using EPS09sNLO1 and three different fragmentation functions. The error bands have been obtained with EPS09sNLOx (x=2,...,31) and fDSS.} 
\label{fig:R_pPb_pi0_y0NLO}
\includegraphics[width=14.5cm]{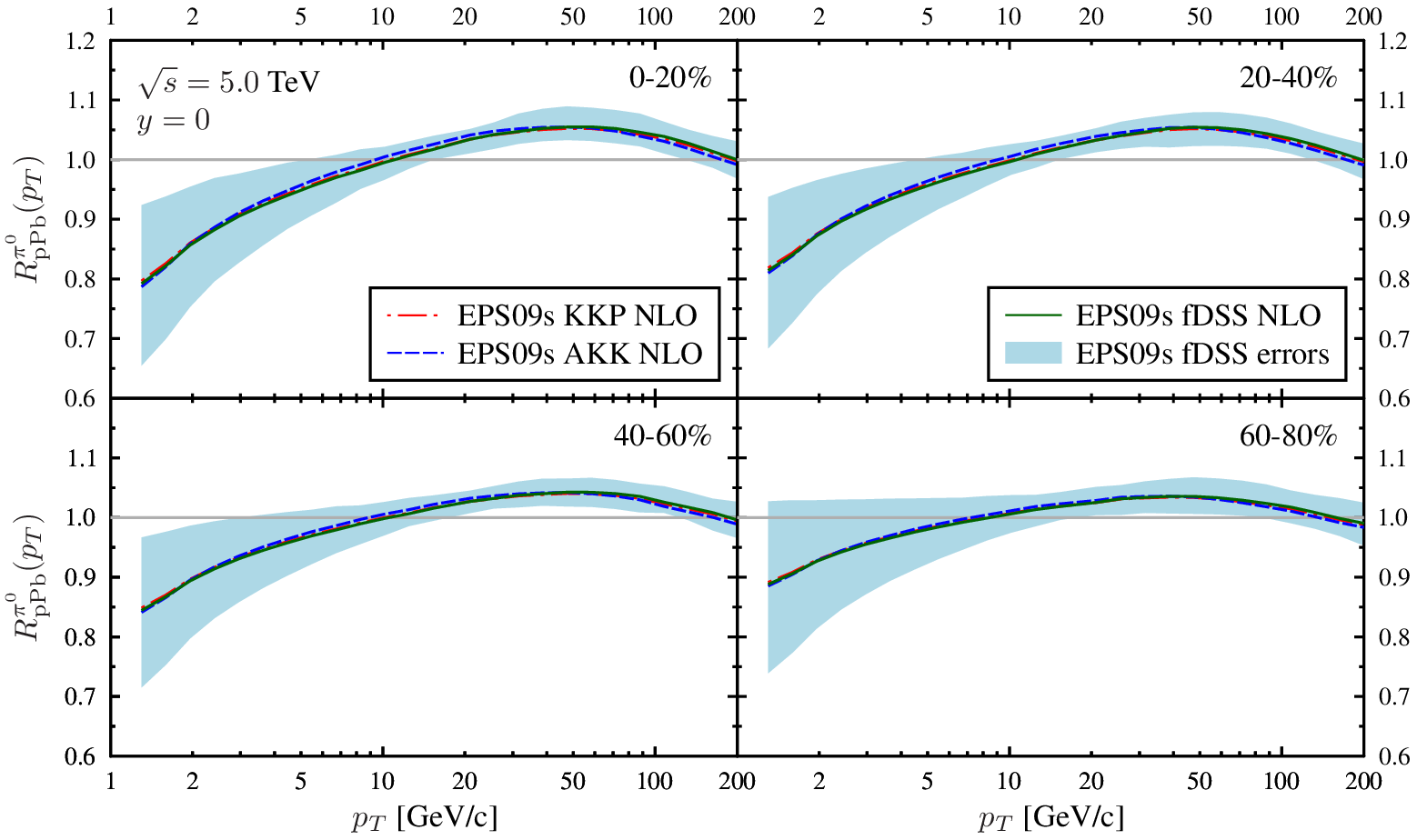}
\caption{The same as Fig.~\ref{fig:R_pPb_pi0_y0NLO} but in a logarithmic scale to emphasize the small-$p_T$ region.}
\label{fig:R_pPb_pi0_y0_logNLO}
\end{center}
\end{figure}

In Fig.~\ref{fig:R_pPb_pi0_y0NLO} we plot our EPS09sNLO results for the nuclear modification factor $R^{\pi^0}_{\rm pPb}(p_T)$ for neutral pion production in p+Pb collisions at $\sqrt{s_{NN}} = 5.0 \text{ TeV}$ at $y=0$
in four different centrality classes.\footnote{Very recently, the LHC moved up to collisions energies $\sqrt{s_{\rm pp}}=8$~TeV, hence we take $\sqrt{s_{NN}} = \sqrt{s_{\rm pp}}\sqrt{Z/A}\approx 5.0$~TeV. Note also that $y$ is the rapidity in the $NN$ cms frame, i.e. we do not include the rapidity shift due to the antisymmetric collision.}
We use again the KKP, AKK and fDSS fragmentation functions here. The uncertainty bands arising from EPS09sNLO are computed using fDSS. The inelastic cross section $\sigma_{inel}^{NN} = 70 \text{ mb}$ for this $\sqrt{s_{NN}}$ is obtained from Fig. 5 of Ref.~\cite{Antchev:2011vs}. This leads to the impact parameter values and the average number of binary collisions for each centrality class given in Table~\ref{tab:pPbparams}. For the projectile proton, we have not assumed any spatial size, so that relative to the deuterium case above, in the collision geometry we replace the thickness function $T_{\rm d}(\textbf{s})$ by $\delta(\textbf{s})$ and the overlap function $T_{{\rm d}A}(\textbf{b})$ by the thickness function $T_{\rm Pb}(\textbf{b})$.

\begin{table}[htb]
\caption{The centrality classes as impact parameter intervals, and average number of binary collisions from optical Glauber model for p+Pb collisions at $\sqrt{s} = 5.0 \text{ TeV}$, with $\sigma_{inel}^{NN} = 70 \text{ mb}$.}
\begin{center}
\begin{tabular}{rccc}
\hline
& $b_1 \textrm{ [fm]}$ & $b_2 \textrm{ [fm]} $ & $\langle N_{bin} \rangle $\\
\hline
$0-20\%$  & 0.0   & 3.471 & 14.24 \\
$20-40\%$ & 3.471 & 4.908 & 11.41\\
$40-60\%$ & 4.908 & 6.012 & 7.663\\
$60-80\%$ & 6.012 & 6.986 & 3.680\\
\hline
\end{tabular}
\end{center}
\label{tab:pPbparams}
\end{table}

As can be seen from Fig.~\ref{fig:R_pPb_pi0_y0NLO}, the nuclear modifications are strongest in the small-$p_T$ region in all centrality classes. To see the behaviour of $R_{\rm pPb}^{\pi^0}$ in this region more clearly, we plot the results also in logarithmic scale in Fig.~\ref{fig:R_pPb_pi0_y0_logNLO}. We again observe the general behavior which follows from the spatial dependence of the nPDFs: the nuclear modifications are stronger in the central collisions and weaker in the peripheral collisions. We also notice that the three fragmentation function sets yield almost identical results.

\begin{figure}[thbp]
\begin{center}
\includegraphics[width=7cm]{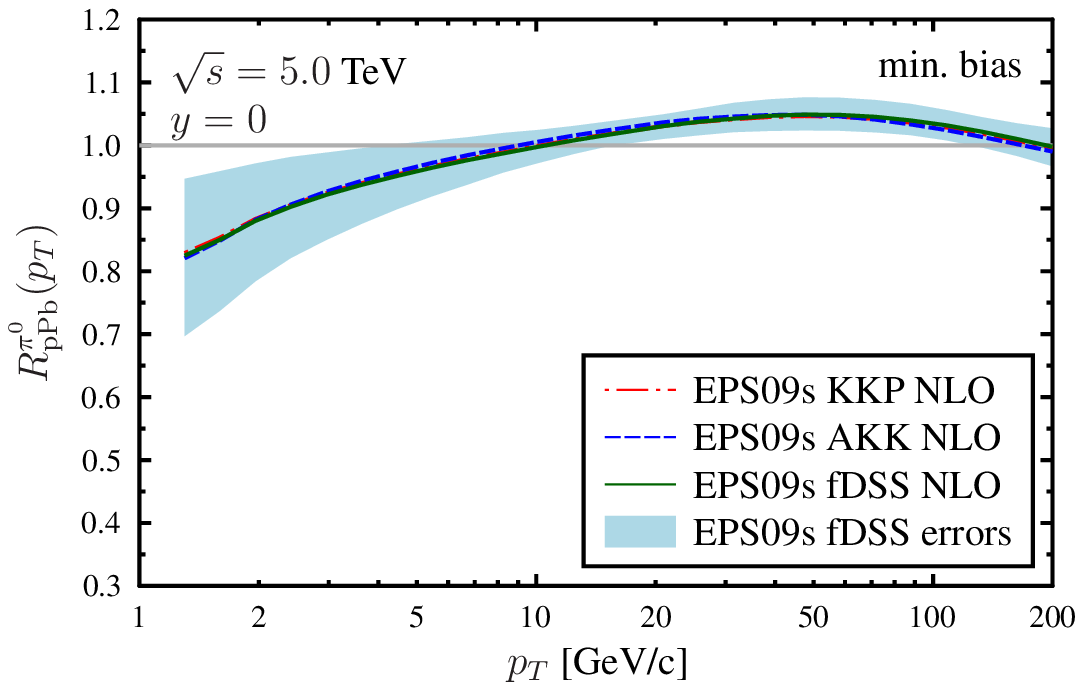} 
\includegraphics[width=7cm]{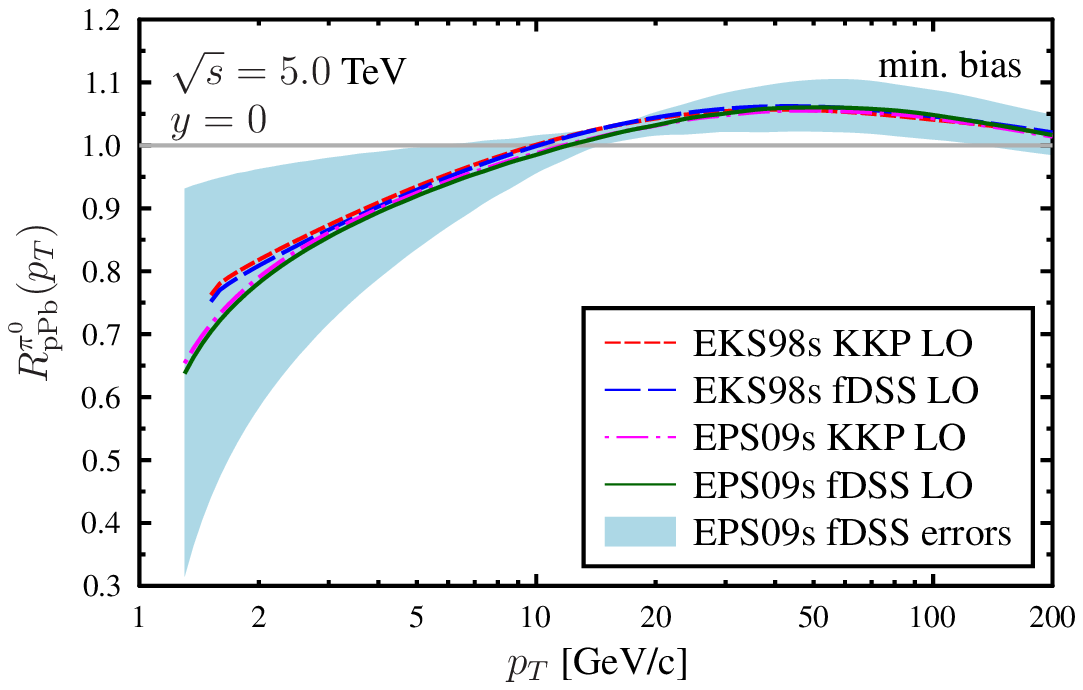}
\caption{\textbf{Left: }The nuclear modification factor $R^{\pi^0}_{\rm pPb}(p_T)$ for $\sqrt{s_{NN}} =5.0 \text{ TeV}$ at $y=0$, computed in NLO pQCD using EPS09sNLO1 (left panel) and three different fragmentation functions. The error band is computed using  EPS09sNLOx (x=2,...,31) and fDSS.
\textbf{Right:} The same but in LO pQCD with EKS98s and EPS09sLO1 with two different fragmentation functions, and 
the error band is for EPS09sLOx (x=2,...,31) with fDSS.}
\label{fig:R_pPb_pi0_y0MB}
\end{center}
\end{figure}

Figure~\ref{fig:R_pPb_pi0_y0MB} shows the corresponding ratio in minimum bias p+Pb collisions, computed both in NLO (left) and in LO (right). Like in the forward-rapidity case at RHIC, and for the same reasons, the EPS09NLO leads to a weaker small-$p_T$ suppression than EPS09LO and EKS98, and the uncertainty band is clearly smaller for the NLO case.  

\begin{figure}[thbp]
\begin{center}
\includegraphics[width = 15cm]{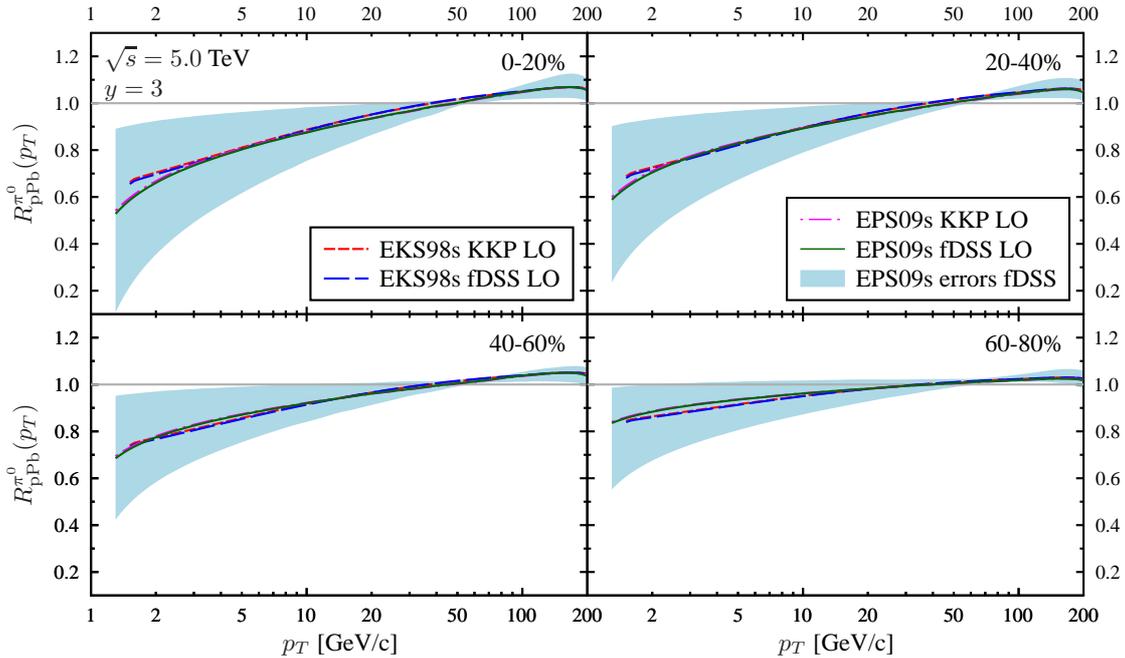}
\caption{The nuclear modification factor $R^{\pi^0}_{\rm pPb}(p_T)$ for $\sqrt{s_{NN}} = 5.0 \text{ TeV}$ at $y=3$ for four different centrality classes, computed in LO pQCD using EPS09sLO1 and two different fragmentation functions. The error band is computed using EPS09sLOx (x=2,...,31) and fDSS.}
\label{fig:R_pPb_pi0_y3_log}
\end{center}
\end{figure}

As a probe of nuclear gluons even deeper in the small-$x$ shadowing region, we plot in Fig.~\ref{fig:R_pPb_pi0_y3_log} our LO results\footnote{For $y=3$, we could not obtain reliable results with INCNLO at $p_T<5$~GeV for this $\sqrt{s_{NN}}$, hence only the LO results are shown here.} 
for $R^{\pi^0}_{\rm pPb}$ at a forward rapidity, $y=3$, for the four centrality classes.
Again the KKP and fDSS fragmentation functions are used, and we see that they yield very similar results. 
We should also point out that the EPS09sLO error band for the peripheral bin in Fig.~\ref{fig:R_pPb_pi0_y3_log} can be regarded as an underestimate in that it has been computed without the error set EPS09sLO7. The reason for this is that the error set EPS09LO7 gives in fact antishadowing at smallest $x$ for the lightest nuclei, and in the EPS09sLO this maps into an antishadowing near the edges of a large nucleus. This unphysical feature can be cured only by redoing the EPS09LO global fit with an improved $A$-dependence of the fit functions. In the meantime, we suggest that  a physically more meaningful upper limit for the LO error band in the small-$x$ region for the peripheral bin can thus be obtained without this LO error set.  

\begin{figure}[thbp]
\begin{center}
\includegraphics[width=7cm]{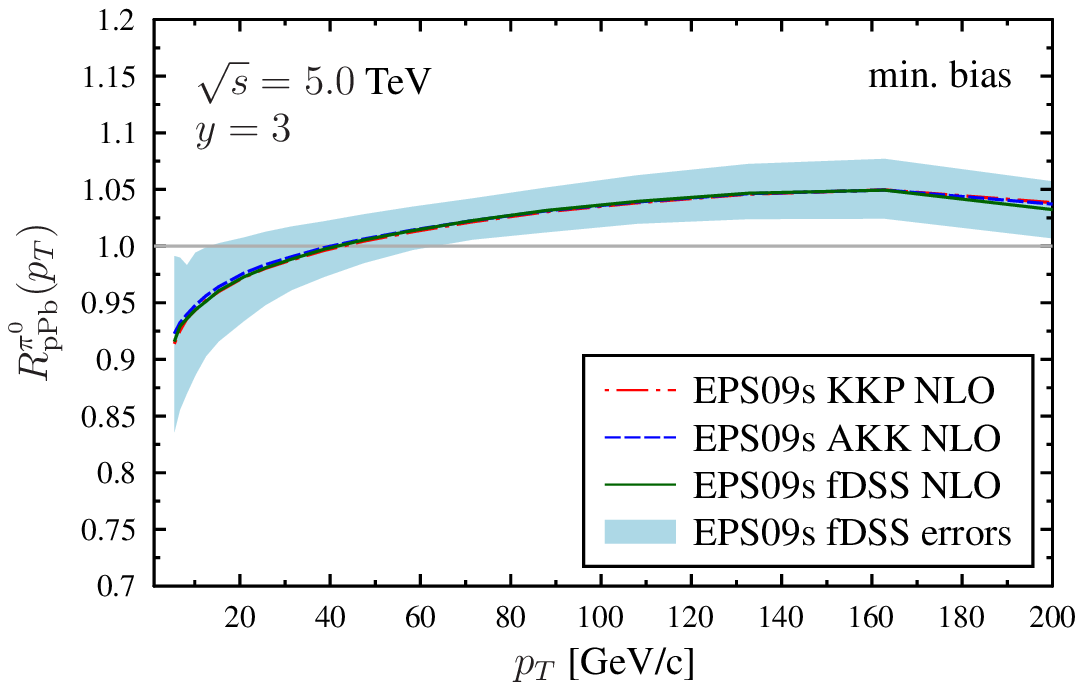} 
\includegraphics[width=7cm]{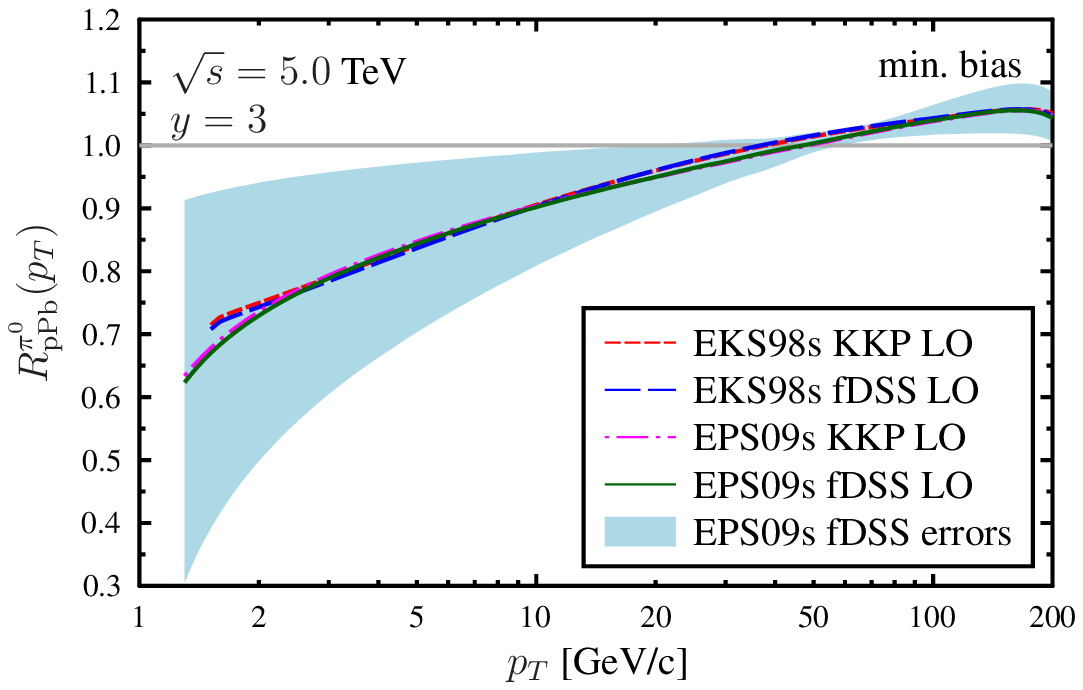}
\caption{Same as Fig.~\ref{fig:R_pPb_pi0_y0MB} but for a forward rapidity $y=3$; the NLO (LO) results are on the left (right).}
\label{fig:R_pPb_pi0_y3MB}
\end{center}
\end{figure}

Finally, in Fig.~\ref{fig:R_pPb_pi0_y3MB} we show the minimum-bias $R^{\pi^0}_{\rm pPb}$ at $y=3$ for the NLO case at $p_T\ge5$ ~GeV and for the LO case starting from $p_T=1.3$~GeV. Note the linear(logarithmic) $p_T$ scale on the left (right). Again we notice the weaker suppression and smaller error bands in the NLO case. Comparing the right panels of Figs.~\ref{fig:R_pPb_pi0_y3MB} and Fig.~\ref{fig:R_pPb_pi0_y0MB}, we see that the smallest-$p_T$ suppressions are of similar magnitude. This is because the ratio $R^{\pi^0}_{\rm pPb}$ in the small-$p_T$ region at the LHC probes already at $y=0$ the flat part of the shadowing assumed as an input in EPS09 (cf. Fig.~\ref{fig:R_g_A_fits}). Hence, a measurement of $R^{\pi^0}_{\rm pPb}$ both in the mid- and forward-rapidities can be expected to serve as a relevant constraint for the smallest-$x$ shadowing region.  

\section{Summary and Conclusions}

We have developed a framework to determine the spatial dependence of the nuclear modifications of PDFs in such a way that the outcome is consistent with the globally analysed EKS98 and EPS09 nPDFs which in turn are DGLAP-based fits to nuclear hard-process data. Both the LO and NLO cases have been considered, and with EPS09 the spatial dependence has been extracted also for all the 30 error sets. Correspondingly, we call the obtained spatially dependent nPDF sets \verb=EPS09s= and \verb=EKS98s=.

The spatial dependence is introduced in terms of powers of the nuclear thickness functions $T_A(\mathbf{s})$. Regarding the power series $r_i^A(x,Q^2,\mathbf{s}) = 1+\sum\limits_{j=1}^{n} c^i_j(x,Q^2)\left[T_A(\mathbf{s})\right]^j$, 
we have shown that the 1-parameter approach ($n=1$, used e.g. in \cite{Eskola:1991ec,Vogt:2004hf}) is not sufficient for reproducing $A$ systematics in the nPDFs, and that we obtain a good overall agreement with the globally analysed averaged nPDFs when we include terms up to $[T_A]^4$. The outcome of the performed fits, the sets of coefficients $\{c^i_j(x,Q^2)\}$ for each parton flavor $i$ at each $x$ and $Q^2$, are tabulated separately for each of the nPDF sets we considered. These tables along with a routine for interpolation and computing the needed thickness functions are downloadable at \cite{downloadpage}. 
 
As a concrete application of our framework, we calculated the nuclear modification factor $R_{AA}^{\rm 1jet}$ for LO primary partonic jet production at different centralities in Au+Au collisions at RHIC and in Pb+Pb at the LHC. We observed that while the central $R_{AA}^{\rm 1jet}$ is quite close to the minimum-bias ratio $\langle R_{AA}^{\rm 1jet}\rangle $ and the peripheral $R_{AA}^{\rm 1jet}$ differs fairly significantly from unity, the central-to-peripheral ratio $R_{CP}^{\rm 1jet}$ differs clearly from the ratio $\langle R_{AA}^{\rm 1jet}\rangle$. 

We also compared our NLO and LO calculations of the nuclear modification factor of neutral pion production in d+Au collisions, $R_{\rm dAu}^{\pi^0}$, in different centrality classes at mid-rapidity with the PHENIX data \cite{Adler:2006wg}. Within all the given errors in the experimental data, the nPDF uncertainties, 
and the possible differences between the experimental and optical Glauber model centrality classes,
the EPS09s results are remarkably consistent with the centrality systematics. To our knowledge, this is the first time this has been demonstrated. Especially, our EPS09s results seem to reproduce the low $p_T$ slope of the data very well in all centrality classes. 

More constraints for the spatial dependence of the nuclear PDFs, and gluons in particular, could be obtained from the scheduled p+Pb collisions at the LHC. We demonstrated this by calculating the NLO and LO predictions from our framework for the ratio $R_{\rm pPb}^{\pi^0}$ in different centrality classes both at mid-rapidity $y=0$ and forward rapidity $y=3$ for $\sqrt{s_{NN}} = 5.0 \text { TeV}$, which corresponds to the recently achieved p+p cms-energy. 

We believe that the nPDF development presented here is an important step forwards, as now a user may for the first time compute the centrality-dependent hard cross-sections more consistently with globally analysed nPDFs.  Our spatially dependent nPDFs should also be applicable in Monte Carlo simulations of nuclear collisions, where the analogues of the thickness functions should be straightforwardly obtainable. In addition, our work should also give an idea how the future global analyses of nPDFs could be constructed so that the spatial dependence would be built in right from the start and not afterwards as has been the case here. 

\section*{Acknowledgements}
We thank M. Wysocki, J. Rak, T. Lappi, T. Renk and H. M\"{a}ntysaari for discussions.
We gratefully acknowledge the following financial support: 
I.H. from the Magnus Ehrnrooth Foundation; 
K.J.E., I.H. and H.H. from the Academy of Finland, K.J.E.'s Project No. 133005.
C.A.S. is supported by the European Research Council grant HotLHC ERC- 2001-StG-279579 and by Ministerio de Ciencia e Innovaci\'on of Spain. C.A.S. is a Ram\'on y Cajal researcher.
This work (H.H.) was supported in part by the U.S. Department of Energy under Grant  DE- FG02-93ER40771.

\appendix

\section{Nuclear Collision Geometry}
\label{sec:geometry}

For clarity, we specify here the modeling and parameters of the nuclear collision geometry, i.e. the nuclear thickness functions $T_A(\mathbf{s})$ and $T_{\rm d}(\mathbf{s})$ (see Refs.~\cite{Eskola:1988yh,Kharzeev:2002ei}) and Glauber modeling (see Refs.~\cite{Miller:2007ri,wong1994intr}), applied in this study. The calculations of $T_A(\mathbf{s})$ and $T_{\rm d}(\mathbf{s})$ are included both in the \verb=EKS98s= and \verb=EPS09s= codes.

\subsection{Nuclear Thickness Functions}
\label{sec:TA}

\subsubsection{Large nuclei}

The total amount of nuclear matter in a colliding nucleus $A$ in the beam direction $z$ at a transverse position $\mathbf{s}$ is given by the nuclear thickness function  
\begin{equation}
T_A(\mathbf{s}) = \int_{-\infty}^{\infty} \mathrm{d}z \, \rho_A(\mathbf{s},z),
\label{eq:T_A}
\end{equation}
where $\rho(\mathbf{s},z)$ is the nucleonic number-density of the nucleus, with a normalization convention 
\begin{equation}
A = \int \mathrm{d}^2\mathbf{s} \,T_A(\mathbf{s}).
\label{eq:ta_normalization}
\end{equation}
In this study we use the standard two parameter Woods-Saxon density profile for $\rho_A$,
\begin{equation}
\rho_A(\mathbf{s},z) = \dfrac{n_0}{1 + \exp \left[\frac{\sqrt{\mathbf{s}^2 + z^2} + R_A}{d}\right]},
\label{eq:rho_A}
\end{equation}
which is a good approximation for nuclei with $A \ge 4$. The parameter values for the Woods-Saxon distribution are
\begin{eqnarray}
d &=& 0.54 \textrm{ fm} \\
R_A &=& 1.12 A^{1/3} - 0.86 A^{-1/3} \textrm{ fm}, 
\end{eqnarray}
and for large nuclei the normalization condition (\ref{eq:ta_normalization}) fixes the constant $n_0$ as
\begin{equation}
n_0 = \frac{3}{4} \frac{A}{\pi R_A^3}\frac{1}{(1+(\frac{\pi d}{R_A})^2)}.
\end{equation}

\subsubsection{Deuterium}
\label{sec:deuterium}
For the thickness function of a deuterium nucleus, the above Woods-Saxon density profile is obviously not applicable anymore. Instead, one may formulate this with the deuteron wavefunction which describes the probability amplitude for the proton and neutron to be separated by a distance $\mathbf{r_{pn}}$. This can be written in terms of the $^3S_1$- and $^3D_1$-wave components as (see e.g. Ref. \cite{Hulthen,Garcon:2001sz})
\begin{equation}
\psi_{M}(\mathbf{r_{pn}}) = \dfrac{u(r_{pn})}{r_{pn}}{\cal Y}_{101}^M(\Omega)
+ \dfrac{w(r_{pn})}{r_{pn}}{\cal Y}_{121}^M(\Omega),
\label{eq:deuteron_wf}
\end{equation}
where the spin-spherical harmonics ${\cal Y}_{JLS}^M(\Omega)$, with $S=1$, consist of three components, 
\begin{equation}
{\bigg[}{\cal Y}_{101}^M(\Omega){\bigg]}_{m_S=\pm 1,0} = \langle \Omega, m_S|L S J M\rangle
= \sum_{M_L,M_S}\langle LSM_LM_S|LSJM \rangle\, Y_{LM_L}(\Omega)\, \delta_{m_SM_S}.
\end{equation}
For the radial parts we use the Hulthen form as in \cite{Hulthen,Kharzeev:2002ei},
\begin{align}
u(r_{pn}) =& N \sqrt{1-\epsilon^2}\left[ 1 - \mathrm{e}^{-\beta(\alpha r_{pn} - x_c)} \right] \mathrm{e}^{\alpha r_{pn}} \theta(\alpha r_{pn} - x_c) \\
w(r_{pn}) =& N \epsilon \left[ 1 - \mathrm{e}^{-\gamma(\alpha r_{pn} - x_c)} \right]^2 \mathrm{e}^{-\alpha r_{pn}} \\
&\left[ 1 + \dfrac{3(1 - \mathrm{e}^{-\gamma \alpha r_{pn}})}{\alpha r_{pn}} + \dfrac{3(1 - \mathrm{e}^{-\gamma \alpha r_{pn}})^2}{(\alpha r_{pn})^2} \right] \theta(\alpha r_{pn} - x_c), \notag
\end{align}
where
$N^2 = \dfrac{2 \alpha}{1 - \alpha \rho},$ in which 
$\alpha^{-1} = 4.316 \, \mathrm{fm}$ is related to the experimentally measured binding energy,
and $\rho$ is fixed by normalization, $\int \mathrm{d}^3 \mathbf{r_{pn}} |\psi_{M}(\mathbf{r_{pn}})|^2=1$. 
For the other parameters, obtained by fitting to experimental data, we
use the "set 1" quoted in \cite{Kharzeev:2002ei}:
\begin{equation}
\begin{array}{rclrcl}
\beta &=& 4.680 &\quad \gamma &=& 2.494  \\
\epsilon &=& 0.03232 &\quad x_c &=& 0
\end{array}
\end{equation}
The angular-averaged radial probability distribution for the proton-neutron distance $r_{pn}$ in deuteron is given by
\begin{equation}
P_{pn}(\mathbf{r_{pn}})=\frac{1}{4\pi}\int d\Omega |\psi(\mathbf{r_{pn}})|^2 = \frac{1}{4\pi}\dfrac{u^2(r_{pn})+w^2(r_{pn})}{r_{pn}^2}.
\end{equation}
For computing the thickness function $T_{\rm d}(\mathbf{s})$ as in Eq.~(\ref{eq:T_A}), we need the nucleon density distribution $\rho_{\rm d}(\mathbf{r})$ at a distance $\mathbf{r}$ from the center of mass of the deuteron. Assuming identical proton and neutron masses, we have $r = r_{pn}/2$. In addition, we require the normalization of $T_{\rm d}$ to be in line with Eq.~(\ref{eq:ta_normalization}). We thus have 
\begin{equation}
T_{\rm d}(\mathbf{s}) = \int_{-\infty}^{\infty} \mathrm{d}z \, \rho_{\rm d}(\mathbf{s},z),
\quad\quad
\rho_{\rm d}(\mathbf{s},z) = 16 P_{pn}(2\mathbf{r}),
\quad
\int \mathrm{d}^2\mathbf{s} \,T_{\rm d}(\mathbf{s}) = 2. 
\end{equation}

\subsection{Optical Glauber Model}
\label{sec:glauber}
Let us then specify the optical Glauber modeling applied for nuclear collisions in this study. For further discussion, see e.g. Refs.~\cite{wong1994intr, Miller:2007ri}. Consider first a nucleon-nucleus ($N$+$A$) collision at an impact parameter \textbf{b}. In the eikonal high collision-energy limit the number of binary inelastic collisions is given by 
\begin{equation}
N_{bin}^{NA}(\mathbf{b}) = T_A(\mathbf{b})\sigma_{inel}^{NN},
\label{eq:nbinNA}
\end{equation}
where $T_A(\mathbf{b})$ is the thickness function defined in Eq.~(\ref{eq:T_A}) and $\sigma_{inel}^{NN}$ is the inelastic nucleon-nucleon cross section. One may interpret $N_{bin}^{NA}(\mathbf{b})/A$ as the probability for an inelastic collision to take place in the $A$ $NN$ collisions that are possible. Consequently, the probability for having no inelastic collisions at all, is 
\begin{equation}
p_{0}(\mathbf{b}) = \left( 1 - \dfrac{1}{A} T_A(\mathbf{b})\sigma_{inel}^{NN} \right)^A\stackrel{A\gg1}{\approx} \mathrm{e}^{-T_A(\mathbf{b})\sigma_{inel}^{NN}},
\end{equation}
and the probabililty for at least one inelastic collision becomes
\begin{equation}
p_{inel}^{NA}(\mathbf{b}) = 1 - p_0(\mathbf{b}) \approx 1 - \mathrm{e}^{-T_A(\mathbf{b})\sigma_{inel}^{NN}}. 
\label{eq:p_inelNA}
\end{equation}
The inelastic cross section for the $N$+$A$ collision we then obtain as 
\begin{equation}
\sigma_{inel}^{NA} = \int {\rm d^2}\mathbf{b}\, p_{inel}^{NA}(\mathbf{b}) 
=  \int {\rm d^2}\mathbf{b}\, \big( 1 - \mathrm{e}^{-T_A(\mathbf{b})\sigma_{inel}^{NN}}\big).
\label{eq:sigma_inelNA}
\end{equation}
As the probability distribution above is expressable in terms of Poissonian probabilities for $n$ inelastic collisions,  
\begin{equation}
1 - \mathrm{e}^{-N_{bin}^{NA}(\mathbf{b})} = \sum_{n=1}^{\infty}\mathrm{e}^{-N_{bin}^{NA} (\mathbf{b})}\frac{[N_{bin}^{NA} (\mathbf{b})]^n}{n!} \equiv  \sum_{n=1}^{\infty} P(n|N_{bin}^{NA} (\mathbf{b})),  
\end{equation}
at a fixed impact parameter we indeed have 
\begin{equation}
\langle n\rangle \equiv  \sum_{n=1}^{\infty} n P(n|N_{bin}^{NA} (\mathbf{b})) = N_{bin}^{NA} (\mathbf{b}).
\end{equation}

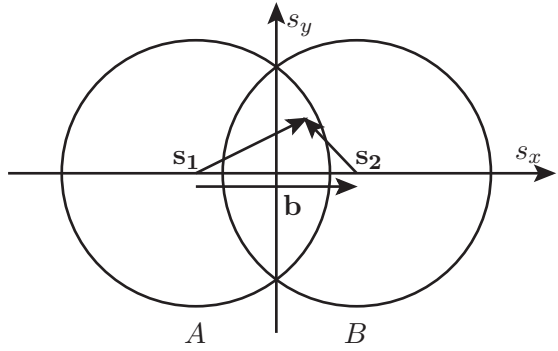
\begin{figure}[htb!]
\begin{center}
\begin{picture}(200,120)(0,0)
\SetWidth{1.0}
\SetColor{Black}
\LongArrow(100,0)(100,120)
\LongArrow(0,60)(200,60)
\LongArrow(130,60)(113,78)
\LongArrow(70,60)(108,79)
\LongArrow(70,55)(126,55)
\Arc(130,60)(50,0,360)
\Arc(70,60)(50,0,360)
\Text(110,52)[rt]{$\mathbf{b}$}
\Text(72,62)[rb]{$\mathbf{s_1}$}
\Text(130,62)[lb]{$\mathbf{s_2}$}
\Text(200,64)[rb]{$s_x$}
\Text(104,120)[lt]{$s_y$}
\Text(70,-4)[cb]{$A$}
\Text(130,-4)[cb]{$B$}
\end{picture}
\caption{Collision geometry in the transverse plane of the two colliding nuclei.}
\label{fig:transversePlane}
\end{center}
\end{figure}

\begin{figure}[tbhp]
\begin{center}
\includegraphics[width=10cm]{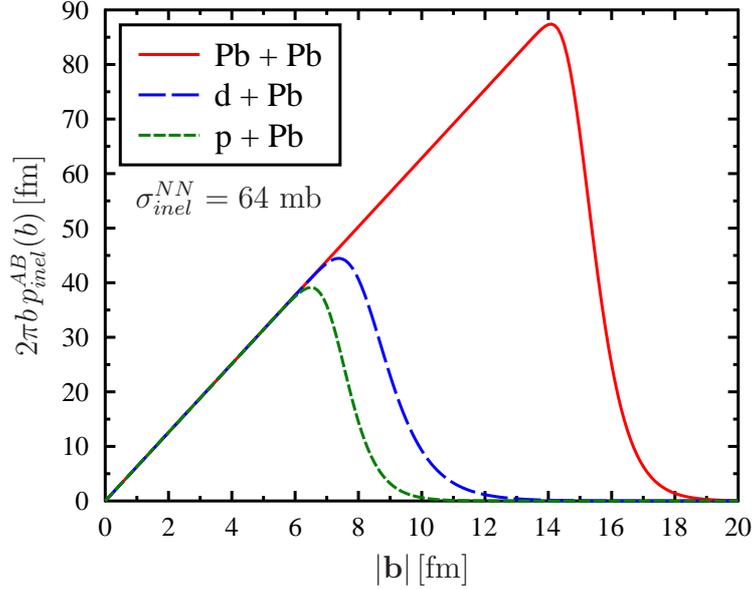}
\caption{The differential inelastic cross section $d\sigma_{inel}^{AB}/db$ as a function of impact parameter $b$
for lead-lead (red solid), deuteron-lead (blue long-dashed), and proton-lead (green dashed) collisions with $\sigma_{inel}^{NN} = 64 \text{ mb}$.}
\label{fig:dsigma_db}
\end{center}
\end{figure}

Let us then consider a nucleus-nucleus ($A$+$B$) collision with collision geometry as in Fig.~\ref{fig:transversePlane}.
A conveninent choice is to take the impact parameter $\mathbf{b}$ along the $x$ axis symmetrically around the origin.
The transverse density of interacting matter at certain impact parameter $\mathbf{b}$ can then be computed from the nuclear overlap function, defined as 
\begin{equation}
T_{AB}(\mathbf{b}) = \int \mathrm{d}^2 \mathbf{s} \,T_A(\mathbf{s_1}) \, T_B(\mathbf{s_2}),
\label{eq:taa}
\end{equation}
where $\mathbf{s_1} = \mathbf{s} + \mathbf{b}/2$ and $\mathbf{s_2} = \mathbf{s} - \mathbf{b}/2$. 
With the normalization for $T_A(\mathbf{s})$ in Eq.~(\ref{eq:ta_normalization}), we have
\begin{equation}
\int \mathrm{d}^2 \mathbf{b}\, T_{AB}(\mathbf{b}) = AB.
\end{equation}
The number of binary collisions at a given impact parameter $\mathbf{b}$ is now
\begin{equation}
N_{bin}^{AB}(\mathbf{b}) = T_{AB}(\mathbf{b})\sigma_{inel}^{NN}.
\label{eq:ncollAB}
\end{equation}
Analogously to the $N$+$A$ case above (see e.g. \cite{wong1994intr}), we may write the probability of an inelastic interaction in an $A$+$B$ collision at an impact parameter $\mathbf{b}$ as 
\begin{equation}
p_{inel}^{AB}(\mathbf{b}) \approx 1 - \mathrm{e}^{-T_{AB}(\mathbf{b})\sigma_{inel}^{NN}},
\label{eq:p_inelAB}
\end{equation}
and the inelastic cross section becomes 
\begin{equation}
\sigma_{inel}^{AB} = \int {\rm d^2}\mathbf{b}\, p_{inel}^{AB}(\mathbf{b}) 
=  \int {\rm d^2}\mathbf{b}\, \big(1 - \mathrm{e}^{-T_{AB}(\mathbf{b})\sigma_{inel}^{NN}}\big).
\label{eq:sigma_inelAB}
\end{equation}
Figure \ref{fig:dsigma_db} shows an example of the probability distributions $p_{inel}^{NA}(\mathbf{b})$ of Eq.~(\ref{eq:p_inelNA}) for p+Pb collisions, and $p_{inel}^{AB}(\mathbf{b})$ of Eq.~(\ref{eq:p_inelAB}) for Pb+Pb and d+Pb collisions for $\sigma_{inel}^{NN}=64$~mb, which corresponds to the cms-energy $\sqrt{s_{NN}}=2.76$~TeV at the LHC. 

The centrality classes in the optical Glauber model can be defined as impact-parameter intervals. The "0-$c$~\% 
central" $A$+$B$ collisions correspond to the most central collisions, $0\le b\le b_c$  which yield $c$~\% of the total inelastic cross section,
\begin{equation}
c\,\% = \dfrac{1}{\sigma_{inel}^{AB}} \int_0^{b_c} \mathrm{d}^2\mathbf{b} \,p_{inel}^{AB}(\mathbf{b})
\equiv 
\dfrac{\sigma_{inel}^{AB}(0,b_c)}{\sigma_{inel}^{AB}}.
\label{eq:centrality}
\end{equation}
The $c_1$-$c_2$~\% centrality class then corresponds to an interval $[b_1,b_2]$ for which
\begin{equation}
(c_2-c_1)\,\% = \dfrac{1}{\sigma_{inel}^{AB}} \int_{b_1}^{b_2} \mathrm{d}^2\mathbf{b} \,p_{inel}^{AB}(\mathbf{b})
= 
\dfrac{\sigma_{inel}^{AB}(b_1,b_2)}{\sigma_{inel}^{AB}}.
\end{equation}
For the studies of different hard-process nuclear modification factors in Sec.~4, we also define the average number of binary collisions in a $c_1$-$c_2$~\% centrality class: 
\begin{equation}
\langle N_{bin}\rangle_{b_1,b_2}^{AB} \equiv \dfrac{\int_{b_1}^{b_2} N_{bin}^{AB}(\mathbf{b})}{\sigma_{inel}^{AB}(b_1,b_2)}
= \dfrac{\int_{b_1}^{b_2} \mathrm{d}^2 \mathbf{b} \, T_{AB}(\mathbf{b})\sigma_{inel}^{NN}}{\int_{b_1}^{b_2} \mathrm{d}^2 \mathbf{b} \, \left[1 - \mathrm{e}^{-T_{AB}(\mathbf{b})\sigma_{inel}^{NN}} \right]},
\label{eq:Nbin_ave} 
\end{equation}
where the denominator is simply $(c_2-c_1)\,\%$ of $\sigma_{inel}^{AB}$. For discussing the analogous centrality classes in $N$+$A$ collisions, we just replace $AB$ by $NA$ in Eqs.~(\ref{eq:centrality}-\ref{eq:Nbin_ave}) above, and also $T_{AB}$ by $T_A$ in Eq.~(\ref{eq:Nbin_ave}).

\section{Nuclear modifications $r_{uv}$ and $r_{us}$}
\label{sec:ruvus}

\begin{figure}[hbtp]
\begin{center}
\includegraphics[width=7cm]{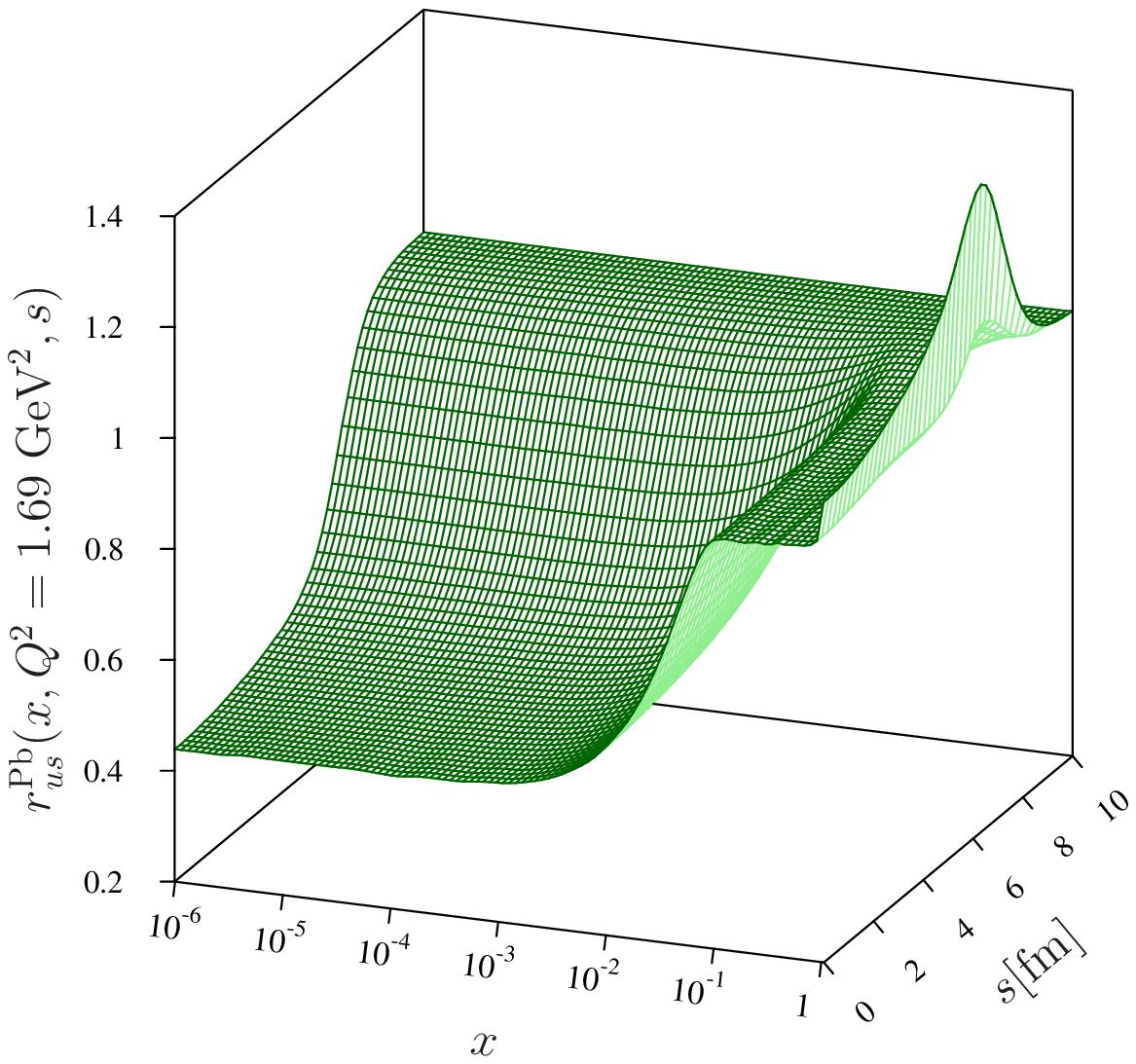}
\includegraphics[width=7cm]{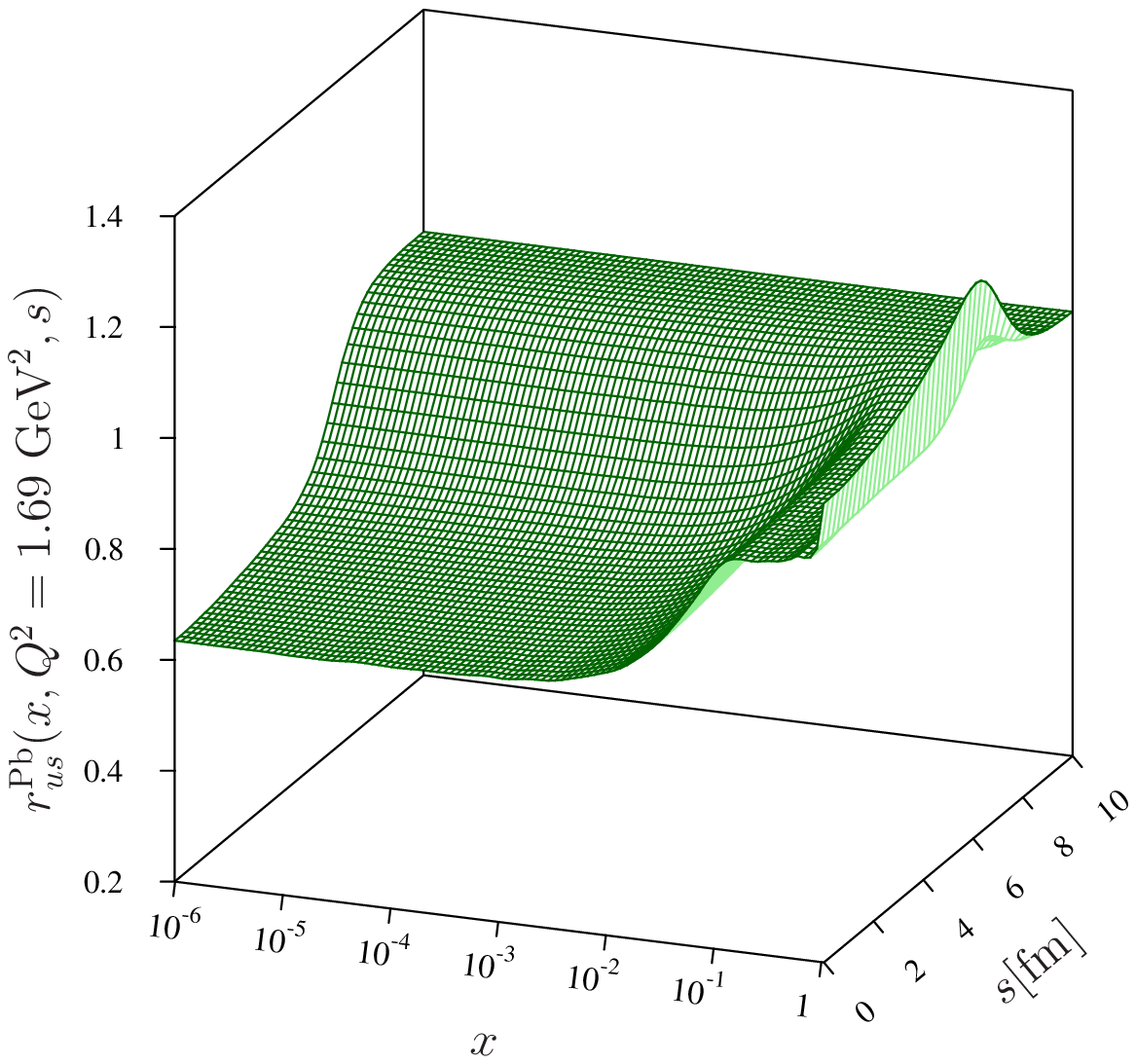}
\includegraphics[width=7cm]{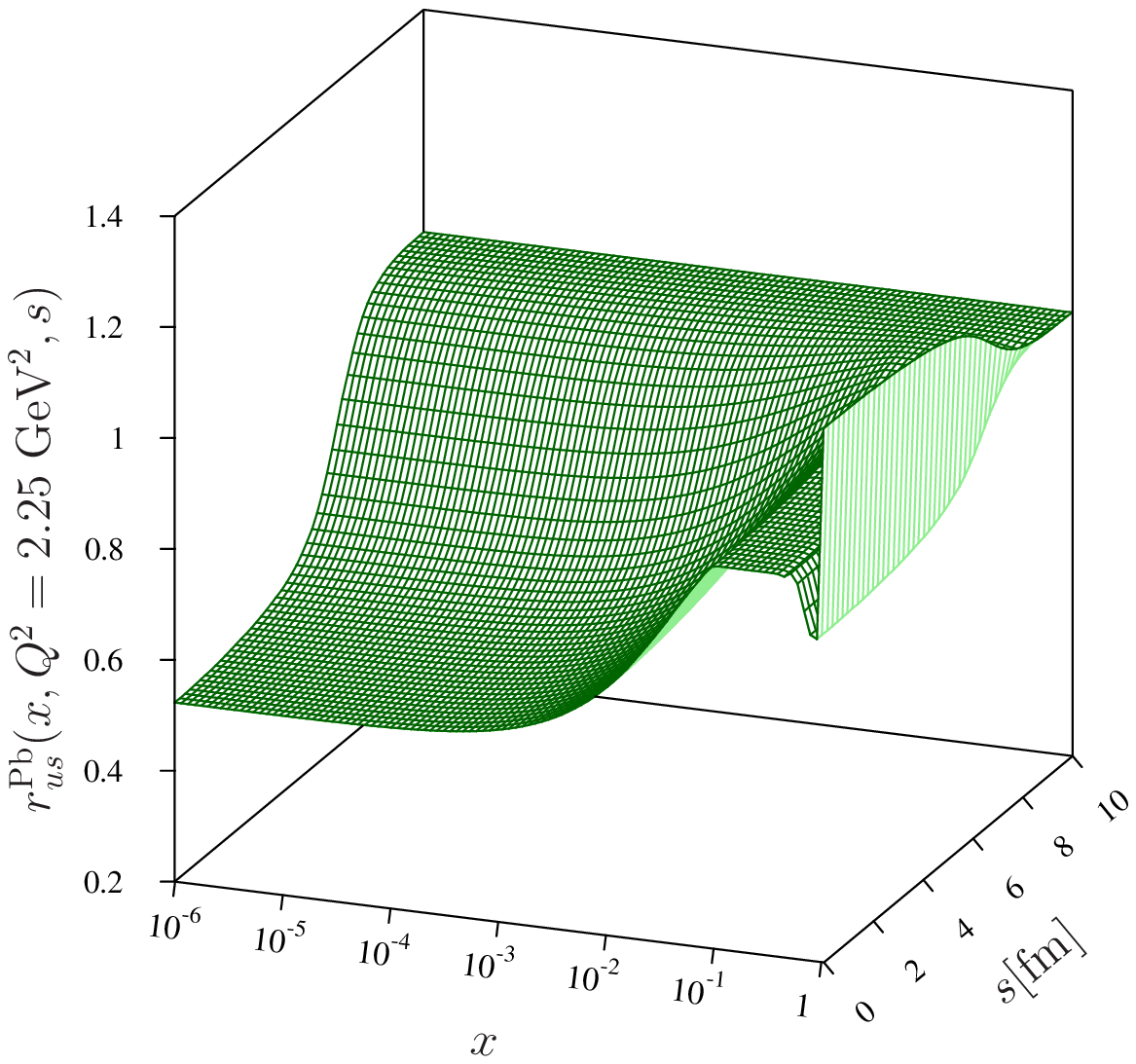}
\caption{The spatially dependent modification of the $\bar u$ distribution in a lead nucleus, $r_g^{\rm Pb}(x,Q^2,s)$, from EPS09sNLO1 (upper left), EPS09sLO1 (upper right) and EKS98s(lower plot) as a function of $x$ and $s$ at the initial scale $Q^2 = 1.69 (2.25) \textrm{ GeV}^2$ of EPS09 (EKS98).}
\label{fig:rxsus3d}
\end{center}
\vspace{-.5cm}
\end{figure}

\begin{figure}[thbp]
\begin{center}
\includegraphics[width=7cm]{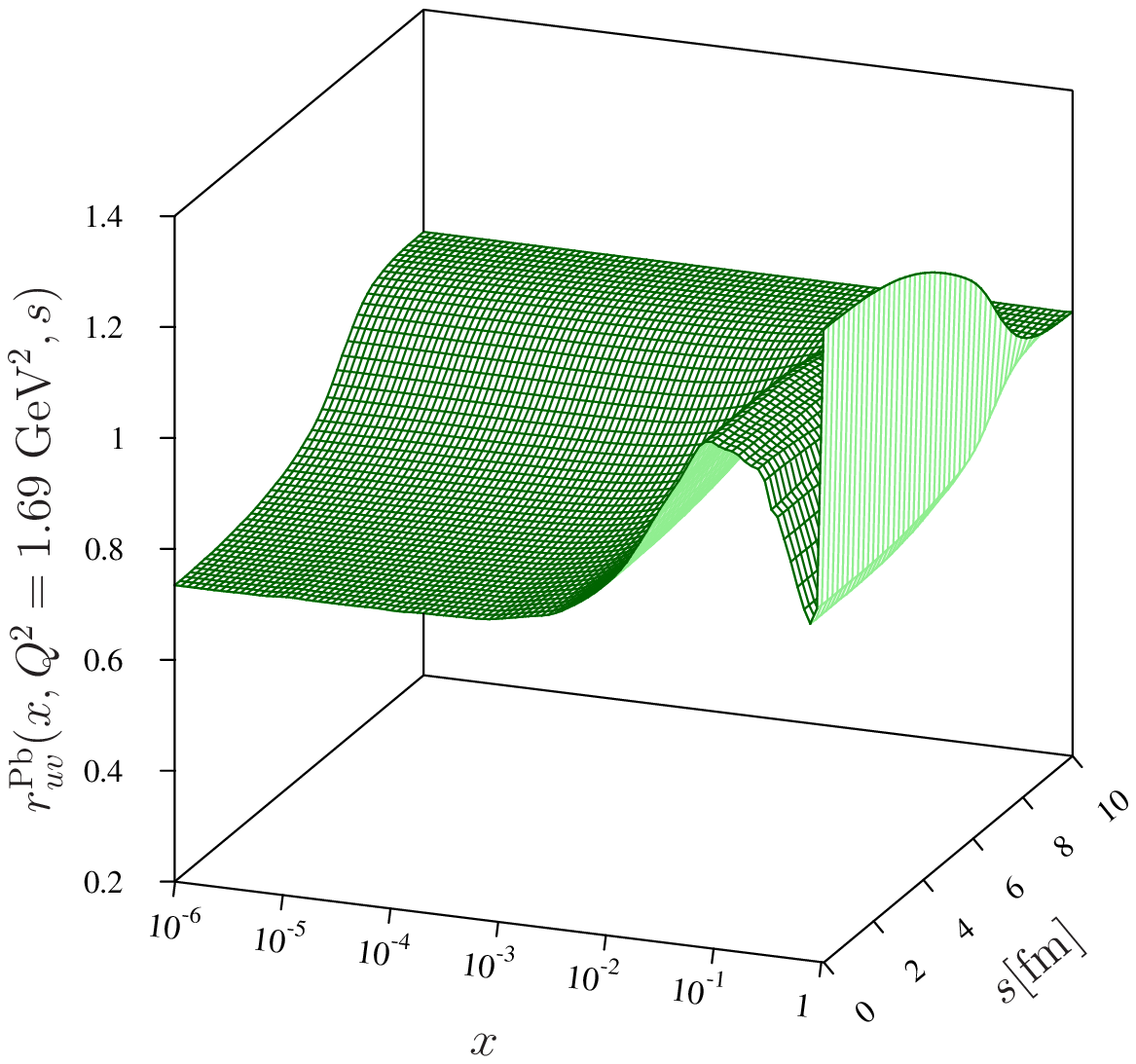}
\includegraphics[width=7cm]{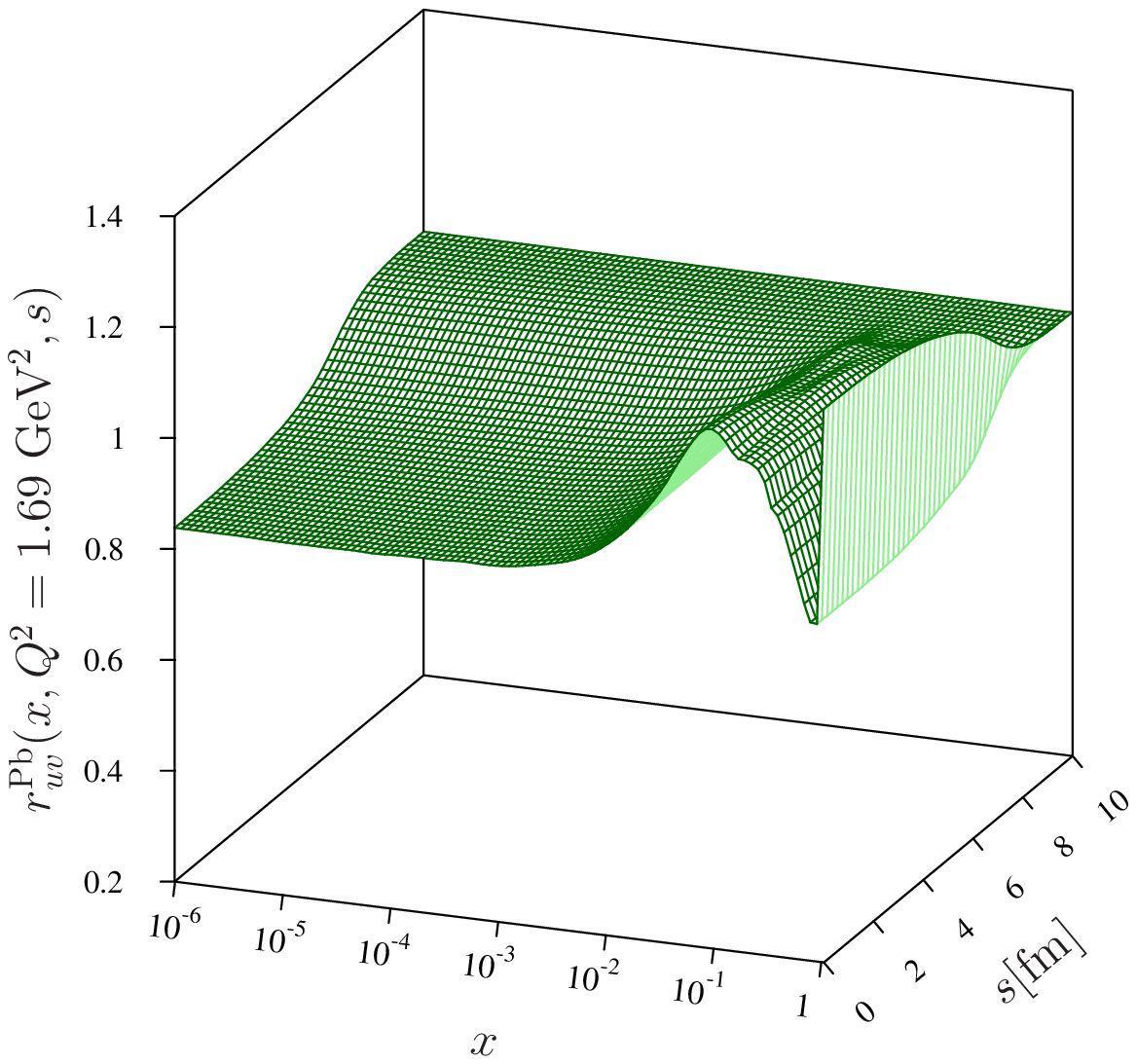}
\includegraphics[width=7cm]{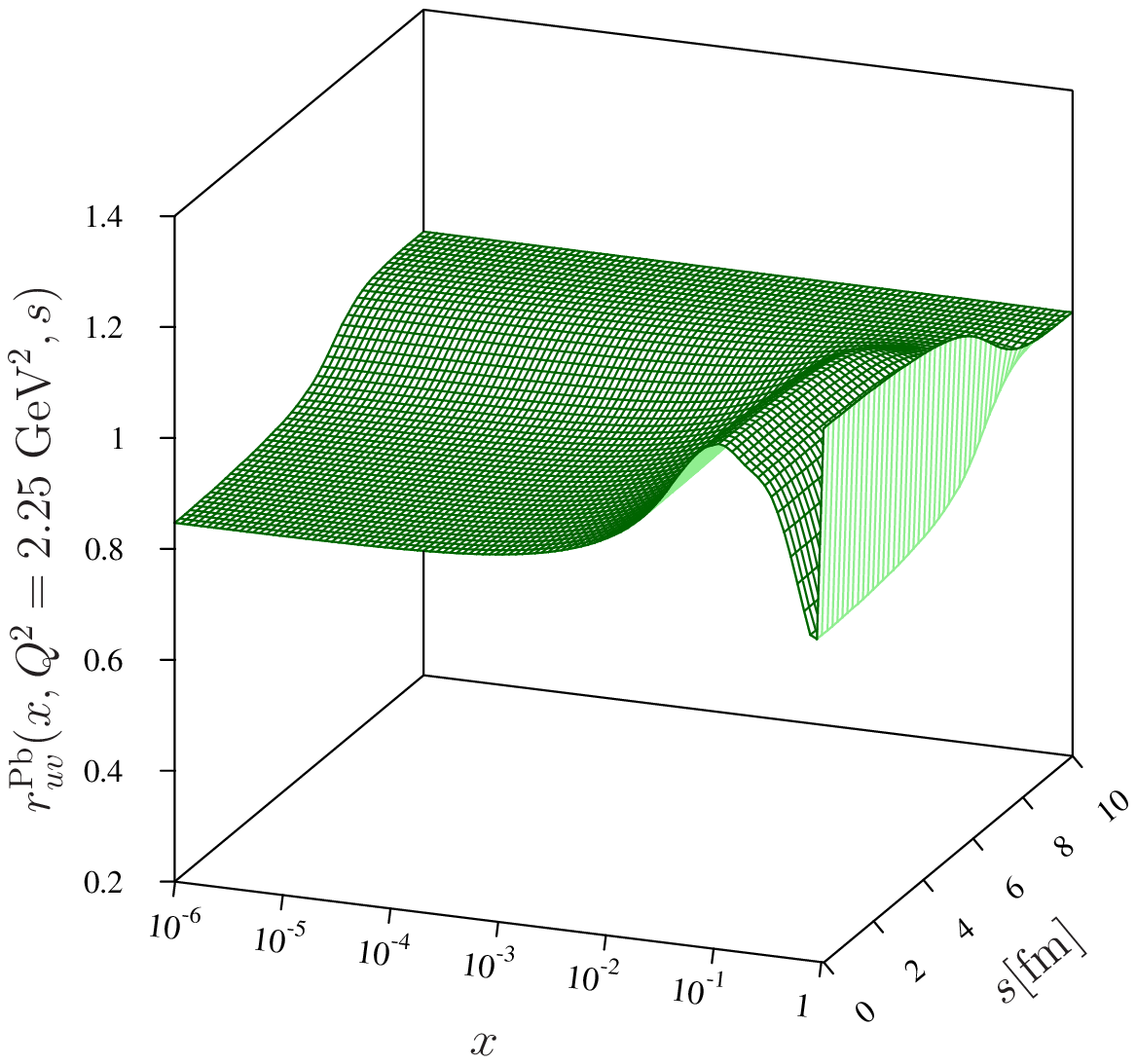}
\caption{The spatially dependent modification of the $u_V$ distribution in a lead nucleus, $r_g^{\rm Pb}(x,Q^2,s)$, from 
EPS09sNLO1 (upper left), EPS09sLO1 (upper right) and EKS98s(lower plot) as a function of $x$ and $s$ at the initial scale $Q^2 = 1.69\, (2.25) \textrm{ GeV}^2$ of EPS09 (EKS98).}
\label{fig:rxsuv3d}
\end{center}
\end{figure}

For completeness, we plot here the nuclear modifications from our fits EPS09sNLO1, EPS09sLO1 and EKS98s in a lead nucleus 
for the $u$ valence quarks in Fig.~\ref{fig:rxsus3d} and $u$ sea quarks in Fig.~\ref{fig:rxsuv3d}. The corresponding modifications for gluons are shown in Fig.~\ref{fig:rxsg3d}.
\vspace{-.0cm}

\newpage
\bibliographystyle{JHEP}
\bibliography{ipdnpdf}

\providecommand{\href}[2]{#2}\begingroup\raggedright\begin{thebibliography}{10}

\bibitem{Collins:1989gx}
J.~C. Collins, D.~E. Soper, and G.~F. Sterman, {\it {Factorization of Hard
  Processes in QCD}},  {\em Adv.Ser.Direct.High Energy Phys.} {\bf 5} (1988)
  1--91, [\href{http://xxx.lanl.gov/abs/hep-ph/0409313}{{\tt hep-ph/0409313}}].
  To be publ. in 'Perturbative QCD' (A.H. Mueller, ed.) (World Scientific
  Publ., 1989).

\bibitem{Brock:1993sz}
{\bf CTEQ Collaboration} Collaboration, R.~Brock et~al., {\it {Handbook of
  perturbative QCD: Version 1.0}},  {\em Rev.Mod.Phys.} {\bf 67} (1995)
  157--248.

\bibitem{Dokshitzer:1977sg}
Y.~L. Dokshitzer, {\it {Calculation of the Structure Functions for Deep
  Inelastic Scattering and e+ e- Annihilation by Perturbation Theory in Quantum
  Chromodynamics.}},  {\em Sov.Phys.JETP} {\bf 46} (1977) 641--653.

\bibitem{Gribov:1972ri}
V.~N. Gribov and L.~N. Lipatov, {\it {Deep inelastic e p scattering in
  perturbation theory}},  {\em Sov.J.Nucl.Phys.} {\bf 15} (1972) 438--450.

\bibitem{Gribov:1972rt}
V.~N. Gribov and L.~N. Lipatov, {\it {e+ e- pair annihilation and deep
  inelastic e p scattering in perturbation theory}},  {\em Sov.J.Nucl.Phys.}
  {\bf 15} (1972) 675--684.

\bibitem{Altarelli:1977zs}
G.~Altarelli and G.~Parisi, {\it {Asymptotic Freedom in Parton Language}},
  {\em Nucl.Phys.} {\bf B126} (1977) 298.

\bibitem{Lai:2010vv}
H.-L. Lai, M.~Guzzi, J.~Huston, Z.~Li, P.~M. Nadolsky, et~al., {\it {New parton
  distributions for collider physics}},  {\em Phys.Rev.} {\bf D82} (2010)
  074024, [\href{http://xxx.lanl.gov/abs/1007.2241}{{\tt arXiv:1007.2241}}].

\bibitem{Martin:2009iq}
A.~D. Martin, W.~J. Stirling, R.~S. Thorne, and G.~Watt, {\it {Parton
  distributions for the LHC}},  {\em Eur.Phys.J.} {\bf C63} (2009) 189--285,
  [\href{http://xxx.lanl.gov/abs/0901.0002}{{\tt arXiv:0901.0002}}].

\bibitem{Ball:2010de}
R.~D. Ball, L.~Del~Debbio, S.~Forte, A.~Guffanti, J.~I. Latorre, et~al., {\it
  {A first unbiased global NLO determination of parton distributions and their
  uncertainties}},  {\em Nucl.Phys.} {\bf B838} (2010) 136--206,
  [\href{http://xxx.lanl.gov/abs/1002.4407}{{\tt arXiv:1002.4407}}].

\bibitem{Eskola:1998df}
K.~J. Eskola, V.~J. Kolhinen, and C.~A. Salgado, {\it {The Scale dependent
  nuclear effects in parton distributions for practical applications}},  {\em
  Eur.Phys.J.} {\bf C9} (1999) 61--68,
  [\href{http://xxx.lanl.gov/abs/hep-ph/9807297}{{\tt hep-ph/9807297}}].

\bibitem{Hirai:2001np}
M.~Hirai, S.~Kumano, and M.~Miyama, {\it {Determination of nuclear parton
  distributions}},  {\em Phys.Rev.} {\bf D64} (2001) 034003,
  [\href{http://xxx.lanl.gov/abs/hep-ph/0103208}{{\tt hep-ph/0103208}}].

\bibitem{Hirai:2004wq}
M.~Hirai, S.~Kumano, and T.~H. Nagai, {\it {Nuclear parton distribution
  functions and their uncertainties}},  {\em Phys.Rev.} {\bf C70} (2004)
  044905, [\href{http://xxx.lanl.gov/abs/hep-ph/0404093}{{\tt
  hep-ph/0404093}}].

\bibitem{deFlorian:2003qf}
D.~de~Florian and R.~Sassot, {\it {Nuclear parton distributions at
  next-to-leading order}},  {\em Phys.Rev.} {\bf D69} (2004) 074028,
  [\href{http://xxx.lanl.gov/abs/hep-ph/0311227}{{\tt hep-ph/0311227}}].

\bibitem{Hirai:2007sx}
M.~Hirai, S.~Kumano, and T.~H. Nagai, {\it {Determination of nuclear parton
  distribution functions and their uncertainties in next-to-leading order}},
  {\em Phys.Rev.} {\bf C76} (2007) 065207,
  [\href{http://xxx.lanl.gov/abs/0709.3038}{{\tt arXiv:0709.3038}}].

\bibitem{Eskola:2009uj}
K.~J. Eskola, H.~Paukkunen, and C.~A. Salgado, {\it {EPS09 - a New Generation
  of NLO and LO Nuclear Parton Distribution Functions}},  {\em JHEP} {\bf 04}
  (2009) 065, [\href{http://xxx.lanl.gov/abs/0902.4154}{{\tt
  arXiv:0902.4154}}].

\bibitem{Schienbein:2009kk}
I.~Schienbein, J.~Y. Yu, K.~Kovarik, C.~Keppel, J.~G. Morfin, et~al., {\it {PDF
  Nuclear Corrections for Charged and Neutral Current Processes}},  {\em
  Phys.Rev.} {\bf D80} (2009) 094004,
  [\href{http://xxx.lanl.gov/abs/0907.2357}{{\tt arXiv:0907.2357}}].

\bibitem{Kovarik:2010uv}
K.~Kovarik, I.~Schienbein, F.~I. Olness, J.~Y. Yu, C.~Keppel, et~al., {\it
  {Nuclear corrections in neutrino-nucleus DIS and their compatibility with
  global NPDF analyses}},  {\em Phys.Rev.Lett.} {\bf 106} (2011) 122301,
  [\href{http://xxx.lanl.gov/abs/1012.0286}{{\tt arXiv:1012.0286}}].

\bibitem{deFlorian:2011fp}
D.~de~Florian, R.~Sassot, P.~Zurita, and M.~Stratmann, {\it {Global Analysis of
  Nuclear Parton Distributions}},
  \href{http://xxx.lanl.gov/abs/1112.6324}{{\tt arXiv:1112.6324}}. 23 pages, 23
  figures.

\bibitem{Eskola:1991ec}
K.~J. Eskola, {\it {Shadowing effects on quark and gluon production in
  ultrarelativistic heavy ion collisions}},  {\em Z.Phys.} {\bf C51} (1991)
  633--642.

\bibitem{Emel'yanov:1999bn}
V.~Emel'yanov, A.~Khodinov, S.~R. Klein, and R.~Vogt, {\it {The Effect of
  shadowing on initial conditions, transverse energy and hard probes in
  ultrarelativistic heavy ion collisions}},  {\em Phys.Rev.} {\bf C61} (2000)
  044904, [\href{http://xxx.lanl.gov/abs/hep-ph/9909427}{{\tt
  hep-ph/9909427}}].

\bibitem{Klein:2003dj}
S.~R. Klein and R.~Vogt, {\it {Inhomogeneous shadowing effects on J / psi
  production in dA collisions}},  {\em Phys.Rev.Lett.} {\bf 91} (2003) 142301,
  [\href{http://xxx.lanl.gov/abs/nucl-th/0305046}{{\tt nucl-th/0305046}}].

\bibitem{Vogt:2004hf}
R.~Vogt, {\it {Shadowing effects on the nuclear suppression factor, R (dAu) ,
  in d+Au interactions}},  {\em Phys.Rev.} {\bf C70} (2004) 064902.

\bibitem{Frankfurt:2011cs}
L.~Frankfurt, V.~Guzey, and M.~Strikman, {\it {Leading twist nuclear shadowing
  phenomena in hard processes with nuclei}},  {\em Phys.Rept.} {\bf 512} (2012)
  255--393, [\href{http://xxx.lanl.gov/abs/1106.2091}{{\tt arXiv:1106.2091}}].

\bibitem{Arsene:2004ux}
{\bf BRAHMS Collaboration} Collaboration, I.~Arsene et~al., {\it {On the
  evolution of the nuclear modification factors with rapidity and centrality in
  d + Au collisions at s(NN)**(1/2) = 200-GeV}},  {\em Phys.Rev.Lett.} {\bf 93}
  (2004) 242303, [\href{http://xxx.lanl.gov/abs/nucl-ex/0403005}{{\tt
  nucl-ex/0403005}}]. Four pages, four figures. Published in PRL. Figures 1 and
  2 have been updated, and several changes made to the text Journal-ref:
  Phys.Rev.Lett. 93 (2004) 242303.

\bibitem{Adams:2004dv}
{\bf STAR Collaboration} Collaboration, J.~Adams et~al., {\it {Pseudorapidity
  asymmetry and centrality dependence of charged hadron spectra in d + Au
  collisions at S(NN)**(1/2) = 200-GeV}},  {\em Phys.Rev.} {\bf C70} (2004)
  064907, [\href{http://xxx.lanl.gov/abs/nucl-ex/0408016}{{\tt
  nucl-ex/0408016}}].

\bibitem{Adler:2006wg}
{\bf PHENIX Collaboration} Collaboration, S.~S. Adler et~al., {\it {Centrality
  dependence of pi0 and eta production at large transverse momentum in
  s(NN)**(1/2) = 200-GeV d+Au collisions}},  {\em Phys.Rev.Lett.} {\bf 98}
  (2007) 172302, [\href{http://xxx.lanl.gov/abs/nucl-ex/0610036}{{\tt
  nucl-ex/0610036}}].

\bibitem{Adler:2006xd}
{\bf PHENIX Collaboration} Collaboration, S.~S. Adler et~al., {\it {Nuclear
  effects on hadron production in d = Au and p + p collisions at s(NN)**(1/2) =
  200-GeV}},  {\em Phys.Rev.} {\bf C74} (2006) 024904,
  [\href{http://xxx.lanl.gov/abs/nucl-ex/0603010}{{\tt nucl-ex/0603010}}].

\bibitem{Adams:2006nd}
{\bf STAR Collaboration} Collaboration, J.~Adams et~al., {\it {Identified
  hadron spectra at large transverse momentum in p+p and d+Au collisions at
  s(NN)**(1/2) = 200-GeV}},  {\em Phys.Lett.} {\bf B637} (2006) 161--169,
  [\href{http://xxx.lanl.gov/abs/nucl-ex/0601033}{{\tt nucl-ex/0601033}}].

\bibitem{Adler:2007aa}
{\bf PHENIX Collaboration} Collaboration, S.~S. Adler et~al., {\it {Centrality
  dependence of charged hadron production in deuteron + gold and nucleon + gold
  collisions at s(NN)**(1/2) = 200-GeV}},  {\em Phys.Rev.} {\bf C77} (2008)
  014905, [\href{http://xxx.lanl.gov/abs/0708.2416}{{\tt arXiv:0708.2416}}].
  330 authors, 15 pages text, 16 figures, 3 tables. Submitted to Phys. Rev.
  Lett. Plain text data tables for the points plotted in figures for this and
  previous PHENIX publications are (or will be) publicly available at
  http://www.phenix.bnl.gov/papers.html.

\bibitem{Abelev:2009hx}
{\bf STAR Collaboration} Collaboration, B.~I. Abelev et~al., {\it {Inclusive
  pi0, eta, and direct photon production at high transverse momentum in p+p and
  d+Au collisions at sqrt(sNN) = 200 GeV}},  {\em Phys.Rev.} {\bf C81} (2010)
  064904, [\href{http://xxx.lanl.gov/abs/0912.3838}{{\tt arXiv:0912.3838}}].

\bibitem{Adare:2010fn}
{\bf PHENIX Collaboration} Collaboration, A.~Adare et~al., {\it {Cold Nuclear
  Matter Effects on $J/\psi$ Yields as a Function of Rapidity and Nuclear
  Geometry in Deuteron-Gold Collisions at $\sqrt{s_{NN}}=200$ GeV}},  {\em
  Phys.Rev.Lett.} {\bf 107} (2011) 142301,
  [\href{http://xxx.lanl.gov/abs/1010.1246}{{\tt arXiv:1010.1246}}].

\bibitem{Adare:2012qf}
A.~Adare, S.~Afanasiev, C.~Aidala, N.~Ajitanand, Y.~Akiba, et~al., {\it
  {Transverse-Momentum Dependence of the J/psi Nuclear Modification in d+Au
  Collisions at sqrt(sNN)=200 GeV}},
  \href{http://xxx.lanl.gov/abs/1204.0777}{{\tt arXiv:1204.0777}}.

\bibitem{Nagle:2010ix}
J.~L. Nagle, A.~D. Frawley, L.~A.~L. Levy, and M.~G. Wysocki, {\it {Theoretical
  Modeling of J/psi Yield Modifications in Proton (Deuteron) - Nucleus
  Collisions at High Energy}},  {\em Phys.Rev.} {\bf C84} (2011) 044911,
  [\href{http://xxx.lanl.gov/abs/1011.4534}{{\tt arXiv:1011.4534}}].

\bibitem{downloadpage}
\url{https://www.jyu.fi/fysiikka/en/research/highenergy/urhic/}.

\bibitem{QuirogaArias:2010wh}
P.~Quiroga-Arias, J.~G. Milhano, and U.~A. Wiedemann, {\it {Testing nuclear
  parton distributions with pA collisions at the TeV scale}},  {\em Phys.Rev.}
  {\bf C82} (2010) 034903, [\href{http://xxx.lanl.gov/abs/1002.2537}{{\tt
  arXiv:1002.2537}}].

\bibitem{Albacete:2010bs}
J.~L. Albacete and C.~Marquet, {\it {Single Inclusive Hadron Production at RHIC
  and the LHC from the Color Glass Condensate}},  {\em Phys.Lett.} {\bf B687}
  (2010) 174--179, [\href{http://xxx.lanl.gov/abs/1001.1378}{{\tt
  arXiv:1001.1378}}].

\bibitem{Salgado:2011wc}
C.~A. Salgado, J.~Alvarez-Muniz, F.~Arleo, N.~Armesto, M.~Botje, et~al., {\it
  {Proton-Nucleus Collisions at the LHC: Scientific Opportunities and
  Requirements}},  {\em J.Phys.G} {\bf G39} (2012) 015010,
  [\href{http://xxx.lanl.gov/abs/1105.3919}{{\tt arXiv:1105.3919}}].

\bibitem{Arleo:2011gc}
F.~Arleo, K.~J. Eskola, H.~Paukkunen, and C.~A. Salgado, {\it {Inclusive prompt
  photon production in nuclear collisions at RHIC and LHC}},  {\em JHEP} {\bf
  1104} (2011) 055, [\href{http://xxx.lanl.gov/abs/1103.1471}{{\tt
  arXiv:1103.1471}}].

\bibitem{Barnafoldi:2011px}
G.~G. Barnafoldi, J.~Barrette, M.~Gyulassy, P.~Levai, and V.~Topor~Pop, {\it
  {Predictions for p+Pb at 4.4A TeV to Test Initial State Nuclear Shadowing at
  energies available at the CERN Large Hadron Collider}},  {\em Phys.Rev.} {\bf
  C85} (2012) 024903, [\href{http://xxx.lanl.gov/abs/1111.3646}{{\tt
  arXiv:1111.3646}}]. Revised version accepted for publication/ Phys. Rev. C,
  in press, 16 pages, 4 figures, text modifications, added references, new
  figure 4, revtex4.

\bibitem{Xu:2012au}
R.~Xu, W.-T. Deng, and X.-N. Wang, {\it {Nuclear modification of high-pT hadron
  spectra in p+A collisions at LHC}},
  \href{http://xxx.lanl.gov/abs/1204.1998}{{\tt arXiv:1204.1998}}.

\bibitem{Gyulassy:1994ew}
M.~Gyulassy and X.-N. Wang, {\it {HIJING 1.0: A Monte Carlo program for parton
  and particle production in high-energy hadronic and nuclear collisions}},
  {\em Comput.Phys.Commun.} {\bf 83} (1994) 307,
  [\href{http://xxx.lanl.gov/abs/nucl-th/9502021}{{\tt nucl-th/9502021}}]. 38
  pages in LaTex, published in Comp. Phys. Comm. 83, 307 (1994). Report-no:
  LBL-34246.

\bibitem{Ferreiro:2009ur}
E.~G. Ferreiro, F.~Fleuret, J.~P. Lansberg, and A.~Rakotozafindrabe, {\it
  {Centrality, Rapidity and Transverse-Momentum Dependence of Cold Nuclear
  Matter Effects on J/Psi Production in d Au, Cu Cu and Au Au Collisions at
  s(NN)**(1/2) = 200 GeV}},  {\em Phys.Rev.} {\bf C81} (2010) 064911,
  [\href{http://xxx.lanl.gov/abs/0912.4498}{{\tt arXiv:0912.4498}}]. 12 pages,
  14 figures, LaTeX. Version to appear in Phys. Rev. C: a few typos corrected
  and one comment about the EPS08 nPDF parametrisation added.

\bibitem{Ferreiro:2012zb}
E.~G. Ferreiro, F.~Fleuret, J.~P. Lansberg, N.~Matagne, and
  A.~Rakotozafindrabe, {\it {Centrality, rapidity, and transverse-momentum
  dependence of gluon shadowing and antishadowing on $J/\psi$ production in
  $d$Au collisions at $\sqrt{s}$=200 GeV}},
  \href{http://xxx.lanl.gov/abs/1201.5574}{{\tt arXiv:1201.5574}}.

\bibitem{Alvero:1998bz}
L.~Alvero, L.~L. Frankfurt, and M.~I. Strikman, {\it {Diffractive production of
  charm and gluon nuclear shadowing}},  {\em Eur.Phys.J.} {\bf A5} (1999)
  97--104, [\href{http://xxx.lanl.gov/abs/hep-ph/9810331}{{\tt
  hep-ph/9810331}}].

\bibitem{Guzey:2009jr}
V.~Guzey and M.~Strikman, {\it {Color fluctuation approximation for multiple
  interactions in leading twist theory of nuclear shadowing}},  {\em
  Phys.Lett.} {\bf B687} (2010) 167--173,
  [\href{http://xxx.lanl.gov/abs/0908.1149}{{\tt arXiv:0908.1149}}].

\bibitem{Eskola:2002kv}
K.~J. Eskola and H.~Honkanen, {\it {A Perturbative QCD analysis of charged
  particle distributions in hadronic and nuclear collisions}},  {\em
  Nucl.Phys.} {\bf A713} (2003) 167--187,
  [\href{http://xxx.lanl.gov/abs/hep-ph/0205048}{{\tt hep-ph/0205048}}].

\bibitem{Pumplin:2002vw}
J.~Pumplin, D.~R. Stump, J.~Huston, H.~L. Lai, P.~M. Nadolsky, et~al., {\it
  {New generation of parton distributions with uncertainties from global QCD
  analysis}},  {\em JHEP} {\bf 0207} (2002) 012,
  [\href{http://xxx.lanl.gov/abs/hep-ph/0201195}{{\tt hep-ph/0201195}}].

\bibitem{Hulthen}
L.~Hulthen and M.~Sugawara, {\em {Handbuch der Physik}}, vol.~39.
\newblock {Springer}, 1957.

\bibitem{Kniehl:2000fe}
B.~A. Kniehl, G.~Kramer, and B.~Potter, {\it {Fragmentation functions for
  pions, kaons, and protons at next-to-leading order}},  {\em Nucl.Phys.} {\bf
  B582} (2000) 514--536, [\href{http://xxx.lanl.gov/abs/hep-ph/0010289}{{\tt
  hep-ph/0010289}}].

\bibitem{Albino:2008fy}
S.~Albino, B.~A. Kniehl, and G.~Kramer, {\it {AKK Update: Improvements from New
  Theoretical Input and Experimental Data}},  {\em Nucl.Phys.} {\bf B803}
  (2008) 42--104, [\href{http://xxx.lanl.gov/abs/0803.2768}{{\tt
  arXiv:0803.2768}}].

\bibitem{deFlorian:2007aj}
D.~de~Florian, R.~Sassot, and M.~Stratmann, {\it {Global analysis of
  fragmentation functions for pions and kaons and their uncertainties}},  {\em
  Phys.Rev.} {\bf D75} (2007) 114010,
  [\href{http://xxx.lanl.gov/abs/hep-ph/0703242}{{\tt hep-ph/0703242}}].

\bibitem{incnlopage}
\url{http://lapth.in2p3.fr/PHOX_FAMILY/readme_inc.html}.

\bibitem{Aversa:1988vb}
F.~Aversa, P.~Chiappetta, M.~Greco, and J.~P. Guillet, {\it {QCD Corrections to
  Parton-Parton Scattering Processes}},  {\em Nucl. Phys.} {\bf B327} (1989)
  105.

\bibitem{Adams:2006uz}
{\bf STAR Collaboration} Collaboration, J.~Adams et~al., {\it {Forward neutral
  pion production in p+p and d+Au collisions at s(NN)**(1/2) = 200-GeV}},  {\em
  Phys.Rev.Lett.} {\bf 97} (2006) 152302,
  [\href{http://xxx.lanl.gov/abs/nucl-ex/0602011}{{\tt nucl-ex/0602011}}].

\bibitem{Adare:2011sc}
{\bf PHENIX Collaboration} Collaboration, A.~Adare et~al., {\it {Suppression of
  back-to-back hadron pairs at forward rapidity in $d+$Au Collisions at
  $\sqrt{s_{NN}}=200$ GeV}},  {\em Phys.Rev.Lett.} {\bf 107} (2011) 172301,
  [\href{http://xxx.lanl.gov/abs/1105.5112}{{\tt arXiv:1105.5112}}].

\bibitem{Antchev:2011vs}
G.~Antchev, P.~Aspell, I.~Atanassov, V.~Avati, J.~Baechler, et~al., {\it {First
  measurement of the total proton-proton cross section at the LHC energy of
  ${\sqrt{s} = 7 \text{ TeV}}$}},  {\em Europhys.Lett.} {\bf 96} (2011) 21002,
  [\href{http://xxx.lanl.gov/abs/1110.1395}{{\tt arXiv:1110.1395}}].

\bibitem{Eskola:1988yh}
K.~J. Eskola, K.~Kajantie, and J.~Lindfors, {\it {Quark and Gluon Production in
  High-Energy Nucleus-Nucleus Collisions}},  {\em Nucl.Phys.} {\bf B323} (1989)
  37.

\bibitem{Kharzeev:2002ei}
D.~Kharzeev, E.~Levin, and M.~Nardi, {\it {QCD saturation and deuteron nucleus
  collisions}},  {\em Nucl.Phys.} {\bf A730} (2004) 448--459,
  [\href{http://xxx.lanl.gov/abs/hep-ph/0212316}{{\tt hep-ph/0212316}}].

\bibitem{Miller:2007ri}
M.~L. Miller, K.~Reygers, S.~J. Sanders, and P.~Steinberg, {\it {Glauber
  modeling in high energy nuclear collisions}},  {\em Ann.Rev.Nucl.Part.Sci.}
  {\bf 57} (2007) 205--243,
  [\href{http://xxx.lanl.gov/abs/nucl-ex/0701025}{{\tt nucl-ex/0701025}}].

\bibitem{wong1994intr}
C.~Y. Wong, {\em {Introduction to High-Energy Heavy-Ion Collisions}}.
\newblock {World Scientific}, 1994.

\bibitem{Garcon:2001sz}
M.~Garcon and J.~W. Van~Orden, {\it {The Deuteron: Structure and
  form-factors}},  {\em Adv.Nucl.Phys.} {\bf 26} (2001) 293,
  [\href{http://xxx.lanl.gov/abs/nucl-th/0102049}{{\tt nucl-th/0102049}}].

\end{thebibliography}\endgroup

\end{document}